\documentclass[a4paper,useAMS,fleqn,usenatbib]{mnras}

\usepackage[T1]{fontenc}
\usepackage{ae,aecompl}

\usepackage{times}
\usepackage{graphicx}	% Including figure files
\usepackage{amsmath}	% Advanced maths commands
\usepackage{amssymb}	% Extra maths symbols

\usepackage{aas_macros}
\usepackage{tikz}
\usepackage{etoolbox}
\usetikzlibrary{shapes}
\usepackage{epsfig}
\usepackage{multirow}
\usepackage{booktabs}
\usepackage{xspace}
\usepackage{lscape}

\newcommand{\lbol}{L$_{\mathrm{bol}}$/L$_{\odot}$ }
\newcommand{\teff}{$T_{\mathrm{eff}}$\xspace}

\newcommand{\lii}{Li\,{\footnotesize I} }
\newcommand{\lsi}{$\mathrel{\hbox{\rlap{\hbox{\lower2pt\hbox{$\sim$}}}\raise2pt\hbox{$<$}}}$}
\newcommand{\gsi}{$\mathrel{\hbox{\rlap{\hbox{\lower2pt\hbox{$\sim$}}}\raise2pt\hbox{$>$}}}$}
\newcommand\nophot{ ~$\cdots$~ }
\newcommand{\ssquare}[1]{\tikz{\node[draw=#1,fill=#1,rectangle,minimum width=0.2cm,minimum height=0.2cm,inner sep=0pt] at (0,0) {};}}

\newcommand{\mycircle}[1]{\tikz{\node[draw=#1,fill=#1,circle,minimum
width=0.2cm,minimum height=0.2cm,inner sep=0pt] at (0,0) {};}}

\newcommand{\mytriangle}[1]{\tikz{\node[draw=#1,fill=#1,isosceles
triangle,isosceles triangle stretches,shape border rotate=90,minimum
width=0.2cm,minimum height=0.2cm,inner sep=0pt] at (0,0) {};}}

\newcommand{\tikzcircle}[2][red,fill=red]{\tikz[baseline=-0.5ex]\draw[#1,radius=#2] (0,0) circle ;}
\newrobustcmd*{\Montriangle}[1]{\tikz{\filldraw[draw=#1,fill=#1] (0,0) -- (0.2cm,0) -- (0.1cm,0.2cm);}}
\newrobustcmd*{\Monsquare}[1]{\tikz{\filldraw[draw=#1,fill=#1] (0,0) rectangle (0.2cm,0.2cm);}}
\newrobustcmd*{\Moncircle}[1]{\tikz{\filldraw[draw=#1,fill=#1] (0,0) circle [radius=0.1cm];}}
\title[T-Tauri stars in Chamaeleon and Rho Ophiuchus]{Fundamental stellar parameters 
for selected T-Tauri stars in the Chamaeleon and Rho Ophiuchus star-forming regions}

\author[D. J. James et al.]
{D.~J. James$^{1}$\thanks{Visiting astronomer, Cerro Tololo Inter-American Observatory, 
National Optical Astronomy Observatory, which is operated by the Association of 
Universities for Research in Astronomy (AURA) under a cooperative agreement with 
the National Science Foundation}, A.~N. Aarnio$^{2}$, A.~J.~W. 
Richert$^{3}$,  P.~A. Cargile$^{4}$ N.~C. Santos$^{5,6}$, C.~H.~F. Melo$^{7}$ and J. Bouvier$^{8}$ \\
$^{1}$Cerro Tololo InterAmerican Observatory, Casilla 603, La Serena, Chile \\
$^{2}$University of Michigan Department of Astronomy, 403 West Hall, 1085 S. 
University Ave, Ann Arbor, MI 48109, USA \\
$^{3}$Department of Astronomy and Astrophysics, The Pennsylvania State University, 525 
Davey Lab, University Park, PA 16802, USA \\
$^{4}$ Harvard-Smithsonian Center for Astrophysics, 60 Garden St, Cambridge, MA 02138, USA \\
$^{5}$ Instituto de Astrof\'isica e Ci\^encias do Espa\c{c}o, Universidade do Porto, CAUP, 
Rua das Estrelas, 4150-762 Porto, Portugal \\
$^{6}$ Departamento de F\'isica e Astronomia, Faculdade de Ci\^encias, Universidade do Porto, 
Rua do Campo Alegre, 4169-007 Porto, Portugal \\
$^{7}$European Southern Observatory (ESO), Casilla 19001, Santiago 19, Chile \\
$^{8}$Laboratoire d'Astrophysique, Observatoire de Grenoble, BP 53, 
38041 Grenoble, France}

\date{Accepted XXX. Received YYY; in original form ZZZ}

\pubyear{2016}

\begin{document}
\label{firstpage}
\pagerange{\pageref{firstpage}--\pageref{lastpage}}
\maketitle

\begin{abstract}

We present the results of an optical photometry and high-resolution 
spectroscopy campaign for a modest sample of X-ray selected stars in 
the Chamaeleon and Rho Ophiuchus star forming regions. With R$\sim$50000 
optical spectra, we establish kinematic membership of the parent association 
and confirm stellar youth for each star in our sample. With the acquisition 
of new standardized $BVIc$ photometry, in concert with near-infrared data 
from the literature, we derive age and mass from stellar positions in 
model-dependent Hertzsprung-Russell diagrams. We compare isochronal ages 
derived using colour-dependent extinction values finding that, within 
error bars, ages are the same irrespective of whether $E(B-V)$, $E(V-Ic)$, 
$E(J-H)$ or $E(H-K)$ is used to establish extinction, although model ages 
tend to be marginally younger for redder $E_{colour}$ values. For Cham I 
and $\eta$ Cham members we derive ages of \lsi 5-6 Myr, whereas our 
three $\eta$ Cha candidates are more consistent with a \gsi 25 Myr 
post-T Tauri star population. In Rho Ophiuchus, most stars in our sample 
have isochronal ages $<$10 Myr. Five objects show evidence of strong infrared 
excess ($Av>5$) in the 2MASS colour-colour diagram, however in terms of 
H $\alpha$ emission, all stars except RXJ1625.6-2613 are consistent with being 
weak-lined T-Tauri stars. Spectral energy distributions (SEDs) over the 
range $\simeq$ 4000\r{A} $< \lambda <$ 1000$\mu$m, show that only one Chamaeleon 
star (RXJ1112.7-7637) and three Rho Ophiuchus stars (ROXR1 13, RXJ1625.6-2613 \& 
RXJ1627.1-2419) reveal substantial departures from a bare photosphere.

\end{abstract}

\begin{keywords}
stars: circumstellar matter, stars: fundamental parameters, 
stars: evolution, stars: Hertzsprung-Russell and colour-magnitude diagrams, 
stars: pre-main-sequence, stars: late-type.
\end{keywords}

\section{Introduction}

Star forming regions (SFRs) and young star clusters seed a variety of empirical and 
theoretical investigations into the fundamental physical processes governing how stars 
form and evolve. Exemplar domains of study include proto-stellar formation and evolution 
(\citealt{Bate98}; \citealt{WT2003}), circumstellar disc dynamics 
(\citealt{HHL2001} ; \citealt{Bate2011}),  extra-solar planet formation 
(e.g. \citealt{Pollack96}; \citealt{Tsukamoto2013}) , and exoplanet 
orbital mechanics (\citealt{Trilling98}; \citealt{Armitage2003}; 
Nelson \& Benz 2003a,b). Observational efforts, through optical and 
infrared photometric and spectroscopic surveys, are required to 
understand such processes and generally necessitate a large sample of 
confirmed members of parent SFRs or young clusters with well-constrained 
fundamental stellar parameters.

Young stellar objects (YSOs) fall into four main classes generally denoted 0$-$III, 
representing an evolutionary sequence from protostar shrouded by natal molecular cloud 
material through to a discless system possibly hosting an evolved planetary system. 
Embedded class 0$-$I objects \citeauthor{lada1987} \citeyear{lada1987}, 
\citeauthor{andre1993} \citeyear{andre1993}) transition to class II when the 
surrounding envelope has cleared. Class II objects correspond with classical T Tauri 
stars (CTTS), defined by their enhanced infrared (disc) emission, excess blue 
emission from accretion luminosity, and numerous spectral features in emission; 
further, broad and complex line morphologies are observed which are due to circumstellar 
material. Class III YSOs show little or no near-infrared excess, and correspond to 
weak-lined T-Tauri stars (WTTS), which also exhibit little spectral line emission 
above expected chromospheric activity levels, consistent with the cessation of 
circumstellar disc accretion as the inner disc has been depleted.

More remains unknown than known concerning the evolution of circumstellar discs; 
in particular the mechanisms most responsible for driving accretion (i.e. 
outward angular momentum transport) and dissipation in Class I and II systems. 
Circumstellar discs are also important testbeds for theories of planet formation 
and migration. Near-infrared spectral energy distributions (SEDs) of so-called 
{\em transition discs} \citep{Forest04} have long been thought to reveal the presence of 
gaps cleared by massive planets, as in type II planet migration (e.g. 
Nelson \& Benz 2003a,b). This has been confirmed in several cases via direct imaging 
and interferometry (e.g. \citealt{ExoPimage}; \citealt{ALMA}). Nearby systems 
showing excess infrared emission consistent with the presence of a disc therefore 
make logical targets for followup imaging to probe disc structure, search for 
signs of protoplanet--disc interactions, and possibly direct detection of embedded 
protoplanets. Measuring stellar ages will therefore be key in observationally 
constraining planet formation timescales.

Circumstellar disc evolution, especially with regard to planet formation and migration, 
is a high priority science driver of the current and next-generation optical, infrared 
and sub-mm observatories such as the Transiting Exoplanet Survey Satellite ({\em TESS}), 
the Atacama Large Millimeter/submillimeter Array (ALMA), and the James Webb Space Telescope 
(e.g. \citealt{wolf2012}; \citealt{Flock2015}; \citealt{planets2025}). Given that 
the relatively bright stars, in nearby SFRs such as Chamaeleon and Rho Ophiuchus\footnote{Rho 
Ophiuchus has already been observed in Field 2 of the Kepler2 mission}, are most likely 
to be included in the {\em TESS} mission input catalogue, it is vitally important to fully 
characterize the fundamental physical properties of their solar-type members. Broadly 
speaking, complete membership catalogues and well-constrained stellar parameters 
-- especially mass and age -- in coeval SFRs and clusters are essential for 
understanding the time evolution of fundamental physics which govern disc evolution, 
and thus planet formation and evolution.

In this manuscript, we detail a photometric and spectroscopic campaign of X-ray 
selected candidate members of the Chamaeleon and Rho Ophiuchus associations, exploiting 
the dataset to characterize their physical properties and evolutionary status. 
In \S~\ref{targetselection}, we discuss target selection, telescope observations 
and data reduction procedures in producing standardized $BVIc$ photometry, 
spectroscopic kinematics and spectral line equivalent widths. We present our 
results and data catalogues in \S~\ref{results}, using them to confirm membership, 
and establish youth, of each star of its parent association. In concert with 
2MASS and WISE near- and mid-infrared photometry, we construct theoretical 
Hertzsprung-Russell diagrams to infer stellar age and mass for each object in 
our sample. In \S~\ref{S_seds}, we detail a programmatic search for the presence 
of circumstellar accretion discs of our sample, exploiting optical and infrared 
photometric data in order to construct SEDs for each candidate member of Chamaeleon 
and Rho Ophiuchus. In concluding, we deliver a discussion and summary of 
our research program in \S~\ref{conclusions}.

\section{Target Selection and Observations}\label{targetselection}

Young, low-mass stellar objects are oftentimes associated with super-solar 
levels of magnetic activity, which is manifest through elevated X-ray and 
radio emission \citep{feig1999,James00,James06,B1xrays}. For our study 
into the properties of both the Chamaeleon and Rho Ophiuchus SFRs, mirroring 
the strategy in our earlier studies (\citealt{James06} - [J06], 
\citealt{Santos2008} - [S08]), we selected candidate {\sc sfr} stars 
within positional error circles of detections in the {\sc rosat} All-Sky 
Survey [{\sc rass}]. For each candidate WTTS, physical properties such 
as proper motions, radial velocities [RVs], Lithium abundances and 
near infrared colours are used to ascertain membership of their parent 
association, details of which are presented in \S~\ref{membership}.

In order to investigate the fundamental properties of each SFR member, 
specifically mass and age, we pursued photometric and spectroscopic 
observing campaigns in the optical to complement existing near- and 
mid-infrared all sky survey photometry now available in the literature. 
Observations, data reduction and analysis for each campaign are 
discussed below.

\subsection{Photometry}\label{S_photometry}

$BVIc$ photometric data for individual fields containing X-ray selected 
candidates of Chamaeleon and Rho Ophiuchus were obtained using the 1.0m 
{\sc smarts} telescope (the former {\sc yalo} telescope) situated at the 
Cerro Tololo InterAmerican Observatory [{\sc ctio}], Chile, on the 
night of UT20070630. The data were acquired with the quad-amplifier 
{\sc y4kcam} {\sc ccd} camera, equipped with Johnson-Cousins filters, 
which with its 15 $\mu$m pixels and 0.289 arcsec/pixel plate scale 
results in an on-sky areal coverage of $19\farcm3\times19\farcm3$. 
We note that during the 2007 observing season, one of the quad-amplifiers 
of {\sc y4kcam} was inoperable in the North-West corner of the array, 
resulting in a reduced field of view per image. An observing log of 
the Chamaeleon and Rho Ophiuchus {\sc y4kcam} observations is presented in 
Table~\ref{YALO-obslog}.

All science and standard star images were processed for overscan region 
subtraction, master-bias subtraction and flat fielding, using twilight 
sky images, employing standard procedures in the {\sc iraf}\footnote{{\sc iraf}, 
in our case through http://iraf.net, is distributed by the National Optical 
Astronomy Observatories, which are operated by the Association of Universities 
for Research in Astronomy, Inc., under cooperative agreement with the 
National Science Foundation.} suite of data reduction algorithms. 
Substantial use was also made of two batch processing {\sc iraf} scripts, 
written by Phil Massey of Lowell Observatory, whose protocol specifically 
handles the {\sc fits} headers and quad-amplifier readout of the {\sc y4kcam} 
{\sc ccd} frames. The low space density of the {\sc sfr}s allows us to employ 
aperture photometry for science images as well as the standard 
star field images. Source searching and aperture photometry, with an 
aperture radius of 13-pixels, were achieved using the {\sc daophot ii} package 
in {\sc iraf} \citep{Stetson90,S93}.

Equatorial $BVIc$ standard stars catalogued in \citet{Landolt92}, 
with a broad dynamic range of photometric magnitude and colour, and 
located at a wide distribution of airmass, were observed in order to 
correct for atmospheric extinction and to transform extinction 
corrected, instrumental magnitudes onto the standard system. Modified 
forms of Bouguer's law, see Eqs.~\ref{phot-CEQ1}$-$\ref{phot-CEQ4}, 
were used to define the relationship between standard and instrumental 
magnitudes for the ensemble of standard stars.

\begin{equation}\label{phot-CEQ1}
V = v + \varepsilon (B-V) + \xi_{v} - \kappa_{v}.X  
\end{equation}

\begin{equation}\label{phot-CEQ2}
V = v + \phi (V-I) + \xi_{v} - \kappa_{v}.X  
\end{equation}

%\begin{equation}\label{phot-eqn}
%(U-B) =  \phi (u-b) + \xi_{ub} - \phi~\kappa_{ub}~X 
%\end{equation}

\begin{equation}\label{phot-CEQ3}
(B-V) =  \mu  (b-v) + \xi_{bv} - \mu~\kappa_{bv}.X  
\end{equation}

\begin{equation}\label{phot-CEQ4}
(V-I)_{c} = \psi (v-i)_{c} + \xi_{vi} - \psi~\kappa_{vi}.X 
\end{equation}

\hspace*{4mm} where
\begin{itemize}
\item
V, (B$-$V) and (V$-$I)$_{c}$ are magnitudes/indices on the $BVIc$ standard system. 
\item
v, (b$-$v), and (v$-$i)$_{c}$ are measured, instrumental magnitudes/indices.
\item
$\kappa_{v}, \kappa_{bv}$ and $\kappa_{vi}$ are wavelength (filter) dependent 
extinction coefficients.
\item
$\varepsilon$, $\phi$, $\mu$ and $\psi$ are colour transformation coefficients. 
\item
$\xi_{v}$, $\xi_{bv}$ and $\xi_{vi}$ are zero-point coefficients.
\item 
X = airmass of target [$\simeq$ sec z - where z is the zenith distance]
\end{itemize}

\begin{table*}
\caption{Calculated extinction, transformation and zero-point coefficients 
on the standard $BVIc$ system for photometric observations using the 
{\sc ctio smarts} 1.0m telescope+Y4Kcam during the night of UT20070630.}
\vspace{3mm} 
\begin{tabular}{lcccccc}
\hline
Colour       & Extinction$^{b}$  & CTC$^{a,b}$  & Zero$^{b}$  & {\sc rms}$^{c}$ & No of.          \\
Equation     & Coefficient       &              & Point       & $\Delta$mag   & Standards$^{d}$ \\
\hline 
$V$ [$B-V$]  & 0.1077            & 0.0366       & -2.0593     & 0.0160        & 84, 69 \\
$V$ [$V-Ic$] & 0.1065            & 0.0277       & -2.0552     & 0.0176        & 81, 70 \\
$B-V$        & 0.1137            & 0.8306       & -0.0857     & 0.0186        & 77, 62 \\
$V-Ic$       & 0.0332            & 1.0076       & ~0.7515     & 0.0158        & 84, 64 \\
\hline
\end{tabular}
\label{CEQ-results} 
\begin{flushleft}
%Notes: \\
a $-$ CTCs are colour transformation coefficients. \\
b $-$ Extinction coefficients (magnitudes/airmass), CTCs and zeropoints, 
as described in Eqs.~\ref{phot-CEQ1}$-$\ref{phot-CEQ4}. \\
%c $-$ Mean difference between measured and published magnitudes of Landolt (1992) standard stars. \\
c $-$ {\sc rms} of differences between measured and published magnitudes of Landolt (1992) standard stars. \\
d $-$ Initial and final numbers of standard stars used in fits to colour equations. \\
\end{flushleft}
\end{table*}

Unknown extinction coefficients, colour transformation coefficients and 
zeropoints were determined by solving the self-similar series of linear 
simultaneous algebraic equations detailed in Eqs.~\ref{phot-CEQ1}$-$\ref{phot-CEQ4}. 
For the ensemble of standard star observations, solutions were derived 
using a least-squares fit algorithm\footnote{The algorithm, {\sc f04amf}, is 
distributed by the Numerical Algorithms Group ({\sc nag})} which processes the 
numeric parameters via an iterative refinement method. Comparing calculated 
magnitudes and colours of Landolt standard stars to their published values 
allows an iterative rejection process to take place, thereby eliminating 
seriously discrepant standard star measurements. Standard star measurements 
were eliminated from the fit if the difference between their calculated and 
published magnitudes were \gsi 3$\sigma$ away from equality, with a final 
solution being accepted if the {\sc rms} of these differences was \lsi 
0.02 magnitudes.  In this way, approximately 70-80 standard stars were 
observed, with 60-70 stars used in the final solution. In order to test 
whether the solutions to the colour equations have parameter dependencies 
on magnitude, colour, airmass or time of observation, we compared calculated 
and published values of the observed Landolt standard star fields searching 
for parameter correlations. None were found.

Solutions to the colour equations are listed in Table~\ref{CEQ-results}. 
Target star photometry is easily determined by substituting the derived 
extinction, transformation and zero-point coefficients, and measured 
instrumental magnitudes into Eqs.~\ref{phot-CEQ1}$-$\ref{phot-CEQ4}. 
Comparing our colour equation coefficients to similar data acquired 
with the exact same instrument set up ({\em c.f.,} \citealt{PACDJJ2010}), 
we find that our colour transformation coefficients agree to within 
$\simeq 1-2$ per cent of the Cargile \& James values, except for the B-V one, 
where the difference is of order 5 per cent. Judging the similarity of extinction 
coefficients and zero-points is very difficult because of the way we solve 
the colour equations, calculating all three coefficients simultaneously, 
allowing the least-squares solution to {\em see-saw} about the colour-transformation 
coefficient as a fulcrum. What we can say is that comparing our coefficient values 
in Table~\ref{CEQ-results} with those in table 2 of the Cargile \& James paper 
shows that there are no obviously deviant coefficient data points. In any case, 
we are able to reproduce the colours and magnitudes of Landolt standard stars 
throughout the UT20070630 night to the $\simeq1-2$ per cent level ({\em e.g.,} see 
{\sc rms} delta-magnitudes in both studies).

Initial stellar coordinates ({\sc ra, dec}) were calculated using 6-coefficient 
fits to {\sc ccd x,y} values using Digitized Sky Survey [{\sc dss}] red 
plates; Object matches were subsequently made with {\sc 2mass} catalogues, 
using a 2-arcsec matching radius, and we employ these {\sc 2mass} J2000 
sexagesimal coordinates for each {\sc sfr} candidate member (columns~4-5 
in table~\ref{YALO-obslog}).

\subsection{Spectroscopy}\label{spectroscopy}

With the express goals of deriving radial velocities and detection of the 
resonance doublet of neutral Lithium at 6708\r{A}, facilitating {\sc sfr} kinematic 
membership and youth assessments, we have obtained high-resolution \'{e}chelle 
spectra of our {\sc sfr} candidate members in Chamaeleon and Rho Ophiuchus. The high 
resolution (R~$\equiv\lambda$/$\Delta\lambda\sim$50000) spectra were acquired in 
service mode with the {\sc uves} spectrograph on the Kueyan 8.2-m telescope located at 
the Paranal site of the European Southern Observatory (programme ID 075.C-0272). 
The observations were made using the Red 580 mode of cross disperser $\#3$, with 
a spectral window of 4800$-$7000\r{A} (with a small gap between 5700$-$5840\r{A}), 
using the {\sc shp}700 filter, employing a 0.9-arcsec slit and 2$\times$2 binning 
of the {\sc mit}/{\sc ll} {\sc eev} detector ({\sc ccd}-44). An observing log of the 
Chamaeleon and Rho Ophiuchus {\sc uves} observations is presented in 
Table~\ref{UVES-obslog}.

Data reduction in the form of removal of instrumental effects (de-biasing, 
flat-fielding), spectral extraction and wavelength calibration (using 
corresponding ThAr lamp spectra) of the {\sc uves} spectra was performed 
using the instrument's dedicated reduction pipelines. Heliocentric 
radial velocities [{\sc rv}s] were calculated by cross-correlation 
techniques \citep{TD79} relative to the slowly-rotating {\sc iau} and 
{\sc gaia} velocity standard star {\sc hd} 76151 \citep{rvstd1,rvstd2} 
using the spectral range 5000-5350\r{A}. This spectral window is ideal 
for {\sc rv} determinations because it contains many deep metal absorption 
lines (especially the Mg {\sc i} triplet) and little telluric contamination. 
Cross correlation of the {\sc hd} 76151 {\sc uves} spectrum against other 
{\sc iau} velocity standards reveal that external errors on the standard 
system are $\simeq0.4$ km~s$^{-1}$. Random errors were calculated in a 
similar manner to the J06 study, and vary from 0.2 km~s$^{-1}$ for slowly 
rotating stars increasing to $\simeq$$2-3$ km~s$^{-1}$ for more rapidly 
rotating objects (\gsi 20 km~s$^{-1}$).

In order to measure equivalent widths [{\sc ew}] of the Li {\sc i} resonance 
doublet at 6708\r{A}, we first normalized target {\sc uves} spectra by 
division of a $\simeq 25^{th}$-order cubic spline fit to the data over 
the 6600-6800 \r{A} spectral window, masking out the Balmer series 
H $\alpha$ region from the fit. We employ both the direct integration 
and Gaussian fitting methods (e.g. see J06 ), so our {\sc ew} values 
encompass contributions from the small Fe {\sc i}+CN lines at 6707.44\r{A}, 
leading to measured Lithium {\sc ew}s that result in a minorly ($10-20$ m\r{A}) 
overestimated photospheric Li presence (Soderblom et al. 1993 report that this 
Fe line blend has an empirical EW = [20($B-V$)$_{0}$ - 3]m\r{A}, for main 
sequence, solar-type stars). In the case of H $\alpha$, we followed the same 
procedure as J06 in that the normalized spectrum of an old, slow rotating, 
minimum-activity standard star of similar spectral type was first subtracted 
from the Chamaeleon and Rho Ophiuchus candidate's spectrum. This methodology 
is powerful because it takes account of (at least to first order) telluric 
contamination and acts to remove the photospheric contribution to the H $\alpha$ 
line. In our case, we used {\sc uves} spectra of the slowly rotating stars {\sc hd} 
76151 (G3V, Prot=15days, \citealt{Barnes2007}), {\sc hd} 196761 (G8V, Prot=31days, 
\citealt{Wright04}) and {\sc hd} 191408 (K2.5V, Prot=45days, \citealt{Barnes2007}), 
whose spectral types were verified by cross-correlation against the digital 
spectral library presented in \citet {SAAOSpTy}.

Finally, we exploit our {\sc uves} spectra in order to calculate spectroscopic 
projected equatorial rotation rates, $\upsilon$sin$i$, for our SFR targets 
by cross correlation with similar spectra of the minimum activity, slowly 
rotating late-type stars listed above ({\em c.f.,} \citealt{JJ97}). Calibration 
of the width of the inverse cross correlation function is achieved by 
artificially broadening (\citealt{Gray92}; limb-darkening fixed at 0.6) a 
high-S/N spectrum of a given minimum activity star, and cross-correlating it 
with its un-broadened original for a series of incremental $\upsilon$sin$i$ 
values. By measuring the width of the central peaks in the resulting inverse 
cross-correlation functions, as a function of rotation rate in the broadening 
kernel, a relationship between rotation rate and cross correlation function 
width is readily produced. Empirical simulations, similar to those detailed 
in J06, were performed to show that {\sc uves} $\upsilon$sin$i$ values are 
reproducible at the $\pm 10$ per cent level down to an upper limit of $\simeq 4-5$ 
km~s$^{-1}$.

\section{Results}\label{results}

\begin{figure}
\centering
\epsfig{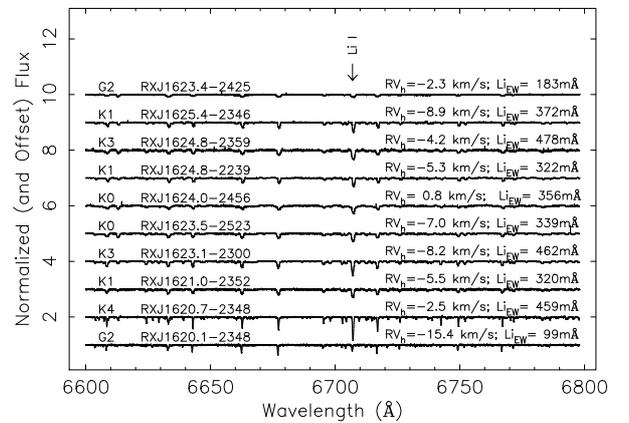}
\caption{Normalized (and offset) rest-frame {\sc uves} spectra for a 
sample of candidate members of the Rho Ophiuchus {\sc sfr} are shown. 
Heliocentric radial velocities and Gaussian-fit Li {\sc i} 6708\r{A} 
{\sc ew}s are annotated for each star (c.f. Table~\ref{SFRcandidates-UVES}).}\label{FeH-UVES-RhoOph}
\end{figure}

Optical $BVIc$ and infrared $JHK$ photometry, from {\sc y4kcam} observations 
and {\sc 2mass}, respectively, are presented for our candidate members of 
the Chamaeleon and Rho Ophiuchus {\sc sfr}s in Table~\ref{SFRcandidates-OPTIRphot}; 
accompanying table notes detail data sources for each object. Data products 
for our {\sc uves} observations are presented in Table~\ref{SFRcandidates-UVES}, 
which include spectral types drawn from the literature. Equivalent widths for the 
H $\alpha$ and \lii 6708\r{A} spectral lines are provided using both Gaussian fit 
and direct integration methods. A representative sample of normalized {\sc uves} 
spectra, in the 6700\r{A} region, for candidate members of Rho Ophiuchus is shown in 
Figure~\ref{FeH-UVES-RhoOph}.

\begin{table*}
%\begin{center}
\vspace{3mm} 
\caption{Optical $BVIc$ (Y4Kcam) and infrared (2MASS) photometry for SFR candidate members}
\begin{tabular}{lrrrcccc}
\hline
~~~~~~~~Object    & $^{a}$V$\pm$err~~~~     &   $^{a}$($B-V$)$\pm$err  &  $^{a}$($V-I$)c$\pm$err  &  $J$$\pm$err       & $H$$\pm$err        & $K'$$\pm$err       & Err  \\
                  &                   &                  &                  &   [2MASS]$^{b}$  &  [2MASS]$^{b}$     &  [2MASS]$^{b}$    & Code$^{b}$ \\ \hline 
Chamaeleon \\ \hline
  RXJ0850.1-7554       &  10.617$\pm$0.001 &  0.747$\pm$0.002 &  0.845$\pm$0.001 &  ~~9.259$\pm$0.026 &  8.848$\pm$0.025 &  8.704$\pm$0.019 & AAA  \\
  RXJ0951.9-7901       &  10.202$\pm$0.001 &  0.831$\pm$0.002 &       ...$^{e}$  &  ~~8.587$\pm$0.032 &  8.138$\pm$0.034 &  8.040$\pm$0.029 & AAA  \\ 
  RXJ1112.7-7637       &  11.583$\pm$0.002 &        ...$^{d}$ &  1.344$\pm$0.003 &  ~~9.275$\pm$0.030 &  8.524$\pm$0.055 &  7.999$\pm$0.031 & AAA  \\ 
  RXJ1129.2-7546       &  12.946$\pm$0.005 &  1.407$\pm$0.023 &  1.774$\pm$0.005 &  ~~9.817$\pm$0.026 &  9.124$\pm$0.021 &  8.878$\pm$0.024 & AAA  \\ 
  RXJ1140.3-8321       &  11.472$\pm$0.002 &  1.142$\pm$0.006 &  1.289$\pm$0.002 &  ~~9.328$\pm$0.023 &  8.709$\pm$0.045 &  8.635$\pm$0.019 & AAA  \\ 
  RXJ1158.5-7754a      &  10.519$\pm$0.001 &  1.179$\pm$0.003 &       ...$^{e}$  &  ~~8.219$\pm$0.029 &  7.556$\pm$0.042 &  7.404$\pm$0.021 & AAA  \\ 
  RXJ1159.7-7601       &  11.181$\pm$0.002 &  1.150$\pm$0.005 &  1.307$\pm$0.002 &  ~~9.140$\pm$0.027 &  8.469$\pm$0.038 &  8.304$\pm$0.027 & AAA  \\ 
  RXJ1201.7-7859$^{c}$ &    8.56$\pm$9.999 &   0.67$\pm$9.999 &   0.75$\pm$9.999 &  ~~7.263$\pm$0.027 &  6.967$\pm$0.044 &  6.848$\pm$0.018 & AAA  \\ 
  RXJ1233.5-7523       &   9.539$\pm$0.001 &  0.738$\pm$0.001 &       ...$^{e}$  &  ~~8.201$\pm$0.020 &  7.883$\pm$0.040 &  7.756$\pm$0.040 & AAA  \\ 
  RXJ1239.4-7502       &  10.344$\pm$0.001 &  0.997$\pm$0.002 &  1.093$\pm$0.001 &  ~~8.434$\pm$0.021 &  7.953$\pm$0.033 &  7.777$\pm$0.021 & AAA  \\  \hline
  Rho Ophiuchus \\ \hline
  RXJ1620.1-2348       &  10.053$\pm$0.001 &  0.666$\pm$0.001 &  0.767$\pm$0.001 &  ~~8.753$\pm$0.024 &  8.381$\pm$0.034 &  8.287$\pm$0.021 & AAA  \\ 
  RXJ1620.7-2348       &  12.668$\pm$0.004 &  1.374$\pm$0.009 &  1.645$\pm$0.005 &  ~~9.867$\pm$0.023 &  9.141$\pm$0.021 &  8.927$\pm$0.019 & AAA  \\ 
  RXJ1621.0-2352       &  10.477$\pm$0.001 &  0.852$\pm$0.001 &  0.935$\pm$0.001 &  ~~8.888$\pm$0.021 &  8.513$\pm$0.045 &  8.393$\pm$0.019 & AAA  \\ 
  RXJ1621.2-2347       &  16.015$\pm$0.026 &  2.282$\pm$0.287 &  2.889$\pm$0.027 & 10.728$\pm$0.024   &  9.498$\pm$0.029 &  8.962$\pm$0.023 & AAA  \\ 
  RXJ1623.1-2300       &  11.920$\pm$0.002 &  1.313$\pm$0.005 &  1.601$\pm$0.002 &  ~~9.042$\pm$0.032 &  8.343$\pm$0.040 &  8.184$\pm$0.024 & AAA  \\ 
  RXJ1623.4-2425       &  12.720$\pm$0.004 &  1.133$\pm$0.014 &  1.692$\pm$0.005 &  ~~9.708$\pm$0.024 &  9.060$\pm$0.024 &  8.762$\pm$0.021 & AAA  \\ 
  RXJ1623.5-2523       &  11.457$\pm$0.001 &  1.268$\pm$0.003 &  1.634$\pm$0.001 &  ~~8.600$\pm$0.024 &  7.979$\pm$0.045 &  7.695$\pm$0.022 & AAA  \\ 
  RXJ1624.0-2456       &  13.140$\pm$0.003 &  1.577$\pm$0.014 &  2.100$\pm$0.004 &  ~~9.445$\pm$0.022 &  8.597$\pm$0.044 &  8.280$\pm$0.020 & AAA  \\   
  RXJ1624.8-2359       &  14.117$\pm$0.005 &  1.810$\pm$0.021 &  2.589$\pm$0.005 &  ~~9.430$\pm$0.027 &  8.275$\pm$0.038 &  7.857$\pm$0.020 & AAA  \\ 
  RXJ1624.8-2239       &   9.781$\pm$0.001 &  0.974$\pm$0.001 &  1.138$\pm$0.001 &  ~~7.779$\pm$0.027 &  7.280$\pm$0.027 &  7.084$\pm$0.018 & AAA  \\
  RXJ1625.0-2508       &  12.276$\pm$0.003 &  1.340$\pm$0.006 &  1.873$\pm$0.004 &  ~~8.992$\pm$0.023 &  8.324$\pm$0.034 &  8.081$\pm$0.024 & AAA  \\ 
  RXJ1625.4-2346       &  11.989$\pm$0.002 &  1.238$\pm$0.005 &  1.707$\pm$0.002 &  ~~8.833$\pm$0.027 &  8.081$\pm$0.021 &  7.822$\pm$0.027 & AAA  \\ 
  RXJ1625.6-2613       &  11.759$\pm$0.002 &  1.283$\pm$0.005 &  1.674$\pm$0.002 &  ~~8.688$\pm$0.019 &  7.947$\pm$0.055 &  7.517$\pm$0.024 & AAA  \\
        ROXR1 13       &  13.881$\pm$0.005 &  2.185$\pm$0.028 &  3.096$\pm$0.005 &  ~~8.090$\pm$0.021 &  6.862$\pm$0.046 &  6.227$\pm$0.018 & AAA  \\ 
  RXJ1627.1-2419       &  14.133$\pm$0.009 &  1.751$\pm$0.033 &  2.823$\pm$0.010 &  ~~8.745$\pm$0.027 &  7.507$\pm$0.038 &  6.719$\pm$0.024 & AAA  \\ 
\hline
\end{tabular}\label{SFRcandidates-OPTIRphot} 
\begin{flushleft}
a $-$ $V$, ($B-V$) and ($V-I$)c data are calculated using our UT20070630 {\sc y4kcam} observations, with internal 
statistical errors annotated. \\ 
b $-$ $JHK$ data are taken from the {\sc 2mass} All-Sky Release Point Source catalogue (March 2003). 
[http://irsa.ipac.caltech.edu/applications/Gator/] \\
c $-$ Optical $BVIc$ data for RXJ1201.7-7859 are taken from Torres et al. (2006).\\
d $-$ The B-filter image had a corrupted readout.\\
e $-$ The star was saturated in the I-filter image.\\
\end{flushleft}
\end{table*}

%\begin{landscape}
\begin{table*}
%\begin{center}
\caption[]{\protect \small {\sc uves} spectroscopic data products for X-ray selected candidate 
members of the Chamaeleon and Rho Ophiuchus SFRs.}
\vspace{3mm} 
\begin{tabular}{lcccclccc}
\hline
~~~~~~~~Target   & {\sc hjd} [days] & Rad Vel.    & $\Delta$RV$^{a}$ &  $\upsilon$sin$i$ & SpTy$^{b}$ & $^{c}$Li {\sc I} 6708\r{A} [m\r{A}] & $^{c}$H $\alpha$ [m\r{A}]       & S/N$^{g}$ \\
                 & (+2450000)     & km~s$^{-1}$ & km~s$^{-1}$      & km~s$^{-1}$       &            & $^{d}$Gauss  $^{e}$Integ.          & ~~~$^{f}$Fit ~~~~ $^{e}$Integ.  & 6700\r{A} \\
\hline 
Chamaeleon \\ \hline
 RXJ0850.1-7554$^{h}$  & 2712.590~~     &  17.1  & ...                                        &  ~~~49      & G6    & 330  ~~~ 319     &   ~497 (G)  ~~~  ~~~493         &  ~~87 \\ 
  RXJ0951.9-7901       & 3452.6444      &  12.0  & $^{-27.8}_{~~32.3}$ , $^{-26.1}_{~~47.1}$  &  ~~~77      & G7    & 287  ~~~ 261     &   1928 (L)  ~~~  1855           &  120  \\
  RXJ1112.7-7637       & 3453.5495      &  15.0  &   1.4, -0.2                                &  ~~~31      & K2    & 417  ~~~ 408     &   2824 (L)  ~~~  2636           &  165  \\
  RXJ1129.2-7546       & 3453.5691      &  15.0  &  -0.9, -0.5                                &  ~~~21      & K3    & 460  ~~~ 451     &   1067 (L)  ~~~  ~~920          &  ~~92 \\
  RXJ1140.3-8321       & 3453.6973      &  12.9  &  -1.1                                      &  ~~~11      & K3/4  & 203  ~~~ 204     &   1167 (G)  ~~~  1395           &  105  \\
  ~RXJ1158.5-7754a     & 3453.7078      &  13.3  &   2.9, 3.2                                 &  ~~~11      & K3    & 466  ~~~ 469     &   1593 (L)  ~~~  1305           &  100  \\
  RXJ1159.7-7601       & 3453.7176      &  14.0  &  -1.1                                      &  ~~~~~9     & K3    & 448  ~~~ 450     &   1043 (G)  ~~~  1274           &  105  \\
  RXJ1201.7-7859       & 3453.7322      &  18.4  &   6.1, 4.8                                 &  ~~~23      & G5    & 238  ~~~ 231     &  ~~325 (L)  ~~~  ~~~259         &  120  \\
  RXJ1233.5-7523       & 3453.7368      &  15.1  &   0.3,-0.9                                 & ~~~~~~~~6.5 & K1    & 125  ~~~ 124     &  ~~193 (G)  ~~~  ~~~170         &  150  \\
  RXJ1239.4-7502       & 3453.7432      &  13.3  &  -1.2,-0.9                                 &  ~~~21      & K2/3  & 412  ~~~ 400     &  ~~808 (G)  ~~~  ~~~827         &  110  \\
  RXJ1303.5-7701       & 3453.7485      & ~~7.9  & 129.0                                      & 118         & early & ~~17  ~~ ~~~13   &  ~~~  ...  \hspace*{11mm}  ...  &  ~~95 \\ \hline
Rho Ophiuchus \\     \hline
  RXJ1620.1-2348       & 3453.8995      &  -15.4 &  ...                                       & ~~~~~8.5    & G2    & ~~99  ~~~ ~~97   &    ~~180 (G) ~~~   ~~138        & ~~70 \\
  RXJ1620.7-2348       & 3456.7326      & ~~-2.5 &  ...                                       & ~~~~~9.5    & K4    & 459  ~~~ 449     &    1745 (G)  ~~~   2145         &  105 \\
  RXJ1621.0-2352       & 3453.9050      & ~~-5.5 &  ...                                       & 29          & K1    & 320  ~~~ 292     &    ~~458 (G) ~~~   ~~436        & ~~65 \\
  RXJ1623.1-2300       & 3456.7538      & ~~-8.2 &  ...                                       & 24          & K3    & 462  ~~~ 448     &    1176 (G)  ~~~   1162         & 110  \\
  RXJ1623.4-2425       & 3456.7774      & ~~-2.3 &  ...                                       & 57          & G2    & 183  ~~~ 168     &    ~~352 (G) ~~~   ~~334        & 125  \\
  RXJ1623.5-2523       & 3466.8534      & ~~-7.0 &  ...                                       & 34          & K0    & 339  ~~~ 319     &    ~~625 (G) ~~~   ~~524        & 105  \\
  RXJ1624.0-2456       & 3466.8706      & ~~~0.8 &  ...                                       & 38          & K0    & 356  ~~~ 338     &    1620 (L)  ~~~   1411         & 100  \\
  RXJ1624.8-2359       & 3481.6751      & ~~-4.2 &  ...                                       & 40          & K3    & 478  ~~~ 452     &    1185 (L)  ~~~   1027         & ~~60 \\
  RXJ1624.8-2239       & 3494.6352      & ~~-5.3 &  ...                                       & 28          & K1    & 322  ~~~ 307     &    1688 (L)  ~~~   1290         & 155  \\
  RXJ1625.0-2508       & 3466.8934      & ~~-3.0 &  ...                                       & 48          & G1    & 167  ~~~ 158     &    ~~250 (G) ~~~   ~~178        & 110  \\
  RXJ1625.4-2346       & 3494.6471      & ~~-8.8 &  ...                                       & 27          & K1    & 392  ~~~ 366     &   ~~~  ...  \hspace*{10mm}  ... & ~~10 \\
  RXJ1625.4-2346       & 3495.6345      & ~~-9.0 &  ...                                       & 27          & K1    & 352  ~~~ 333     &    ~~614 (G) ~~~   ~~513        & 130  \\
  RXJ1625.6-2613       & 3476.7809      & ~~-3.8 &  -0.3                                      & 17          & K7Ve  & 479  ~~~ 471     &   $>$6400 (L*)    5976          & 100  \\
  ROXR1 13             & 3505.7105      & ~~-7.3 &   ...                                      & 68          & K0    & 320  ~~~ 284     &    ~~667 (G) ~~~   ~~712        & ~~55 \\
  ROXR1 13             & 3505.7991      &  -12.2 &   ...                                      & 69          & K0    & 296  ~~~ 268     &    1298 (L)  ~~~   1250         & ~~65 \\
  RXJ1627.1-2419       & 3505.6780      & ~~-5.0 &  -2.3                                      & 60          & G1    & 194  ~~~ 174     &    3521 (L)  ~~~   2943         & ~~50 \\
\hline
\end{tabular}\label{SFRcandidates-UVES} 
\begin{flushleft}
%Notes: \\
a $-$ We compare our {\sc uves} RVs to those detailed in J06 [{\sc uves} - J06 values]. Two values 
listed represent two epochs in the J06 study, which for RXJ0951.9-7901, include its status as 
a double-lined spectroscopic binary (J06). \\
b $-$ Spectral types for Chamaeleon stars are taken from Covino et al. (1997) and 
Alcal\'{a} et al. (2000), whereas Mart\'{\i}n et al. (1998) data were used for 
Rho Ophiuchus stars, except for RXJ1625.6-2613, whose spectral type was reclassified 
as K7Ve in Torres et al. (2006). \\ 
c $-$ Li/residual H $\alpha$ features were measured for EWs; by {\em residual}, we mean 
that the normalized spectrum of an old, slowly rotating minimum-activity standard 
star, of similar spectrum type, broadening if the target has $\upsilon$sin$i$ $>$ 10 km~s$^{-1}$, 
was first subtracted from the normalized spectrum of each target. \\
d $-$ EWs measured using a Gaussian fit between the limits of where the Li feature 
approaches unity on either side of its central absorption feature. \\
e $-$ EWs measured using direct integration between the same wavelength limits as 
for the Gaussian fitting method. \\
f $-$ EWs of residual H $\alpha$ emission features were measured with either a Gaussian 
(G) or Lorentzian (L) function, with limits placed where the feature approaches the 
continuum level on the red and blue wings. For the Rho Ophiuchus target, RXJ1625.6-2613, the 
residual emission profile presented substantial non-uniform morphology and sub-structure, 
which was not adequately fitted by either a Gaussian or Lorentzian function.\\
g $-$ Approximate $S/N$ of continuum region around 6700\r{A}. \\
h $-$ UVES data were not obtained for this target; for reference, we reproduce here 
the measurements of J06. \\
%  Delta Vsini(UVES)-Vsini(J06) = -0.27 +/- 0.46 km/s (1sig=1.74 km/s; n=15)
\end{flushleft}
\end{table*}

Each subset of the observational data can be engaged in deriving fundamental 
properties for each candidate {\sc sfr} member and of the parent {\sc sfr} 
itself. We will investigate five cornerstone characteristics of the 
observational data in the following manner. {\bf First}, our {\sc uves} spectra 
provide radial velocities and Lithium detections for each star, which assist 
us in assigning kinematic membership of each {\sc sfr} and in assessing 
stellar youth. {\bf Second}, optical and infrared photometry, in concert 
with spectral types, allows us to measure colour-dependent reddening vectors, 
which contribute to our spectral energy distribution analysis of Chamaeleon 
and Rho Ophiuchus members suspected of being classical T Tauri stars (see 
\S~\ref{S_seds}). {\bf Third}, {\sc jhk} photometry can be exploited to search 
for excess infrared emission, indicative of the presence of a circumstellar 
accretion disc around young stars, and represents one of the key demarcation 
signatures between classical and weak-line T Tauri stars. Extremely high 
levels of H $\alpha$ emission can also be indicative of circumstellar 
material, allowing us to use our {\sc uves} spectra to correlate infrared 
excess with strong H $\alpha$ emission. {\bf Fourth}, employing effective 
temperatures and bolometric corrections (for bolometric luminosity 
determinations) derived from spectral types, facilitating the construction 
of theoretical Hertzsprung-Russell diagrams [HRDs] leading to determinations 
of stellar mass and age. {\bf Fifth}, combining optical and near-to-mid infrared 
photometry with radiative transfer models, we construct spectral energy 
distributions [SEDs] for each candidate {\sc sfr} member observed to search 
for, and characterize, signature evidence of circumstellar discs and 
blueward excesses indicative of ongoing accretion. The following sections 
are dedicated to describing each observational dataset in terms of the 
fundamental nature of our target stars, culminating in a commentary for 
each of these five investigative avenues.

\subsection{Membership of the parent SFR}\label{membership}

An understanding of the age-rotation-magnetic activity paradigm 
(e.g. \citealt{Hemp95}, \citealt{neuhauser1995}, \citealt{AL1996}, 
J06, \citealt{B1xrays}) already implies that these low-mass 
{\sc sfr} candidate members, being optical counterparts to {\sc rass} 
X-ray detections, are either very young and rapidly rotating, 
are short-period multiple stars systems, or are both. Irrespective 
of whether each Chamaeleon and Rho Ophiuchus candidate is a classical 
or weak-lined T Tauri star, we can assign membership of each star 
relevant to their parent associations through a careful consideration 
of each one's astrometric and kinematic properties.

\begin{figure*}
\centering
\epsfig{figure=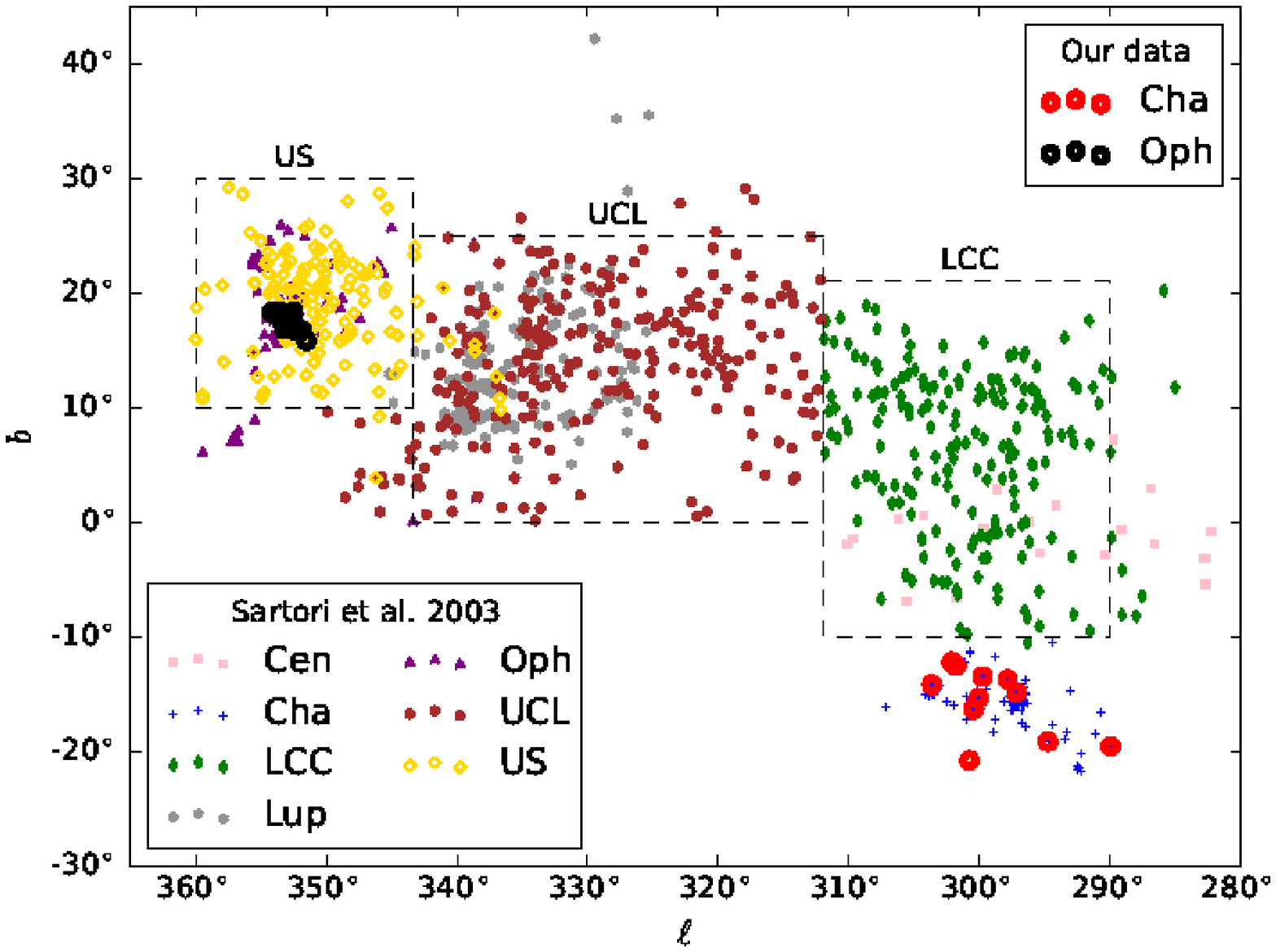,width=84.2mm}
\hspace*{1mm}\epsfig{figure=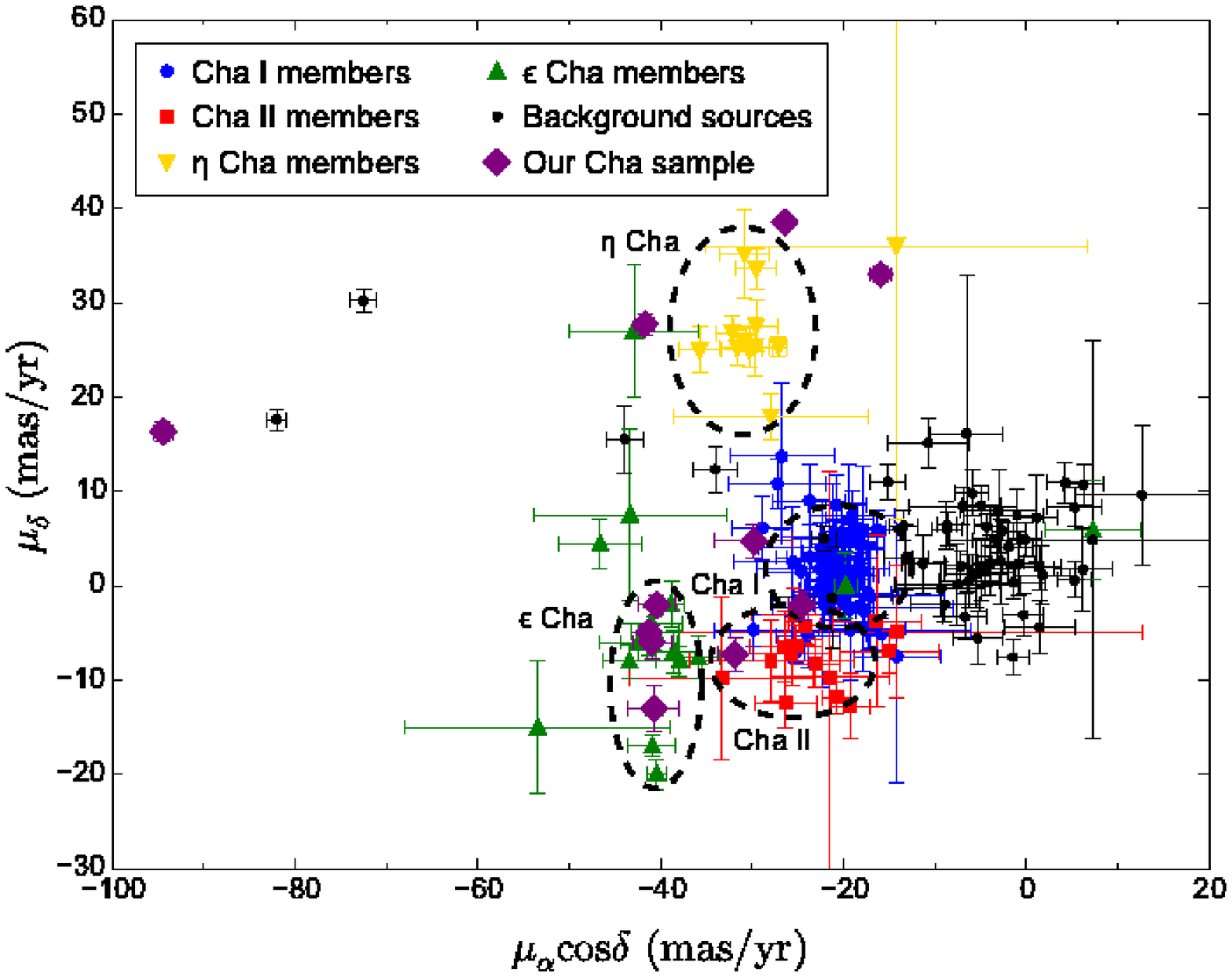,width=90.8mm}
\caption{Left-hand panel: In units of degrees, Galactic longitude ($l$) and 
latitude ($b$) are plotted for our Chamaeleon ({\bf {\textcolor{red}{$\bullet$}}}) and Rho Ophiuchus 
({\bf {\textcolor{black}{$\bullet$}}}) candidate SFR members. Comparable data for several well-described 
SFRs, taken from \citet{Sartori03}, are included for reference (US $-$ Upper Scorpius; 
UCL $-$ Upper Centaurus Lupus; Lup $-$ Lupus; LCC $-$ Lower Centaurus Crux). Right-hand 
panel: In concert with data for several sub-groups of the Chamaeleon SFR (\citealt{PMsChamI}), 
proper motion vectors (in mas/yr) are plotted for stars in our Chamaeleon sample. The proper 
motion data for RXJ1233.5-7523 (middle-left of the plot) show that this star does not 
belong to any of the main sub-groups of the Chamaeleon SFR.}
\label{Kinematics-ChamRhoOph}
\end{figure*}

In Figure~\ref{Kinematics-ChamRhoOph}, we display astrometric data for 
each star in our sample to ascertain whether our targets are consistent 
with membership of any particular young SFR in their vicinity. In the 
left-hand plot, we show that all of our targets, as expected and without 
exception, fall into the expected regions of ($l,b$), Galactic coordinate 
space, for both the Chamaeleon and Rho Ophiuchus SFRs (based on previous 
work and examples by \citealt{Sartori03} and \citealt{Aarnio2008}).

While SPM4 (\citealt{SPM4}) and/or UCAC4 (\citealt{UCAC4}) proper motion 
data exist for about half of our Rho Ophiuchus sample, they are oftentimes 
neither precise enough, nor unambiguous enough, to provide us with a 
definitive membership indication of their parent association 
(c.f. \citealt{mama08}, \citealt{PMs-PMS}). Only one Rho Ophiuchus 
object is a clear proper motion non-member of the SFR, RXJ1620.1-2348, 
for which we show in \S~\ref{LiMEM} is a probable field star. For the 
Chamaeleon sample however, a summary of high quality proper motion 
vectors are available in the literature (e.g. \citealt{PMsChamI}; 
\citealt{PMsChamII}), which we plot in the right-hand panel of 
Figure~\ref{Kinematics-ChamRhoOph}. A comparison to proper motion 
vectors of young stars in the vicinity allows us to allocate 
sub-group Chamaeleon membership for our targets.

Interestingly, this proper motion vector phase diagram reveals 
the following: Three of our Chamaeleon sample, RXJ0850.1-7554, 
RXJ0951.9-7901 and RXJ1140.3-8321, have two-dimensional 
kinematics broadly consistent with being members of the Eta 
Chamaeleon population, in agreement with the findings of \citet{PMsChamI}. 
On the other hand, \citet{elliott2014} assert that RXJ0951.9-7901 and 
RXJ1140.3-8321 are in fact members of the older ($30-40$ Myr, \citealt{Kraus2014}; 
45 Myr, \citealt{Bell2015}) Tucana-Horologium [Tuc-Hor] association. We 
further note that the position of RXJ0850.1-7554 in proper motion 
space lies several sigma away from the main Eta Cha clump, and in 
fact \citet{elliott2014} report that RXJ0850.1-7554 is a 
member of the 30 Myr Carina association. 

A further four objects, RXJ1158.5-7754a, RXJ1159.7-7601, RXJ1201.7-7859 
and RXJ1239.4-7502 are kinematic members of Eps Chamaeleon (in agreement 
with the recent Eps Cha membership allocations given by 
\citealt{epsCha2013}). Two stars, RXJ1112.7-7637 and RXJ1129.2-7546, 
have proper motion vectors consistent with the Cham I SFR (in accord with 
\citealt{PMsChamI}), although we note that there is some overlap in this 
region of the kinematic phase diagram with the Cham II cloud and a wide 
slew of background sources. We further note that RXJ1303.5-7701 is 
kinematically consistent with membership of the Cham II association. 
Finally, RXJ1233.5-7523 has a very discrepant proper motion vector 
compared to the any of the Chamaeleon sub-groups, and indeed many of 
the background sources in the region. We therefore judge this star 
to be a member of the field population, (concordant with the \citealt{PMsChamI} 
findings) and do not consider it further in our analysis.

\subsection{UVES Radial Velocities}\label{RVs}

One-dimensional kinematic membership of parent SFRs can be judged through 
assessment of their {\sc rv} vectors, which are provided in Table~\ref{SFRcandidates-UVES}. 
For the Chamaeleon candidates, except for RXJ1303.5-7701, this 
is a simple process because there is a clear grouping of values $\simeq$15 
km~s$^{-1}$, within a scatter of a few km~s$^{-1}$. This velocity clumping 
correlates extremely well with extant {\sc rv} measurements of objects in the 
Chamaeleon region (J06, \citealt{Guenther07}, \citealt{Nguyen2012}; 
\citealt{Biazzo12}; \citealt{Frasca2015}). We note that the Carina and 
Tuc-Hor associations have systemic RVs of $9.7\pm0.5$ and $20.9\pm0.9$ 
km~s$^{-1}$ respectively (\citealt{Kraus2014}, \citealt{Malo2013}). For the 
Chamaeleon candidates RXJ0850.1-7554, RXJ0951.9-7901 \& RXJ1140.3-8321, which 
are suspected of being members of either the Carina or Tuc-Hor associations instead 
of the $\eta$ Cham associations, our {\sc uves} RVs cannot unambiguously distinguish 
between the three associations. 

For RXJ1303.5-7701 however, this object is either a non-member of one of the 
Chamaeleon regions or a binary system, or both. Comparison of its {\sc uves} RV 
to the one reported by J06 shows that the object is binary in nature 
(RV variant). Moreover, our {\sc uves} spectrum confirms the James et al. finding 
that the object is of early-type spectral-type, having few metal lines and very 
broad H $\alpha$ and H$\beta$ Balmer series line profiles, when indeed a G7 
spectrum was expected \citep{AL2000}. A zoom-in of a red-plate {\sc dss} image 
shows that, in fact, there is a visual double in the {\sc rass} X-ray error circle 
of RXJ1303.5-7701, one star extremely bright, the other one less so. Our 
spectroscopic observations (J06 and {\sc uves} data presented herein) may 
therefore not be for the correct optical counterpart in the {\sc rass} X-ray 
error circle, and in a case of {\em mistaken identity} of the brighter star with the 
fainter one, the true source of the X-ray emission remains elusive. 

However for the Rho Ophiuchus stars in our sample, a 1-d kinematic analysis 
is a little more difficult to interpret $-$ essentially through a paucity of 
existing data for our stars and the nature of their single-epoch measurements. 
However, while there are not extensive, programmatic velocity surveys of the 
Rho Ophiuchus region yet published, the 1-d radial velocity of its low-mass 
members is now quite well established to be $-7.0\pm0.7$ km~s$^{-1}$ 
(\citealt{Kurosawa2006}, \citealt{Guenther07}, \citealt{Prato2007}). With 
the obvious exceptions of RXJ1620.1-2348 and RXJ1624.0-2456, most of our 
Rho Ophiuchus stars do indeed have {\sc uves} {\sc rv}s clustered around the {\sc sfr} 
systemic velocity of $-7.0$ km~s$^{-1}$, albeit with a scatter of $\sim 4$ 
km~s$^{-1}$.

\subsubsection{Radial Velocities: Comparison to J06}\label{RVs-J06}

Because our {\sc uves} spectra were acquired approximately two years after the 
{\sc feros} data presented in the J06 study, a comparison of their time-separated 
RVs allows us to identify obvious binary stars in our sample. In order to facilitate 
such a comparison, we have included a separate column in Table~\ref{SFRcandidates-UVES} 
detailing a star-by-star RV difference between the two studies. From two closely-spaced 
epochs in our {\sc uves} data, the RVs of ROXR1 13 are indicative of orbital motion 
over the course of about 2~hours, albeit that the per-datum precision is quite low 
(2-3 km~s$^{-1}$) due to the high rotation velocity of the object. For the two 
Rho Ophiuchus stars common to our study and the J06 one, there is no strong evidence 
of RV variation.

Three Chamaeleon objects appear to be RV variable. The early-type star RXJ1303.5-7701 
shows a very large $\Delta$RV value and is clearly a spectroscopic binary star, although 
there is no evidence or a second or third set of spectral lines in our {\sc uves} spectra. 
Based on our studies alone, RXJ1201.7-7859 is consistent with being a single-lined spectroscopic 
binary star [SB1], with three RVs quite widely separated from each other, at values 4-5 times 
the per-datum measurement error. Combined with literature RV values however (\citealt{Mala06}; 
\citealt{Torres2006}; \citealt{PMsChamI}; \citealt{epsCha2013}; \citealt{elliott2014}), its 
SB1 is confirmed.

One of the more interesting curiosities is that of RXJ0951.9-7901. Its J06 entry shows 
that it is a rapidly-rotating, double-lined spectroscopic binary [SB2], whose double-line 
features were detected over two different epochs of observation (separated by about 6-days). 
It is worth noting that K-band speckle interferometry and direct imaging of the system 
show that it has no nearby companions in the 0.13-6.0 arcsecond range (\citealt{kohler01}), 
although its J06 K$_{1}$ and K$_{2}$ velocities are large enough that both components of 
an SB2 system at the distance of the Chamaeleon region would easily lie inside an 
0.13$^{"}$ annulus. Our {\sc uves} spectrum only shows a single peak in the cross-correlation 
function, which can be readily explained as having been observed at inferior or superior 
conjunction of an SB2 system. Based on a further three epochs of observations, \cite{Cov97} 
and \cite{Guenther07} list this star as RV-constant, whose RV values are within $\sim$ 
1-2 km~s$^{-1}$ of our {\sc uves} value. Finally, \cite{Torres2006} report four epochs of RV 
measurements for this star, close to our RV value, albeit with quite a large scatter 
(6.1 km~s$^{-1}$); for such a rapidly rotating star, such a scatter is not particularly 
unusual. Its J06 SB2 nature may therefore be in doubt, and further RV observations are 
necessary in order for us to confirm or refute its multiplicity status.

\vspace*{4mm}

\subsection{Lithium abundance of young stars}\label{LiMEM}

Fortunately, we can call on Lithium detection as a criterion in 
assessing youth, and by inference, membership of a young SFR. As 
both Chamaeleon and Rho Ophiuchus are considered to be active, 
or at least very recent, sites of star formation, one would expect 
that insufficient time has passed for substantial depletion of natal 
Lithium (by proton burning - \citealt{Lidepletion}) to have 
occurred. Members of young {\sc sfr}s should therefore contain 
Lithium abundance levels far in excess of their typically-much 
older field star counterparts in the Galaxy.

\begin{figure}
\centering
\epsfig{figure=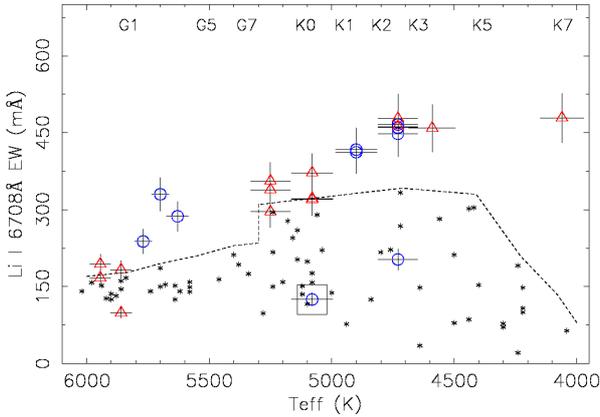,width=85mm}
\caption{Lithium 6708\r{A} equivalent width is plotted against 
effective temperature for our candidate members of the Chamaeleon 
({\textcolor{blue}{$\circ$}}) and Rho Ophiuchus ({\textcolor{red}{$\triangle$}}) 
SFRs. The boxed data-point represents the proper motion non-member of 
the Chamaeleon regions, RXJ1233.5-7523. The hand-drawn dashed black line 
indicates a by-eye estimate of the upper boundary of Lithium for their 
similar-T$_{eff}$ counterparts in the 125~Myr Pleiades open cluster 
({\textcolor{black}{$\ast$}}; \citealt{soderblom1993}).}
\label{Li-UVES-ChamRhoOph}
\end{figure}

Because determining precise and accurate ages of field stars is 
problematic at best, sometimes yielding different ages for different 
methods (e.g. \citealt{Barnes2007}, \citealt{PACDJJ2010}), we choose 
to compare Lithium levels in our young {\sc sfr} candidate members to a 
main sequence open cluster of known age, such as the 125-Myr Pleiades 
(\citealt{Basri96}; \citealt{Plei-LDB}). Plotted in 
Figure~\ref{Li-UVES-ChamRhoOph}, we show the Lithium EWs of Chamaeleon 
and Rho Ophiuchus candidates ({\em c.f.,} Table~\ref{SFRcandidates-UVES}), 
as a function of effective temperature (using the spectral type-temperature 
relation for main sequence stars reported by \citealt{kenyon1995} - 
[KH95]); also included, are corresponding data for the Pleiades 
cluster \citep{soderblom1993}. Clearly, most of the Chamaeleon and 
Rho Ophiuchus candidates are indeed Lithium rich, containing at least 
as much Lithium as their Pleiades counterparts, oftentimes far more. 

There are three notable exceptions to this statement. In Rho 
Ophiuchus, RXJ1620.1-2348 exhibits a Lithium {\sc ew} of only 
99~m\r{A}, considerably below its Pleiades counterparts. 
Given that its {\sc rv} is also considerably discrepant from the 
systemic 1-d velocity of Rho Ophiuchus, we postulate that this 
star is more likely to be a non-member of its parent {\sc sfr} 
than a member, and do not consider it in further in our 
analyses. In the case of the two low-Lithium Chamaeleon 
candidates, simple explanations present themselves readily.

The proper motion data for RXJ1233.5-7523 (\citealt{PMsChamI} -- see 
also Figure~\ref{Kinematics-ChamRhoOph}) show that this star does not 
belong to any of the main sub-groups of the Chamaeleon SFR, and its 
low Lithium EW is indicative of its field star status. Star 
RXJ1140.3-8321 however is a proper motion member of the Eta Cha 
region (6-7 Myr old, \citealt{epsCha2013}), has a set of invariant 
radial velocities consistent with Eta Cha membership, and has a 
\lii 6708 \r{A} EW within $\pm 10$ per cent of the J06 study. However, 
as we have discussed previously in this section, there is evidence to 
suggest that this particular star in fact belongs to the considerably 
older (45 Myr) Tuc-Hor association. While \cite{IC4665RDJ09} warn that 
pre-main sequence [PMS] Lithium depletion in solar-type stars cannot be 
confidently employed as a {\em precise} age indicator in young kinematic 
groups, the Lithium abundance of RXJ1140.3-8321 strongly indicates a 
post-T Tauri star evolutionary status ({\em c.f.,} figure 6 of 
\citealt{IC4665RDJ09} \& figure 7 of \citealt{Kraus2014}). We henceforth 
consider RXJ1140.3-8321 as a Tuc-Hor candidate member.

In the cases of RXJ0850.1-7554 and RXJ0951.9-7901, adding in Lithium into 
the membership consideration, we still cannot distinguish between their 
belonging to $\eta$ Cha or Carina/Tuc-Hor respectively. This is because 
even after 30-40 Myr of stellar evolution post birthline, mid-G stars 
still nave not undergone substantial Lithium depletion, and their position 
in an abundance-temperature plot cannot be differentiated from that of 
very young T-Tauri stars \citep{IC4665RDJ09}. We return to their membership 
status in \S~\ref{HRDs}~\&~\ref{AGES}.

\subsubsection{Lithium: Comparison to J06}\label{Li-J06}

\begin{figure}
\centering
\epsfig{figure=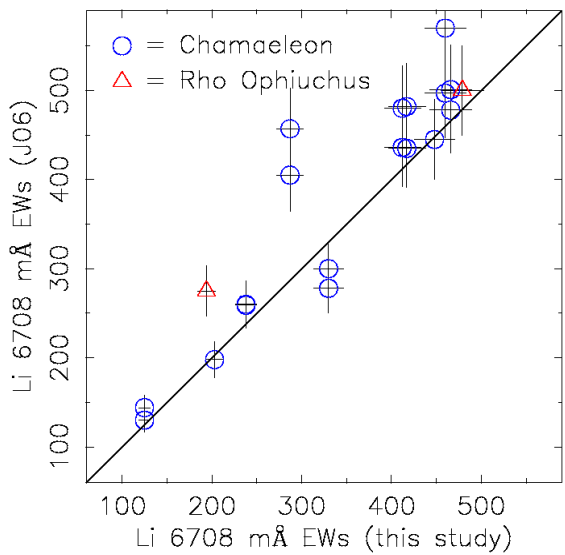,width=80mm,angle=-90}
\caption{Comparison of Li I 6708 \r{A} equivalent widths for stars 
in common between the present study and that of J06. Some stars 
are shown twice, representing multiple epochs of measurement in the 
J06 study. The solid line represents equality between the two studies and 
is not a fit to the data.}
\label{LiUVES-J06}
\end{figure}

Naturally, we ought to compare the variation in \lii 6708\r{A} EWs for those stars 
common to the present study and that of J06 (see Figure~\ref{LiUVES-J06}), because 
outwith the different instruments used, the formalism of the data analysis procedures 
is identical. With the exception of a few systems, there is good agreement between the 
two campaigns. However in the case of the deviant Rho Ophiuchus datum (RXJ1627.1-2419) 
at $\simeq$ 250 m\r{A}, the disagreement can likely be explained by comparison to the 
low quality spectrum (S/N=14) in the J06 study, manifested as a poor Gaussian fit to the 
data and/or poor continuum placement. For the two most deviant Cham stars, one is again 
likely due to a poor quality (S/N=11) J06 spectrum of the Cham I star RXJ1129.2-7546. The 
other, $\eta$ Cham candidate RXJ0951.9-7601 (although, see also \S~\ref{AGES} for details 
of its candidacy of the Tuc-Hor group), it is listed as a double-lined spectroscopic binary 
star in the J06 study, and moreover, of the late-type stars in our samples it is the fastest 
rotator. Potentially, this is an interesting system for a time-domain Lithium variation 
follow-up campaign, although line-blending due to rapid rotation and/or binarity make it 
a problematic system for detailed analysis.

\vspace*{4mm}

\subsection{Reddening Vectors and Extinction}\label{Red-Av}

By comparing our target optical and near-infrared photometry (see 
Table~\ref{SFRcandidates-OPTIRphot}) to theoretical colours based 
on their spectral type (see Table~\ref{SFRcandidates-UVES}), we can 
determine photometric colour-dependent reddening and extinction vectors 
for each {\sc sfr} candidate member; results of this analysis are presented 
in Table~\ref{Reddening-Extinction-Table}. We note that a small handful 
of stars yield astrophysically unrealistic negative reddening values, which 
may result from incorrect spectral type classifications and/or variable 
photometric data (presumably resulting from heightened stellar activity), 
and we decline to calculate extinction values for them. Furthermore, we 
remind the reader that our reddening/extinction analyses are founded upon 
KH95 model parameters, which are based on main sequence stellar properties. 
For stellar objects in young star formation regions, such an approach may 
not be entirely appropriate. In Appendix~\ref{PMS-types}, we re-perform 
our reddening/extinction analyses using the recent pre-main sequence 
model parameters presented by \citet{PM13}, with relevant discussion 
included therein. 

\begin{table*}
\begin{center}
\caption[]{\protect \small Reddening and Extinction Vectors for Chamaeleon and Rho Ophiuchus Targets}
\vspace{3mm} 
\begin{tabular}{lcccccccc}
\hline
~~~~~~~~Target        &  $E(B-V)$$^{a}$  & $E(V-Ic)$$^{a}$ & $E(J-H)$$^{a,b}$ & $E(H-K)$$^{a,b}$ & $A_{v}$$^{c}$ &  $A_{v}$$^{c}$ & $A_{v}$$^{c}$ & $A_{v}$$^{c}$ \\
               &                &               &                &                & [$E(B-V)$] & [$E(V-Ic)$] & [$E(J-H)$] & [$E(H-K)$] \\
\hline 
Chamaeleon \\ \hline
RXJ0850.1-7554        &   0.067 &  0.075  &  0.064 &  0.056 &   0.208  &  0.186    &  0.567   &  0.918   \\
RXJ0951.9-7901        &   0.121 &  ...    &  0.112 &  0.010 &   0.375  &  ...      &  0.991   &  0.167   \\
RXJ1112.7-7637        &   ...   &  0.334  &  0.306 &  0.407 &   ...    &  0.828    &  2.711   &  6.648   \\
RXJ1129.2-7546        &   0.437 &  0.694  &  0.209 &  0.128 &   1.355  &  1.721    &  1.850   &  2.095   \\
RXJ1140.3-8321$^{d}$  &   0.172 &  0.209  &  0.135 & -0.044 &   0.533  &  0.518    &  1.194   &  ...     \\
RXJ1140.3-8321$^{d}$  &   0.072 &  0.139  &  0.096 & -0.054 &   0.223  &  0.345    &  0.847   &  ...     \\
RXJ1158.5-7754a       &   0.209 &  ...    &  0.179 &  0.034 &   0.648  &  ...      &  1.584   &  0.561   \\
RXJ1159.7-7601        &   0.180 &  0.227  &  0.187 &  0.047 &   0.558  &  0.563    &  1.655   &  0.773   \\
RXJ1201.7-7859        &   0.01  & -0.01   & -0.012 &  0.031 &   0.031  &  ...      &  ...     &  0.510   \\
%RXJ1233.5-7523        &  -0.112 &  9.999  & -0.098 &  0.009 &          &  9.999    &          &  0.147   \\
RXJ1239.4-7502$^{d}$  &   0.107 &  0.083  &  0.036 &  0.058 &   0.332  &  0.206    &  0.319   &  0.952   \\
RXJ1239.4-7502$^{d}$  &   0.027 &  0.013  & -0.003 &  0.058 &   0.084  &  0.032    &  ...     &  0.952   \\ \hline
Rho Ophiuchus \\ \hline
RXJ1620.7-2348        &   0.304 &  0.495  &  0.203 &  0.086 &   0.942  &  1.228    &  1.795   &  1.410   \\
RXJ1621.0-2352        &   0.002 &  0.005  & -0.041 &  0.002 &   0.006  &  0.012    &   ...    &  0.039   \\
RXJ1621.2-2347        &   1.312 &  1.809  &  0.746 &  0.418 &   4.067  &  4.486    &  6.608   &  6.828   \\
RXJ1623.1-2300        &   0.343 &  0.521  &  0.215 &  0.041 &   1.063  &  1.292    &  1.903   &  0.675   \\
RXJ1623.4-2425        &   0.513 &  0.962  &  0.379 &  0.220 &   1.590  &  2.386    &  3.361   &  3.594   \\
RXJ1623.5-2523        &   0.448 &  0.784  &  0.225 &  0.176 &   1.389  &  1.944    &  1.994   &  2.878   \\
RXJ1624.0-2456        &   0.757 &  1.250  &  0.452 &  0.209 &   2.347  &  3.100    &  4.005   &  3.416   \\
RXJ1624.8-2239        &   0.124 &  0.208  &  0.083 &  0.078 &   0.384  &  0.516    &  0.739   &  1.279   \\
RXJ1624.8-2359        &   0.840 &  1.509  &  0.671 &  0.300 &   2.604  &  3.742    &  5.943   &  4.902   \\
RXJ1625.0-2508        &   0.740 &  1.153  &  0.409 &  0.165 &   2.294  &  2.859    &  3.626   &  2.696   \\
RXJ1625.4-2346        &   0.388 &  0.777  &  0.336 &  0.141 &   1.203  &  1.927    &  2.981   &  2.307   \\
RXJ1625.6-2613        &  -0.097 &  0.074  &  0.139 &  0.273 &   ...    &  0.184    &  1.233   &  4.447   \\
ROXR1 13              &   1.365 &  2.246  &  0.832 &  0.527 &   4.232  &  5.570    &  7.372   &  8.606   \\
RXJ1627.1-2419        &   1.151 &  2.103  &  0.979 &  0.710 &   3.568  &  5.215    &  8.676   & 11.591   \\ \hline
\end{tabular}\label{Reddening-Extinction-Table} 
\end{center}
\begin{flushleft}
%Notes: \\
a $-$ Reddening vectors are calculated by subtracting appropriate photometric colours for 
each target (see Table~\ref{SFRcandidates-OPTIRphot}) from corresponding theoretical values 
for dwarf stars (from KH95) based on their spectral types (see Table~\ref{SFRcandidates-UVES}). \\
b $-$ $E(J-H)$ and $E(H-K)$ vectors are based on KH95 JHK data which had been first transformed onto 
the {\sc 2mass} system using the \citet{Carpenter2001} relations. \\
c $-$ Extinction vectors [$A_{v}$], based on four colour-dependent reddening vectors, are calculated 
to be: For optical data, Av=3.1$\times$$E(B-V)$ and Av=2.48$\times$$E(V-Ic)$ (from \citealt{BB88}); 
For infrared data, Av=8.86$\times$$E(J-H)$ and Av=16.32$\times$$E(H-K)$ (from \citealt{RM2005}). \\
d $-$ Two sets of reddening and extinction values are calculated, one for each of the two 
spectral sub-types. \\
\end{flushleft}
\end{table*}

For our targets in the Chamaeleon sub-regions, most stars exhibit modest $A_{v}$ 
values ($A_{v}$$\ll1$) derived from optical and near-infrared photometry. Two 
objects however, RXJ1112.7-7637 \& RXJ1129.2-7546, show elevated levels 
of extinction ($A_{v}$$>$1) across the optical and near-infrared range. 
The case for Rho Ophiuchus is considerably more interesting however, 
with $>90$ per cent of the sample showing elevated reddening 
and extinction values across the optical and near-infrared colours; 
such a result is hardly surprising when one considers the abundantly 
clear evidence of interstellar gas and dust to be found in the magnificent 
{\em true colour} images of the Rho Ophiuchus region (e.g. Hatchell et al. 2012). 
Four Rho Ophiuchus stars, RXJ1621.2-2347, RXJ1624.8-2359, ROXR1 13 \& RXJ1627.1-2419, 
exhibit in high ($A_{v}$ $>$ 5) extinction values.

We note that irrespective of whether main-sequence or pre-main sequence theoretical 
data are used, the same stars in Chamaeleon and Rho Ophiuchus are shown to have high 
($A_{v,\lambda}$ $>$ 5) extinction values (c.f. Tables~\ref{Reddening-Extinction-Table} 
\& \ref{Reddening-Extinction-PMSTable}). We posit that these high reddening/extinction 
stars are either {\bf (i)} situated in heavy extinction regions of their parent clouds and/or 
{\bf (ii)} systems hosting circumstellar accretion/debris discs, which act to mimic the 
effects of high reddening/extinction, especially in the near and mid-infrared. Further 
discussion and analysis of the high extinction stars are presented in 
\S~\ref{JHK-excess} \& \S~\ref{S_seds}.

\subsection{Classical or Weak-Lined?}\label{JHK-excess}

\begin{figure}
\centering
\epsfig{figure=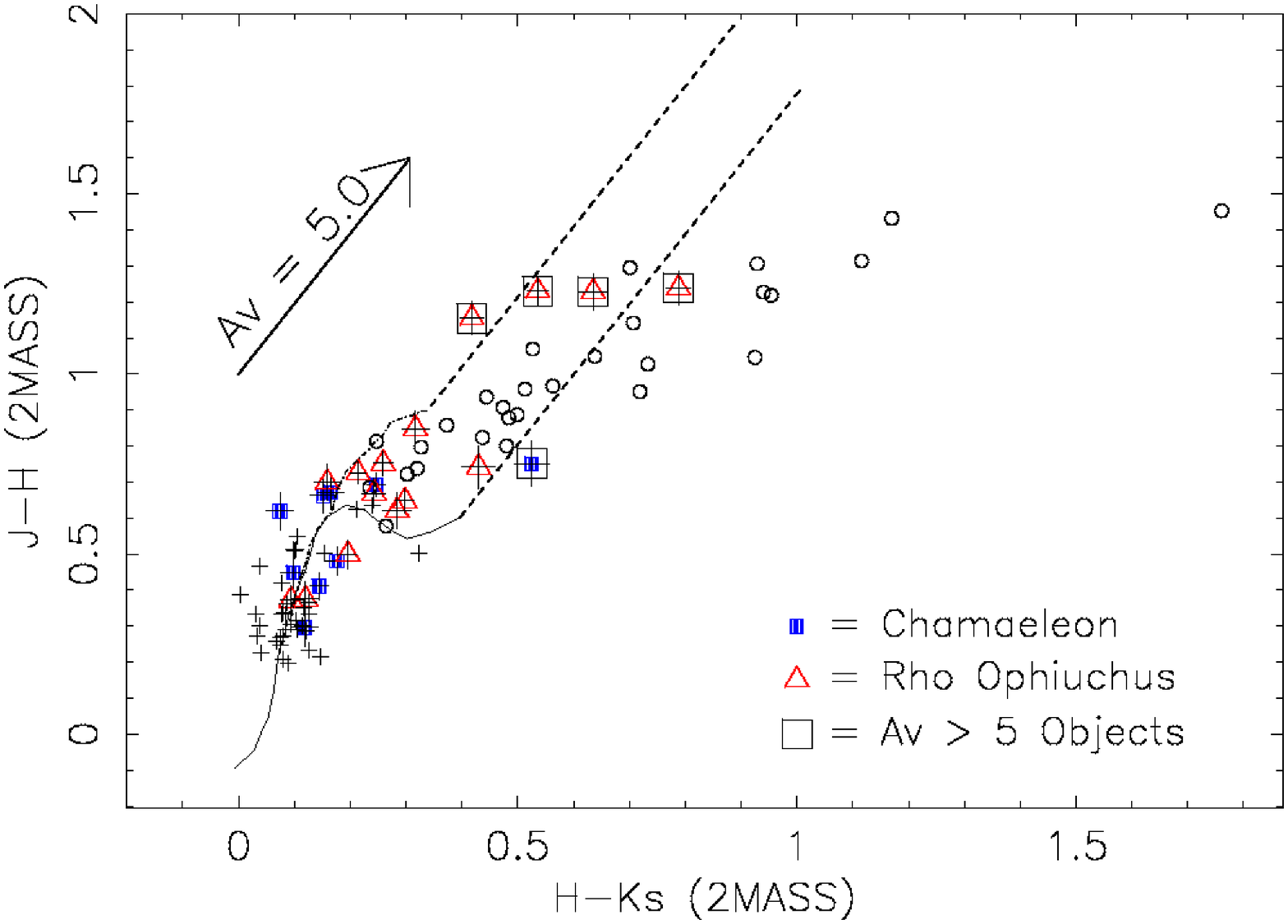,width=85mm}
\caption{Colour-colour diagram using {\sc 2mass} {\sc jhk} photometry 
representing our SFR candidate members. We include for comparison data 
representing field dwarfs (small crosses - \citealt{nidever2002}), 
and Classical T Tauri stars in the Orion SFR (open circles - 
\citealt{neuhauser1995}). The solid and dot-dashed lines represent 
$J-H$/$H-K$ loci for the intrinsic colours of field dwarfs and giants 
(as detailed by \citealt{BB88}), transformed onto the {\sc 2mass} 
JHKs system. The two dashed lines represent reddening vectors 
originating from the extrema of the Bessel and Brett dwarf and 
giant sequences.}
\label{UVES-master-JHHK}
\end{figure}

Exploiting near-infrared photometry can be an efficient, powerful, didactic 
method for triaging classical and weak-lined stars. A quick visual inspection 
of Figure~\ref{UVES-master-JHHK} allows such a methodology to be easily 
understood, although caution must be used in interpreting high interstellar 
extinction cases. In this colour-colour diagram, we plot {\sc jhk} photometry 
for our candidate members of the Chamaeleon and Rho Ophiuchus  {\sc sfr}s, as 
well as loci of field dwarfs and giants, and by extension, their high 
extinction loci. For comparison, we also plot the data for a sample of 
F$\rightarrow$ M-dwarfs on the main sequence \citep{nidever2002}, as well 
as data for a sample of known Classical T Tauri stars in the very young 
($\simeq 1-2$ Myr) Orion Nebula Cluster (e.g. \citealt{neuhauser1995}).

This colour-colour diagram, and the data presented in 
Table~\ref{Reddening-Extinction-Table}, show that one star in 
Chamaeleon and four stars in Rho Ophiuchus exhibit substantial infrared 
excess, having data positions well-displaced from un-reddened main-sequence 
late-type dwarf and giant loci. Interestingly, these five candidates, 
RXJ1112.7-7637 (Cham), ROXR1 13 ($\rho$ Oph), RXJ1621.2-2347 ($\rho$ Oph), 
RXJ1624.8-2359 ($\rho$ Oph) and RXJ1627.1-2419 ($\rho$ Oph) yield $E(J-H)$ 
and/or $E(H-K)$ derived extinctions of $A_{v}$ $>$ 5. The remainder of our 
sample exhibits more moderate extinction, consistent with their having 
little or moderate infrared excess emission. A handful of stars have infrared 
colours that lie close to the giant locus, and in-between the giant and dwarf 
loci, which if confirmed as young members of their parent {\sc sfr}s, one can 
imagine interpreting them as being sub-giant-like stars still in the very 
early pre-main sequence phase ($<$10Myr), having somewhat inflated radii 
thereby mimicking giant-like stellar properties.

These five high extinction objects ($A_{v}$$>$5), are Lithium rich 
(see Table~\ref{SFRcandidates-UVES} and \citealt{M98}), and are 
therefore good candidates for classical T-Tauri star status. 
Historically, albeit with some definition differences among various 
authors and specific SFRs, classical T-Tauri stars invariably 
exhibit very strong Balmer series H $\alpha$ emission 
(EW$_{H\alpha}$ $> 5-10$\r{A} $-$ \citealt{Barrado2003}, \citealt{M98}). 
Only one of our targets, RXJ1625.6-2613, shows H $\alpha$ emission 
at these CTTS-like levels (see also \S~\ref{CTTS-notWTTS}), and in 
fact, none of the five high extinction stars have EW$_{H\alpha}$ $> 5$\r{A}. 
The four high extinction Rho Ophiuchus stars are listed as being either WTTS 
or young solar-type stars in \citet{M98}. While several objects in our sample 
exhibit H $\alpha$ EWs of order $1-3$\r{A}, these activity emission levels are 
consistent with rapidly rotating late-type stars early on the {\sc pms} and zero-age 
main sequence (e.g. \citealt{SSHJ93}, \citealt{JJ97}, \citealt{M98}, 
\citealt{N2547-RDJ2000}, \citealt{IC4665RDJ09}).

\subsubsection{H $\alpha$ Activity: Comparison to J06}\label{Ha-J06}

\begin{figure}
\centering
\epsfig{figure=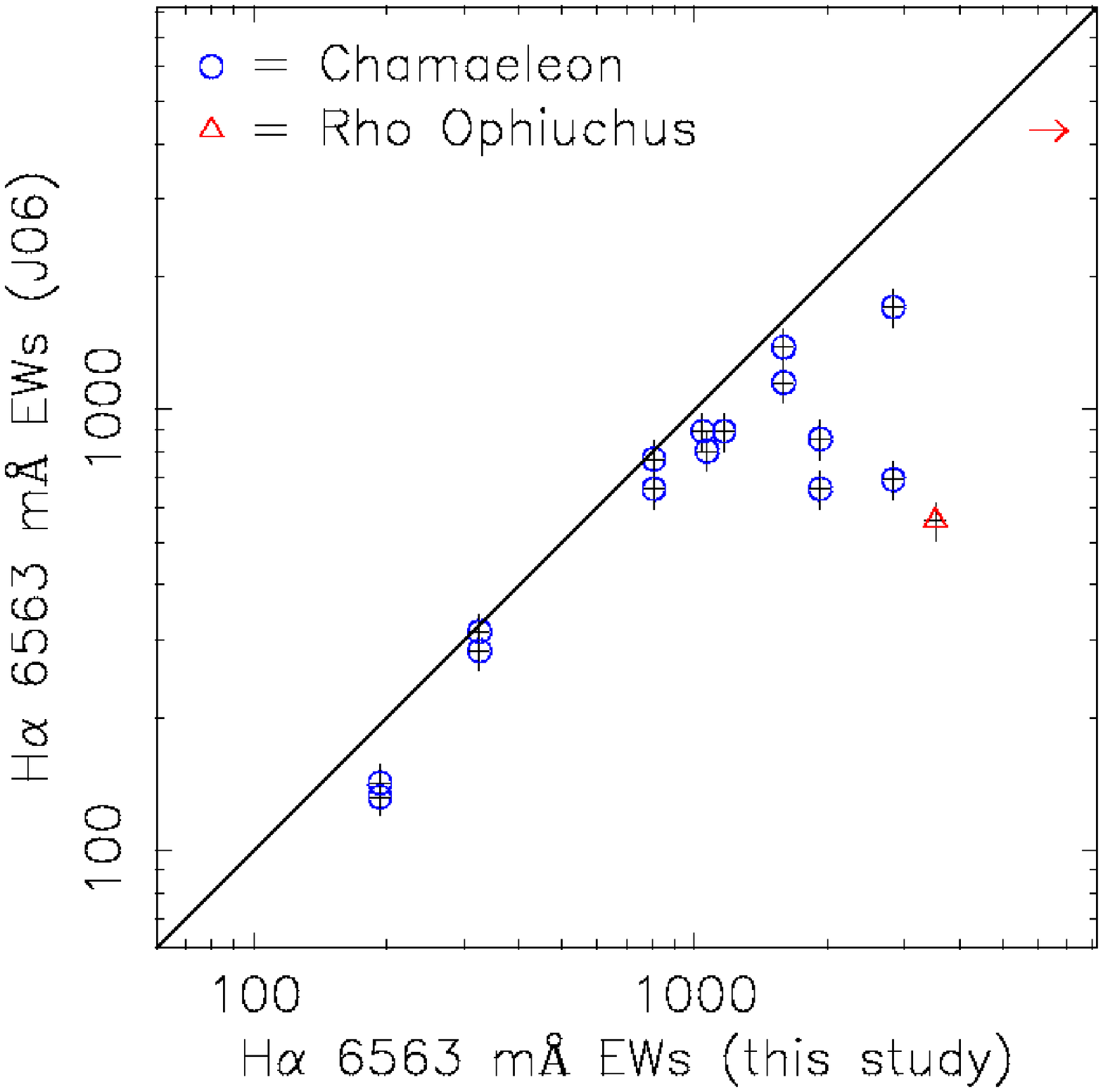,width=80mm,angle=-90}
\caption{Comparison of Balmer series H $\alpha$ 6563\r{A} equivalent widths 
for stars in common between the present study and that of J06. Some stars 
are presented twice, representing multiple epochs of measurement in the 
J06 study. The solid line represents equality between the two studies and 
is not a fit to the data.}
\label{HaUVES-J06}
\end{figure}

The most likely scenario where activity in the Balmer series H $\alpha$ 6563 \r{A} 
line is temporally variable for our targets is two-pronged. For such young stars, some 
combination of circumstellar disc-accretion (\citealt{ONC-Ha}; \citealt{NGC2264-Ha}) 
and/or chromospheric magnetic activity (\citealt{Eibe98}; \citealt{Barnes01}; 
\citealt{Bell2012}) can readily explain short- and long-term spectral line variations 
in the Balmer series H $\alpha$ species.

Comparing H $\alpha$ emission in the present sample to the J06 one (see 
Figure~\ref{HaUVES-J06}) reveals several interesting objects. In the two 
Rho Oph cases, there is convincing empirical evidence to suggest that both 
stars have circumstellar accretion discs (see \S~\ref{S_seds} and 
Figure~\ref{F_juicyseds}), and high levels of extinction (at least $A_{v}>4$), 
which naturally provides a physical environment with which to explain their 
Balmer series H $\alpha$ variable emission over widely-separated epochs.

In the case of Chamaeleon candidates, there are two clear cases of variable 
H $\alpha$ emission between the J06 and present study. The first arises from the 
very rapid rotator RXJ0951.9-7901, which is listed as a double-lined spectroscopic 
binary in J06, and will be re-classified in this study as a 45 Myr Tuc-Hor object 
(see discussion in \S~\ref{AGES}). Given its rapid rotation, modest extinction 
(see Table~\ref{Reddening-Extinction-Table}), post-T Tauri age and clean bare 
photosphere {\sc sed} (see \S~\ref{S_seds} and Figure~\ref{F_bareSEDsCham}), its 
time variable H $\alpha$ emission is most likely due to a chromospherically-induced 
temporal event such as stellar activity $-$ and its cycle $-$ and/or a magnetic activity 
flare. The second is a Cham I object (RXJ1112.7-7637), which exhibits high levels of 
extinction ($A_{v}>5$) and shows clear evidence of near- and mid-infra red excess in 
its {\sc sed} (see Figure~\ref{F_juicyseds}). In this case, we feel comfortable 
attributing its H $\alpha$ emission variability to physical processes associated 
with its circumstellar accretion disc.

\subsection{Hertzsprung-Russell Diagrams: Mass, Age }\label{HRDs}

Placing observational parameters onto the theoretical plane is a 
well-established and well-calibrated method of determining fundamental 
stellar characteristics. Converting photometry and spectral types into 
bolometric luminosity and effective temperatures allows the construction 
of HRDs, which when combined with theoretical stellar models, yields 
mass and age determinations for a sample of stars. 

\begin{figure}
\epsfig{figure=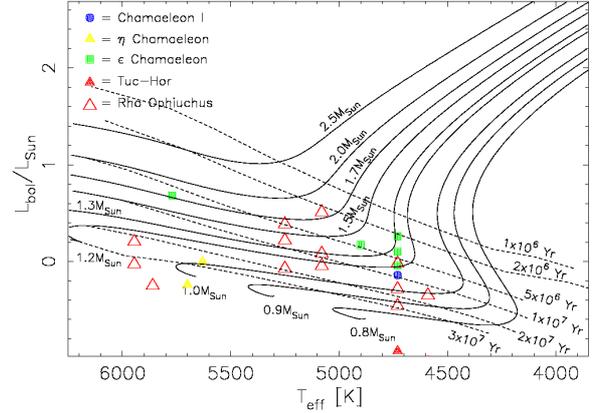,angle=270,width=84mm}
\caption{An Hertzsprung-Russell diagram, using $A_{v}$ extinction values 
derived from $B-V$ colours (see Table~\ref{Reddening-Extinction-Table}), 
is plotted for Cham I (\tikzcircle[blue, fill=blue]{2pt}), $\eta$ Cham 
({\Montriangle{yellow}}), $\epsilon$ Cham ({\Monsquare{green}}), Tuc-Hor 
({\Montriangle{red}}) \& Rho Ophiuchus  ({\textcolor{red}{$\triangle$}}) 
candidates. Effective 
temperatures, bolometric corrections and reddening vectors are based 
on KH95 colour-spectral type relationships for dwarf stars, with distances 
to individual stellar groups cited in Table~\ref{AGEtable}. Stellar isochrones 
(dashed tracks) and mass tracks (solid lines) are computed from {\sc parsec} 
solar metallicity, theoretical stellar models \citep{PARSEC}.}\label{KH95HRD-AvBV}
\end{figure}

\begin{figure}
\epsfig{figure=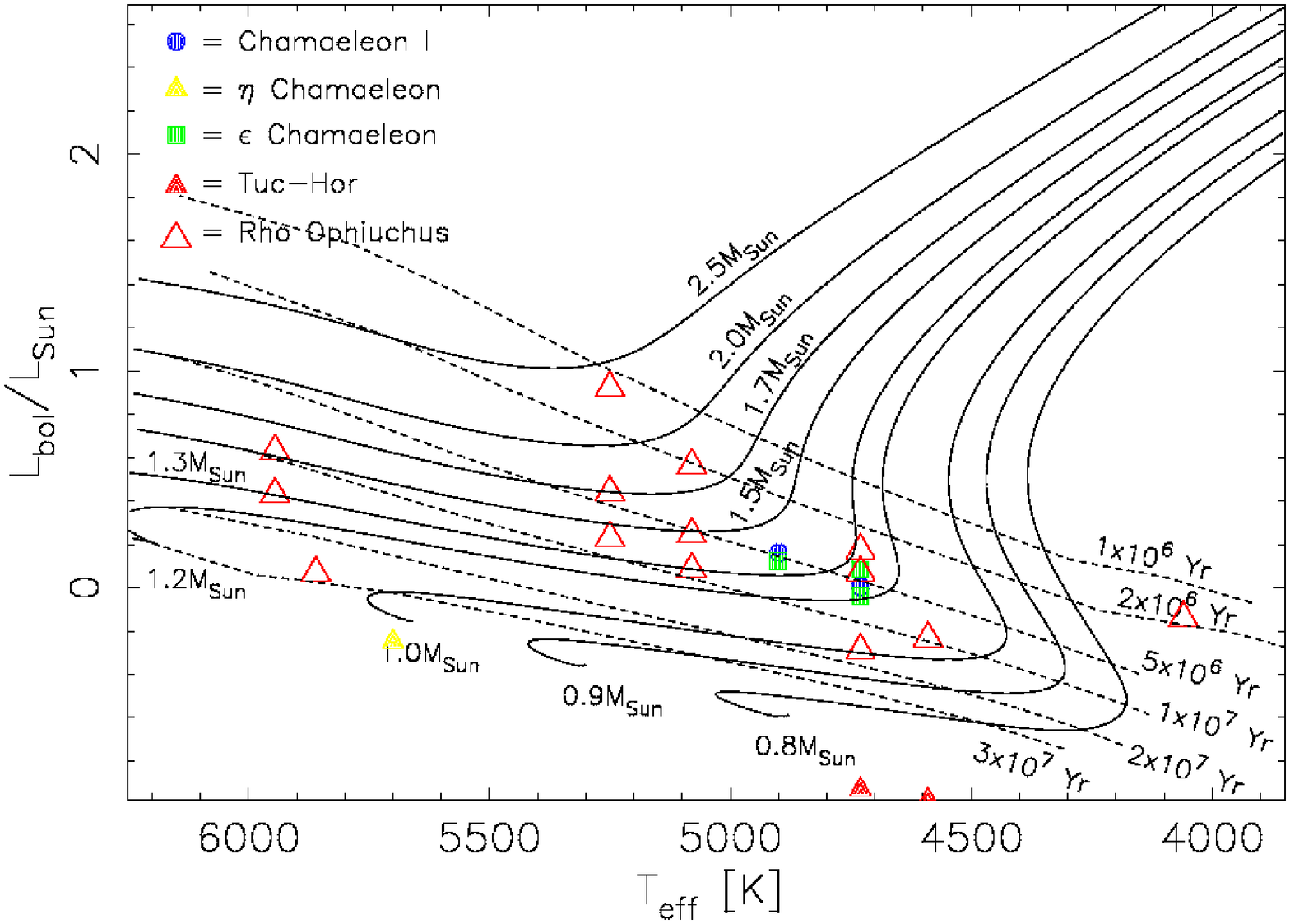,angle=270,width=84mm}
\caption{{\em Idem} to Figure~\ref{KH95HRD-AvBV}, except $A_{v}$ extinction 
values are derived from $V-Ic$ colours.}
\label{KH95HRD-AvVI}
\end{figure}

\begin{figure}
\epsfig{figure=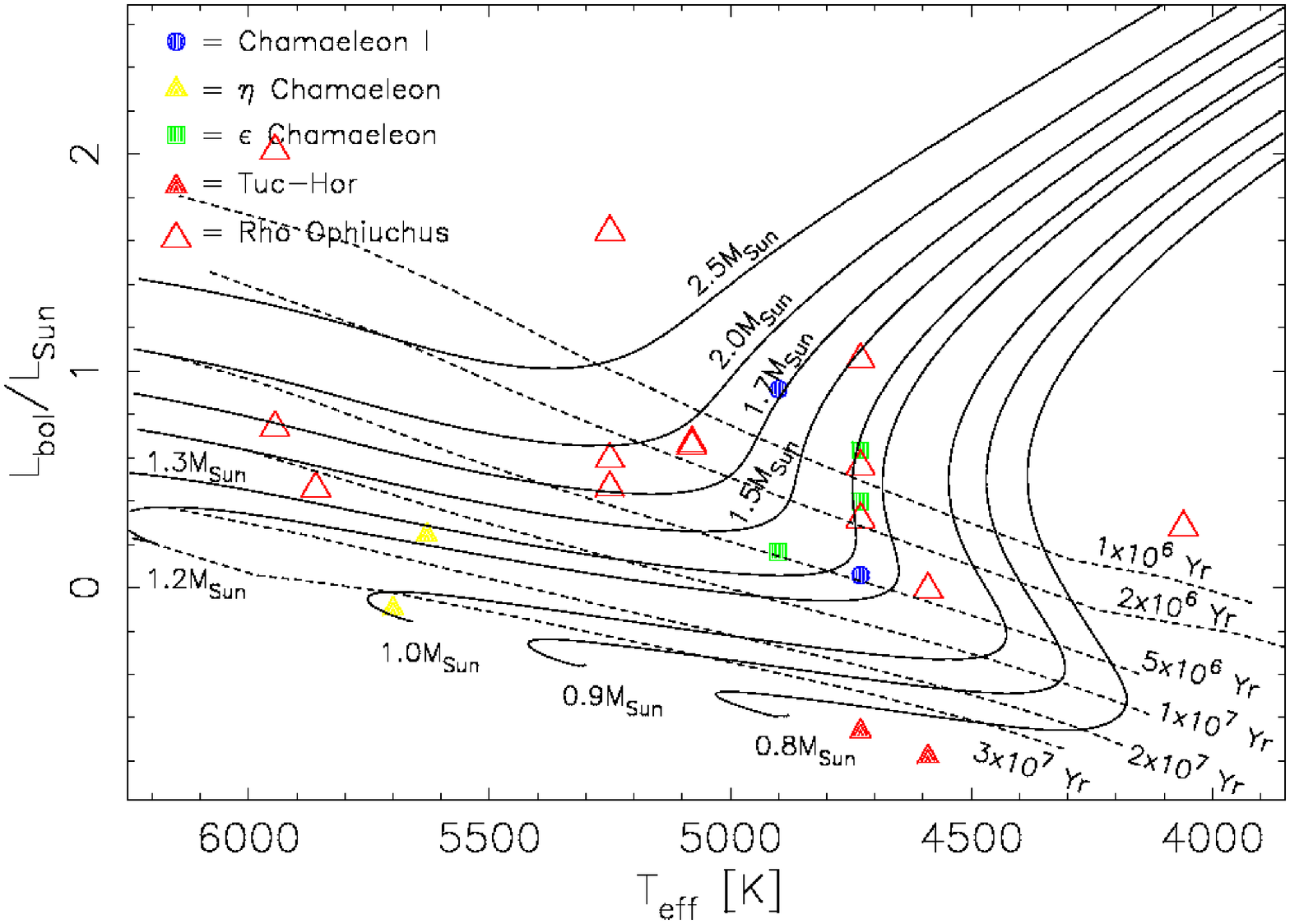,angle=270,width=84mm}
\caption{{\em Idem} to Figure~\ref{KH95HRD-AvBV}, except $A_{v}$ extinction 
values are derived from $J-H$ colours.}
\label{KH95HRD-AvJH}
\end{figure}

\begin{figure}
\epsfig{figure=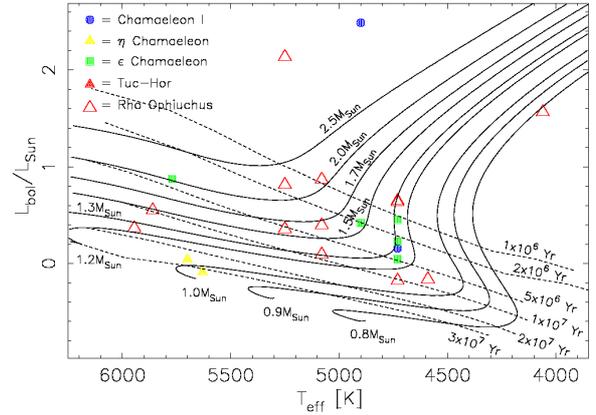,angle=270,width=84mm}
\caption{{\em Idem} to Figure~\ref{KH95HRD-AvBV}, except $A_{v}$ extinction 
values are derived from $H-Ks$ colours.}
\label{KH95HRD-AvHK}
\end{figure}

Using main sequence bolometric corrections and spectral type-photometric 
colours \& effective temperature relations from KH95, we calculate bolometric 
luminosity for each of our targets employing their V-magnitudes detailed 
in Table~\ref{SFRcandidates-OPTIRphot} and four separate extinction values 
($A_{v}$), based on $B-V$, $V-I$, $J-H$, and $H-K$ colour (see 
Table~\ref{Reddening-Extinction-Table}). Distances employed for luminosity 
calculations are detailed in the footnotes to Table~\ref{AGEtable}, and all 
theoretical models are based on solar metallicity PAdova and TRieste Stellar 
Evolution Code [{\sc parsec} - \citep{PARSEC}] stellar evolution models in our 
analysis, a pre-main sequence Deuterium burning phase with initial abundance 
of $2.12\times10^{-5}$, and incorporating an initial helium fraction of Y$=$0.249. 
The results of our HRD analyses are detailed in Tables~\ref{AGEtable}~\&~\ref{MASStable}, 
and graphically represented in Figures~\ref{KH95HRD-AvBV}$-$\ref{KH95HRD-AvHK}. 

Based on proper motion and radial velocity vectors (see \S~\ref{membership}), 
we triage our Chamaeleon candidates into their parent kinematic group. However, 
in the case of RXJ1140.3-8321, in going forward we consider it more likely a 
member of the older $\simeq 40$ Myr Tuc-Hor association (\citealt{Bell2015}; 
\citealt{Kraus2014}; \citealt{elliott2014}), and use the \citet{elliott2014} 
distance of 48pc in calculating its bolometric luminosity. For RXJ0850.1-7554 
and RXJ0951.9-7901, we start with the assumption that the objects are candidate 
members of the $\eta$ Cha association. However, in light of the arguments presented 
in \S~\ref{membership}$-$ \S~\ref{LiMEM}, after having considered their HRD properties, 
we will re-evaluate their membership properties at the end of \S~\ref{AGES}.

It has occurred to us that we are using main sequence theoretical models to determine 
effective temperatures and bolometric corrections, which may not be wholly appropriate 
for stars in young SFRs. While we continue with this analysis, we also perform an HRD 
age and mass analysis using pre-main sequence empirically calibrated stellar parameters, 
the results of which are presented in \S~\ref{PMS-types}.

\subsubsection{HRD: Stellar Ages}\label{AGES}

While small number statistics for the individual Chamaeleon sub-groups, 
as well as the Rho Ophiuchus sample, precludes a detailed analysis of their 
ages, a colour-dependent HRD analysis can be informative in the assessment 
of stellar youth. One {\em caveat lector} to bear in mind is that five stars, 
RXJ1112.7-7637 (Cham), ROXR1 13 ($\rho$ Oph), RXJ1621.2-2347 ($\rho$ Oph), 
RXJ1624.8-2359 ($\rho$ Oph) and RXJ1627.1-2419 ($\rho$ Oph) show evidence 
of elevated extinction ($A_{v}$ $>$ 5), as well as exhibiting strongly 
colour-dependent extinction vectors (see Table~\ref{Reddening-Extinction-Table}), 
characteristics likely to produce colour-dependent ages and masses in the HRDs.

HRDs ages of our limited numbers of stars across the Cham {\footnotesize I} and 
$\epsilon$ Chamaeleon regions are broadly consistent with young SFRs having ages 
\lsi 5-10 Myr. Both Cham {\footnotesize I} stars have mean ages $<$ 5Myr, which are 
perfectly aligned with the results of \citet{LuhmanChaI} and \citet{Spina2014ChaI}. 
Interpreting the mean age determinations for our four $\epsilon$ Cham stars is also 
straightforward, as they all return consistent values of \lsi 5 Myr, in agreement with 
the \citet{feigl2003} and \citet{epsCha2013} studies. 

For the $\eta$ Cham candidates, both RXJ0850.1-7554 and RXJ0951.9-7901 appear to 
be considerably older than the 5-10 Myr age of the group (\citealt{LawsonEtaCha}, 
\citealt{LuhmanEtaCha}), having ages at least 25-30 Myr, more appropriate to those 
of post T-Tauri stars. This supposition for both of these objects is consistent with 
the \citet{elliott2014} assertion that they are members of older associations (30 Myr 
Carina in the case of RXJ0850.1-7554, and the $\simeq 40$ Myr Tuc-Hor group in the 
case of RXJ0951.9-7901). For RXJ1140.3-8321, which was already assumed to be a 
Tuc-Hor member, its K95 HRD analysis unsurprisingly yields isochronal ages of 
$\sim 50$ Myr for both measurements of its K3 and K4 spectral types.

In the case of Rho Ophiuchus stars, their HRD mean ages show clear evidence 
of a group of young SFR objects broadly consistent with a $\simeq$ 
1-10 Myr distribution, supporting conclusions found by \citet{LuhmanRhoOph99}, 
\citet{Wilking2005} \& \citet{AdO2010}. Several stars exhibit a 
spread in ages, based on different colour extinction relations in 
the \lbol calculation, of an order-of-magnitude or more, always 
decreasing in age as the extinction relationships derived go from blue 
to red reddening vectors ({\em i.e.,} $E(B-V)$ $\rightarrow$ $E(H-K)$). 
Of these targets, RXJ1621.2-2347, RXJ1623.4-2425, RXJ1624.8-2359, 
RXJ1625.4-2346, RXJ1627.1-2419, three exhibit $A_{v}$$>$ 5 values 
(see \S~\ref{Red-Av}). As to why the age determinations of RXJ1623.4-2425 
especially, and RXJ1625.4-2346, change so markedly as a function 
of colour-dependent extinction we cannot explain. We also note that 
RXJ1623.4-2425 and RXJ1625.0-2508 appear to be considerably older 
than the remainder of the Rho Ophiuchus sample.

\begin{table*}
\vspace{3mm}
\caption[]{\protect \small Isochronal ages for Chamaeleon and Rho Ophiuchus candidates 
using four separate colour-dependent extinction laws (based on KH95 effective 
temperatures and bolometric corrections for dwarf stars) and {\sc parsec} stellar models.}
\begin{tabular}{lcccccccccc}
\hline
~~~~~~~~~Target$^{a}$              & $T_{\mathrm{eff}}$$^{b}$ & [\lbol ]$^{c}$  & Age$^{a}$  & [\lbol]$^{c}$  & Age$^{a}$     & [\lbol]$^{c}$   & Age$^{a}$    & [\lbol]$^{c}$   & Age$^{a}$       & Mean age             \\
                                   &[K]   &           & [Myr]            &          & [Myr]   &           & [Myr]  &           & [Myr]     & [Myr]                \\ \hline 
                                   &      & \multicolumn{2}{c}{$^{b}$$E(B-V)$}  & \multicolumn{2}{c}{$^{b}$$E(V-I)$} & \multicolumn{2}{c}{$^{b}$$E(J-H)$} &\multicolumn{2}{c}{$^{b}$$E(H-K)$}  &  \\
% \cmidrule(lr){1-1}                            \cmidrule(lr){3-4}               \cmidrule(lr){5-6}           \cmidrule(lr){7-8}             \cmidrule(lr){9-10} 
                                               \cmidrule(lr){3-4}               \cmidrule(lr){5-6}           \cmidrule(lr){7-8}             \cmidrule(lr){9-10} 
Chamaeleon I   (\mycircle{blue})   &      &           &                  &          &         &           &        &           &           &                      \\
 \cmidrule(lr){1-1} 
RXJ1112.7-7637                     & 4900 &  ~~9.999  &   ...            & ~~0.161  &  5.0    & ~~0.915   & $<1$   & ~~2.489   & $<1$      &   $<2.3$             \\
RXJ1129.2-7546                     & 4730 &  $-0.141$ &   9.5            & ~~0.005  &  5.5    & ~~0.057   & 5.0    & ~~0.155   &  3.5      &   5.9 $\pm$ 1.3      \\
$\eta$ Cham  (\mytriangle{yellow}) &      &           &                  &          &         &           &        &           &           &                      \\
 \cmidrule(lr){1-1}
RXJ0850.1-7554                     & 5700 &  $-0.242$ &  $>30$           & $-0.251$ &  $>30$  & $-0.099$  &  $>30$ & ~~0.042   &   27      &   $>29$              \\ 
RXJ0951.9-7901                     & 5630 &  $-0.005$ &     26           & ~~9.999  &  ...    & ~~0.241   &    15  & $-0.089$  & $>30$     &   $>24$              \\
Tuc Hor  ({\Montriangle{red}})     &      &           &                  &          &         &           &        &           &           &                      \\
 \cmidrule(lr){1-1}
RXJ1140.3-8321$^\dagger$           & 4730 &  $-0.926$ &   $>30$          & $-0.932$ &  $>30$  & $-0.662$  & $>30$  & ~~9.999   &  ...      &  $>30$               \\
RXJ1140.3-8321$^\ddagger$          & 4590 &  $-1.030$ &   $>30$          & $-0.981$ &  $>30$  & $-0.780$  & $>30$  & ~~9.999   &  ...      &  $>30$               \\
$\epsilon$ Cham (\ssquare{green})  &      &           &                  &          &         &           &        &           &           &                      \\
 \cmidrule(lr){1-1}
RXJ1158.5-7754a                    & 4730 &  ~~0.260  &    2.0           & ~~9.999  &  ...    & ~~0.634   &   $<1$ & ~~0.225   &  2.5      &   $<1.8$             \\
RXJ1159.7-7601                     & 4730 &  $-0.041$ &    6.5           & $-0.039$ &  6.5    & ~~0.398   &    1.5 & ~~0.045   &  5.0      &   4.9 $\pm$ 1.2      \\
RXJ1201.7-7859                     & 5770 &  ~~0.681  &    6.5           & ~~9.999  &  ...    & ~~9.999   &   ...  & ~~0.872   &  4.5      &   5.5 $\pm$ 1.0      \\
RXJ1239.4-7502$^\dagger$           & 4900 &  ~~0.172  &    4.5           & ~~0.121  &  5.5    & ~~0.166   &   5.0  & ~~0.420   &  2.0      &   4.3 $\pm$ 0.8      \\
RXJ1239.4-7502$^\ddagger$          & 4730 &  ~~0.104  &    4.0           & ~~0.084  &  4.5    & ~~9.999   &   ...  & ~~0.452   &  1.0      &   3.2 $\pm$ 1.1      \\
Rho Ophiuchus ({\textcolor{red}{$\triangle$}}) &  &   &                  &          &         &           &        &           &           &                      \\
 \cmidrule(lr){1-1}
RXJ1620.7-2348                     & 4590 &  $-0.349$ &   15             & $-0.234$ &  9.5    & $-0.008$  &   4.0  & $-0.162$  &  7.5      &  9.0 $\pm$ 2.3       \\ 
RXJ1621.0-2352                     & 5080 &  ~~0.081  &    9.5           & ~~0.084  &  9.0    & ~~9.999   &   ...  & ~~0.094   &  9.0      &  9.2 $\pm$ 0.2       \\ 
RXJ1621.2-2347                     & 4730 &  $-0.458$ &   29             & $-0.290$ &   16    & ~~0.559   &   $<1$ & ~~0.647   & $<1$      &  $<12$               \\ 
RXJ1623.1-2300                     & 4730 &  $-0.021$ &    6.5           & ~~0.070  &  4.5    & ~~0.315   &   2.0  & $-0.176$  &  12       &  6.3 $\pm$ 2.1       \\ 
RXJ1623.4-2425                     & 5860 &  $-0.250$ &  $>30$           & ~~0.068  &   28    & ~~0.458   &    12  & ~~0.551   &   10      &  $>20$               \\ 
RXJ1623.5-2523                     & 5250 &  ~~0.218  &    8.5           & ~~0.440  &  4.5    & ~~0.460   &   4.5  & ~~0.814   &  2.0      &  4.9 $\pm$ 1.3       \\
RXJ1624.0-2456                     & 5250 &  $-0.072$ &   19             & ~~0.230  &  8.0    & ~~0.592   &   3.0  & ~~0.356   &  5.5      &  8.9 $\pm$ 3.5       \\
RXJ1624.8-2239                     & 5080 &  ~~0.511  &    2.5           & ~~0.564  &  2.0    & ~~0.653   &   1.5  & ~~0.869   & $<1$      &  $<1.8$              \\ 
RXJ1624.8-2359                     & 4730 &  $-0.284$ &   16             & ~~0.172  &  3.5    & ~~1.052   &   $<1$ & ~~0.636   & $<1$      &  $<5.4$              \\ 
RXJ1625.0-2508                     & 5945 &  ~~0.205  &     23           & ~~0.431  &   15    & ~~0.738   &   8.0  & ~~0.366   &   17      &  15.8 $\pm$ 3.1      \\
RXJ1625.4-2346                     & 5080 &  $-0.045$ &   15             & ~~0.245  &  5.5    & ~~0.666   &   1.5  & ~~0.397   &  3.5      &  6.4 $\pm$ 3.0       \\
RXJ1625.6-2613                     & 4060 &  ~~9.999  &    ...           & $-0.140$ &  2.0    & ~~0.279   &   $<1$ & ~~1.565   & $<1$      &  $<1.3$              \\
ROXR1 13                           & 5250 &  ~~0.386  &    5.0           & ~~0.921  &  1.0    & ~~1.642   &   $<1$ & ~~2.136   & $<1$      &  $<2.0$              \\
RXJ1627.1-2419                     & 5945 &  $-0.028$ &  $>30$           & ~~0.630  &  5.5    & ~~2.015   &   $<1$ & ~~3.181   & $<1$      &  $\simeq 9$ (?)      \\
\hline
\end{tabular}\label{AGEtable}
\begin{flushleft}
%Notes: \\
$a -$ Isochronal age determinations are segregated into stars comprising disparate young {\sc sfr} regions, with symbols 
\mycircle{blue}, \mytriangle{yellow}, {\Montriangle{red}}, \ssquare{green} \& {\textcolor{red}{$\triangle$}}, 
matching those data presented in Figures~\ref{KH95HRD-AvBV}$-$\ref{KH95HRD-AvHK}, using solar metallicity {\sc parsec} 
models \citep{PARSEC}. \\
$b -$ \lbol~ data are calculated using \teff and reddening vectors based on KH95 colour-spectral types relationships 
for dwarf stars (see also Tables~\ref{SFRcandidates-UVES}~\&~\ref{Reddening-Extinction-Table}). \\
$c -$ Distances used for \lbol ~calculations are as follows: Cham I (\mycircle{blue}) d=160pc \citep{Whittet97}; 
$\eta$ Cham (\mytriangle{yellow}) d=97pc (\citealt{epsCha2013}; \citealt*{Bell2015})  ; $\epsilon$ Cham (\ssquare{green}) 
d=115pc \citep{epsCha2013}; Tuc-Hor ({\Montriangle{red}}) d=48pc \citep{elliott2014}; 
Rho Ophiuchus ({\textcolor{red}{$\triangle$}}) d=131pc \citep{mama08}. \\
$\dagger$ $-$ for a given star having two spectral types, calculations are made using earlier spectral type/higher \teff. \\
$\ddagger$ $-$ {\em Idem}, calculations are made using later spectral type/lower \teff. \\
\end{flushleft}
\end{table*}

\subsubsection{HRD: Stellar Masses}

With the exception of three stars, RXJ1112.7-7637, ROXR1 13 and RXJ1627.1-2419, 
stellar masses derived from the HRDs are remarkably consistent (see 
Table~\ref{MASStable}), and show that our sample is essentially composed of 
stars $1-2$ times the mass of the Sun. The masses for one Cham I, and three 
Rho Ophiuchus objects, are not constrained, which except for our coolest K7e 
star, is caused by strong ($A_{v}$$>$5) and variable extinction across the 
four photometric colours we consider.

\begin{table*}
\vspace{3mm}
\caption[]{\protect \small {\sc parsec} model masses for Chamaeleon and Rho Ophiuchus candidates 
using four separate colour-dependent extinction laws (based on KH95 effective 
temperatures and bolometric corrections for dwarf stars).}
\begin{tabular}{lcccccc}
\hline
~~~~~~~~~Target                    & $T_{\mathrm{eff}}$ &   Mass              & Mass                & Mass              & Mass               &  Mean             \\
                                   &  [K]               &  [M$_{\odot}$]      & [M$_{\odot}$]       & [M$_{\odot}$]     & [M$_{\odot}$]      &  Mass [M$_{\odot}$]    \\
                                   &                    &   $E(B-V)$          & $E(V-I)$            & $E(J-H)$          & $E(H-K)$          &                   \\
                                                          \cmidrule(lr){3-3}  \cmidrule(lr){4-4}   \cmidrule(lr){5-5}  \cmidrule(lr){6-6} 
Chamaeleon I   (\mycircle{blue})   &                    &                     &                     &                   &                    &                   \\
 \cmidrule(lr){1-1}   
RXJ1112.7-7637                     & 4900               &    ...              &   1.4             &  1.7               &  $>2.5$             &   $>1.9$          \\
RXJ1129.2-7546                     & 4730               &    1.1              &   1.2             &  1.2               &  1.3                &   1.2 $\pm$ 0.04  \\
$\eta$ Cham  (\mytriangle{yellow}) &                    &                     &                   &                    &                     &                   \\
 \cmidrule(lr){1-1}
RXJ0850.1-7554                     & 5700               &    1.0              &   1.0             &  1.0               &  1.0                &   1.0  $\pm$ ...  \\
RXJ0951.9-7901                     & 5630               &    1.0              &   ...             &  1.2               &  1.0                &   1.07 $\pm$ 0.07 \\
Tuc Hor  ({\Montriangle{red}})     &                    &                     &                   &                    &                     &                   \\
 \cmidrule(lr){1-1}
RXJ1140.3-8321$^\dagger$           & 4730               &    ...              &   ...             &  ...               &  ...                &   ...             \\
RXJ1140.3-8321$^\ddagger$          & 4590               &    ...              &   ...             &  ...               &  ...                &   ...             \\
$\epsilon$ Cham (\ssquare{green})  &                    &                     &                   &                    &                     &                   \\
 \cmidrule(lr){1-1}
RXJ1158.5-7754a                    & 4730               &    1.3              &   ...             &  1.3               &  1.3                &   1.3  $\pm$ ...  \\
RXJ1159.7-7601                     & 4730               &    1.4              &   1.2             &  1.3               &  1.2                &   1.28 $\pm$ 0.05 \\
RXJ1201.7-7859                     & 5770               &    1.7              &   ...             &  ...               &  2.0                &   1.85 $\pm$ 0.15 \\
RXJ1239.4-7502$^\dagger$           & 4900               &    1.4              &   1.3             &  1.4               &  1.5                &   1.40 $\pm$ 0.04 \\
RXJ1239.4-7502$^\ddagger$          & 4730               &    1.4              &   1.3             &  ...               &  1.3                &   1.33 $\pm$ 0.03 \\
Rho Ophiuchus ({\textcolor{red}{$\triangle$}}) &              &                     &                   &                    &                     &                   \\
 \cmidrule(lr){1-1}
RXJ1620.7-2348                     & 4590               &    1.0              &   1.0             &  1.0               &  1.0                &   1.0  $\pm$ ...  \\
RXJ1621.0-2352                     & 5080               &    1.3              &   1.3             &  ...               &  1.3                &   1.3  $\pm$ ...  \\
RXJ1621.2-2347                     & 4730               &    0.9              &   1.0             &  1.3               &  1.3                &   1.13 $\pm$ 0.10 \\
RXJ1623.1-2300                     & 4730               &    1.2              &   1.2             &  1.3               &  1.1                &   1.20 $\pm$ 0.04 \\
RXJ1623.4-2425                     & 5860               &    ...              &   1.1             &  1.4               &  1.5                &   1.33 $\pm$ 0.12 \\
RXJ1623.5-2523                     & 5250               &    1.4              &   1.5             &  1.7               &  2.2                &   1.70 $\pm$ 0.18 \\
RXJ1624.0-2456                     & 5250               &    1.0              &   1.2             &  1.9               &  1.6                &   1.43 $\pm$ 0.20 \\
RXJ1624.8-2239                     & 5080               &    1.7              &   1.8             &  1.9               &  2.0                &   1.85 $\pm$ 0.06 \\
RXJ1624.8-2359                     & 4730               &    1.0              &   1.3             &  1.5               &  1.3                &   1.28 $\pm$ 0.10 \\
RXJ1625.0-2508                     & 5945               &    1.1              &   1.3             &  1.7               &  1.2                &   1.33 $\pm$ 0.13 \\
RXJ1625.4-2346                     & 5080               &    1.3              &   1.5             &  1.9               &  1.6                &   1.58 $\pm$ 0.13 \\
RXJ1625.6-2613                     & 4060               &    ...              &   $<0.8$          &  $<0.8$            &  0.8                &   $<0.8$          \\
ROXR1 13                           & 5250               &    1.6              &   2.4             &  $>2.5$            &  $>2.5$             &   $>2.3$          \\
RXJ1627.1-2419                     & 5945               &    1.1              &   1.5             &  $>2.5$            &  $>2.5$             &   $>1.9$          \\ \hline
\end{tabular}\label{MASStable}
\begin{flushleft}
Notes: \\
$a -$ {\sc parsec} model mass determinations are segregated into stars comprising disparate young {\sc sfr} regions, 
with symbols \mycircle{blue}, \mytriangle{yellow}, {\Montriangle{red}}, \ssquare{green} \& {\textcolor{red}{$\triangle$}}, 
matching those data presented in Figures~\ref{KH95HRD-AvBV}$-$\ref{KH95HRD-AvHK}. \\
$b -$ \lbol data are calculated using \teff and reddening vectors based on KH95 colour-spectral types 
relationships for dwarf stars (see also Tables~\ref{SFRcandidates-UVES}~\&~\ref{Reddening-Extinction-Table}). \\
$c -$ Distances used for \lbol calculations are the same as those presented in the footnotes to 
Table~\ref{AGEtable}. \\
$\dagger$ $-$ for a given star, calculations are made using earlier spectral type/higher \teff. \\
$\ddagger$ $-$ for a given star, calculations are made using later spectral type/lower \teff. 
\end{flushleft}
\end{table*}

The observed age and mass dependence for some of our objects on the colour used to 
derive $A_{v}$ is likely the product of excess emission from circumstellar discs. For 
instance, a target with excess emission in the red/infrared will appear older when 
using bluer colours, i.e. $B-V$ and $V-I$, due to a downward shift on the diagram. 
To further examine this disc hypothesis, in \S~\ref{S_seds} compare our Y4KCam and 
2MASS photometry for each of our objects, as well as publically available infrared 
photometry from several other instruments, with theoretical spectral energy distributions 
for young stellar objects with discs.

\subsubsection{HRD: Comparing Stellar Models:}

\begin{figure}
\centering
\epsfig{figure=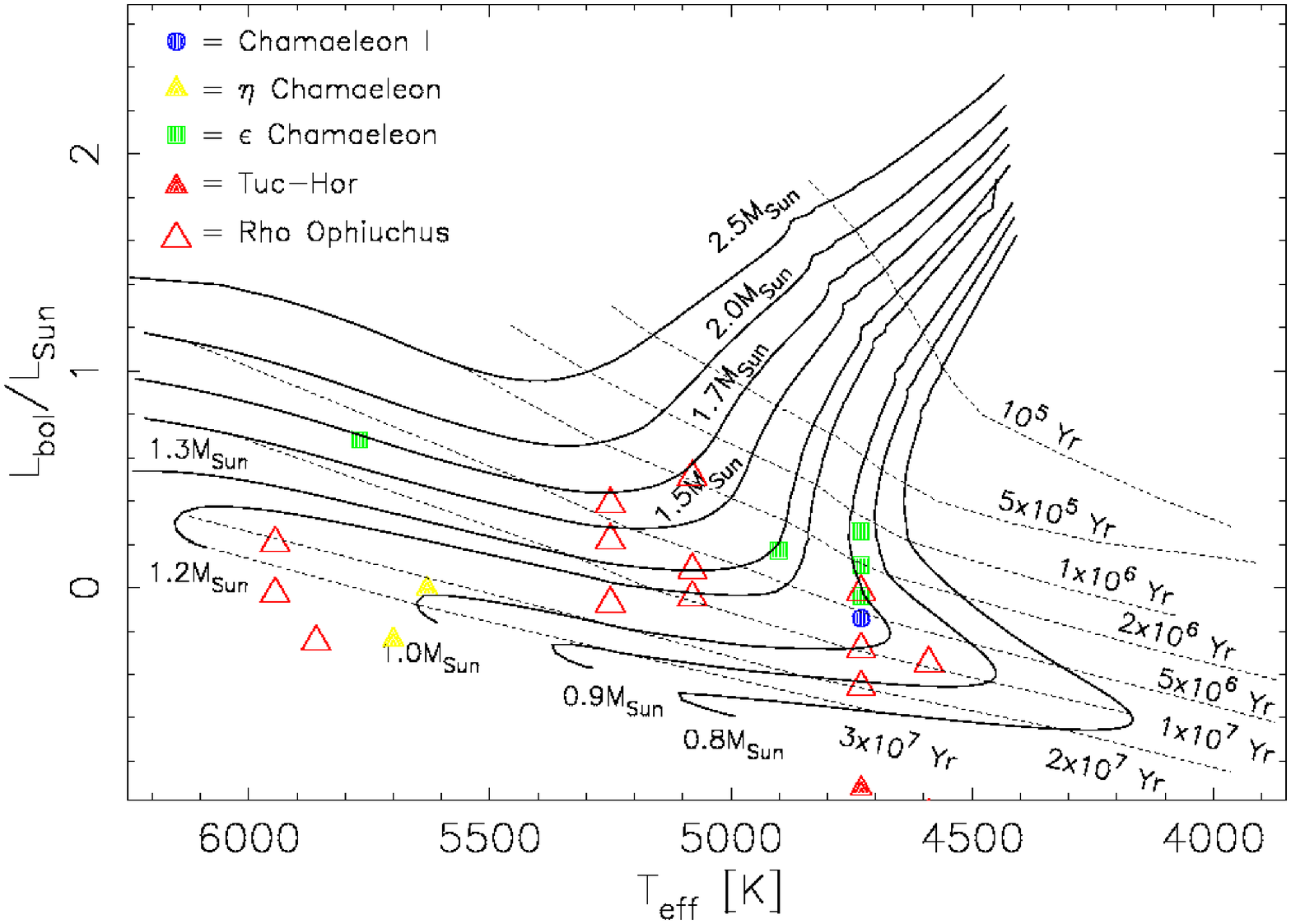,width=85mm}
\epsfig{figure=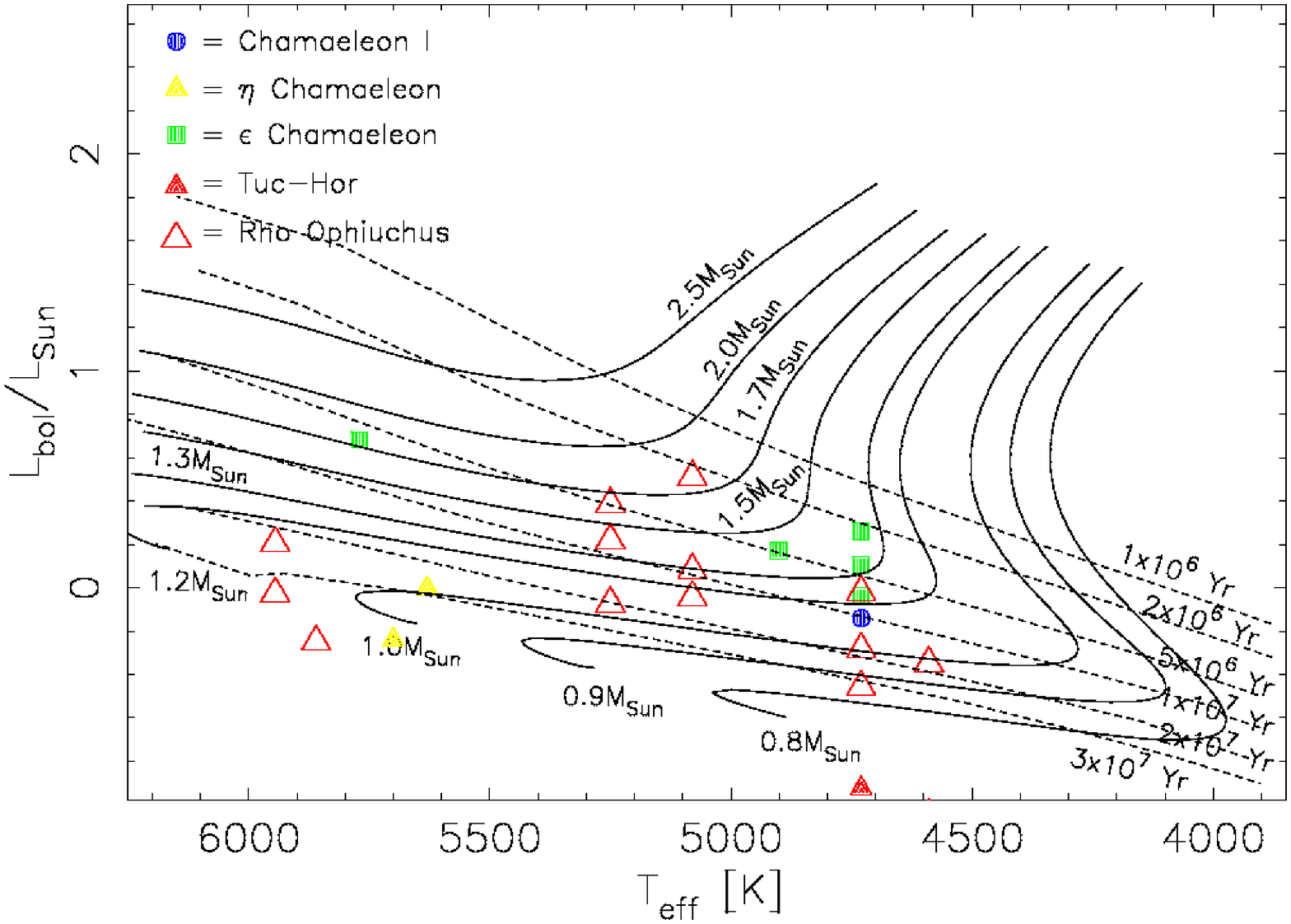,width=85mm}
\epsfig{figure=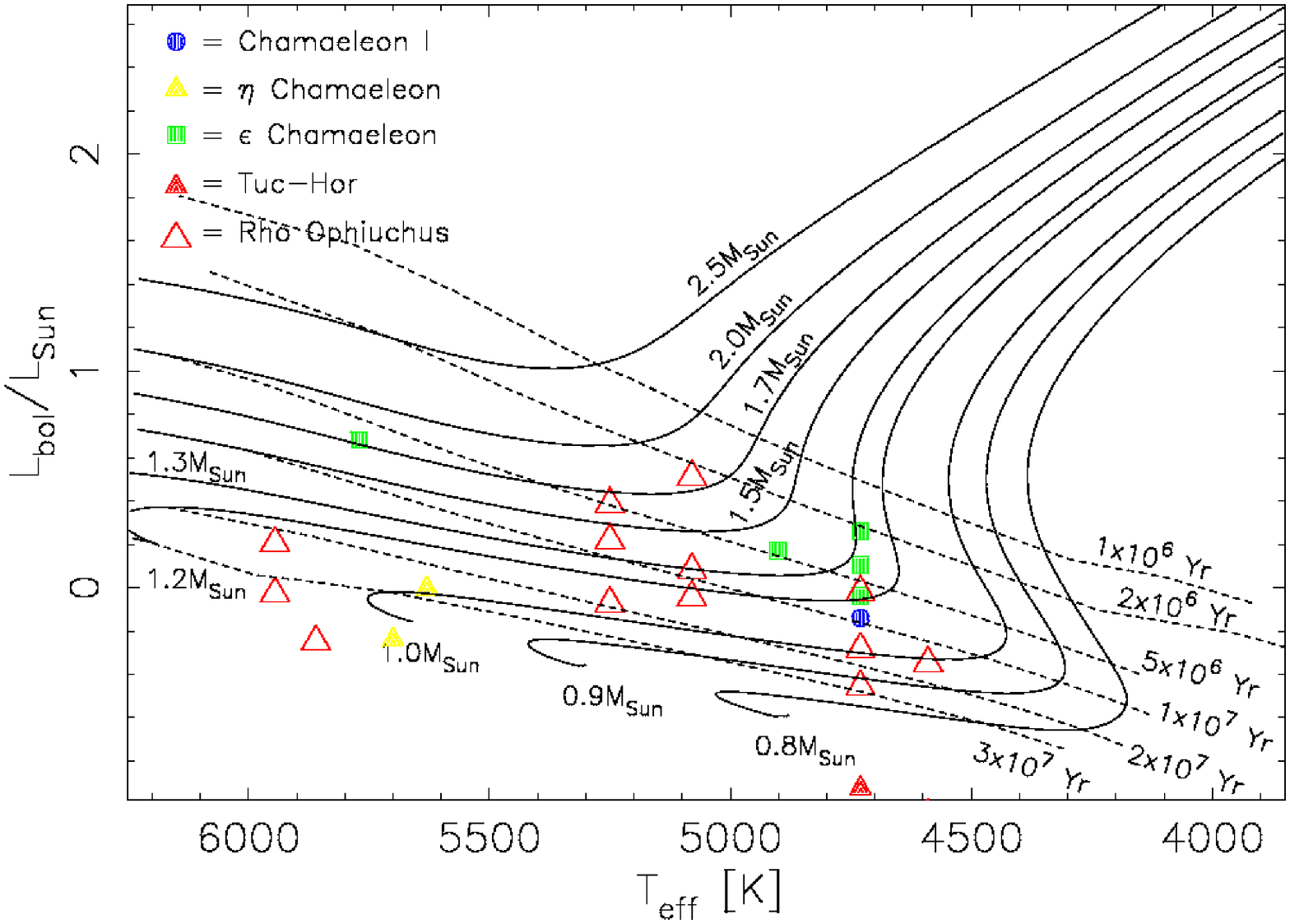,width=58.5mm,angle=-90}
\caption{A triumvirate of Hertzsprung-Russell diagrams, using $A_{v}$ extinction values 
derived from $B-V$ colours (see Table~\ref{Reddening-Extinction-Table}), 
is plotted for Cham I (\tikzcircle[blue, fill=blue]{2pt}), $\eta$ Cham 
({\Montriangle{yellow}}), $\epsilon$ Cham ({\Monsquare{green}}), Tuc-Hor 
({\Montriangle{red}}) \& Rho Ophiuchus  ({\textcolor{red}{$\triangle$}}) 
candidates. Effective temperatures, bolometric corrections and reddening vectors 
are based on KH95 colour-spectral type relationships for dwarf stars, with 
distances to individual stellar groups cited in Table~\ref{AGEtable}. 
Stellar isochrones (dashed tracks) and mass tracks (solid lines) are computed 
from theoretical pre-main sequence, solar metallicity stellar models by 
D'Antona \& Mazzitelli (1997 - top plot), the Dartmouth group (\citealt{D08} - 
middle plot) and the {\sc parsec} group (\citealt{PARSEC} - lower plot).}
\label{HRDs-compareFIG}
\end{figure}

In an attempt to quantify the dependence of derived stellar masses and ages 
on our choice of stellar evolutionary model, we can easily compare our 
{\sc parsec} results with those calculated using other stellar models. For 
convenience, we have chosen the isochrones and mass tracks of the Dartmouth 
group \citet{D08} and the \citet{DAM97} [DAM97] team, both of which are of 
solar-metallicity. The former has an initial He content of Y=0.274 and does 
not explicitly include Deuterium burning during the pre-main sequence, while 
the latter does include Deuterium burning on the {\sc pms} starting with an 
initial abundance of $^{2}_{1}$H of $4.5\times10^{-5}$, and incorporates an 
initial He content of Y=0.280. We also note that the DAM97 model also relies 
upon a gray atmosphere approximation as an exterior boundary condition, which 
will have different effects on stellar effective temperatures and luminosities 
compared to the D08 and {\sc parsec} models. For brevity, we restrict ourselves at 
the present time to a comparison of the models containing \lbol calculations 
using $A_{v}$ extinction values derived from B-V colours and effective 
temperatures derived from KH95 main-sequence spectral types. The graphical 
and numerical results of their inter-comparability are presented in 
Figure~\ref{HRDs-compareFIG} and Table~\ref{HRDs-compareTABLE}.

Our analysis shows that for extinction based on B-V colours, the Dartmouth group 
isochrones return stellar ages that are on average 0.68 Myr ($\pm$ 0.21) older 
than the {\sc parsec} ones, whereas the DAM97 ones are 2.86 Myr ($\pm$  0.58) younger. 
For the Cham I, $\epsilon$ Cha and Rho Ophiuchus stars, such a large {\sc parsec}/DAM97 
age differential actually represents a substantial fraction of the system age 
($\simeq 30-50$ per cent). For extinction based on other colours (V-Ic, J-H, H-Ks), 
we find that Dartmouth models remain closely matched to their {\sc parsec} counterparts 
being $0.68\pm1.06$ Myr younger, $0.08\pm0.29$ Myr older and $0.03\pm0.44$ Myr 
younger respectively. For the DAM97 model comparisons however, the other three 
colour-based extinctions again yield younger ages albeit at a lower level than 
their B-V colours ($1.32\pm2.06$, $0.68\pm0.98$ and $1.50\pm1.56$ Myr younger 
for V-Ic, J-H and H-Ks colours respectively).

In terms of masses, the Dartmouth models are essentially identical to the 
{\sc parsec} ones across all four colour-based extinctions ($0.00 \pm 0.02$, 
$0.03\pm0.09$, $0.02\pm0.06$ and $0.03\pm0.04$ M$_{\odot}$ - the latter 
three being more massive). For the DAM97 models, all four extinctions yield 
slightly lower mass HRD values on average but only at the $\simeq$0.1 M$_{\odot}$ 
level ($0.08\pm0.03$, $0.06\pm0.15$, $0.15\pm0.12$ and $0.13\pm0.13$ M$_{\odot}$).

\begin{table*}
\vspace{3mm}
\caption[]{\protect \small A comparison of stellar ages and masses for our candidate sample, using using $A_{v}$ extinction values 
derived from $B-V$ colours, for the {\sc parsec}, Dartmouth group and D'Antona \& Mazzitelli stellar models.}
\begin{tabular}{lcccccccccc}
\hline
~~~~~~~~~~Target$^{a}$             & $T_{\mathrm{eff}}$$^{a}$ & [\lbol ]$^{a,b}$ & Age$^{c}$ & Mass$^{c}$    & Age$^{d}$ & Mass$^{d}$     & Age$^{e}$ & Mass$^{e}$    & $\Delta$Age$^{f}$ & $\Delta$Mass$^{f}$  \\
                                   &   [K]                    &                & [Myr]     & [M$_{\odot}$] &   [Myr]   & [M$_{\odot}$]  &    [Myr]  & [M$_{\odot}$] & [Myr]       & [M$_{\odot}$] \\ \hline 
                                   &   &  & \multicolumn{2}{c}{{\sc parsec}$^{c}$} & \multicolumn{2}{c}{Dartmouth08$^{d}$} & \multicolumn{2}{c}{DAM97$^{e}$}  &  &  \\
                                               \cmidrule(lr){4-5}               \cmidrule(lr){6-7}           \cmidrule(lr){8-9}            
Chamaeleon I   (\mycircle{blue})   &                          &                &           &               &           &                &           &               &             &               \\
 \cmidrule(lr){1-1} 
RXJ1112.7-7637                     & 4900                     &    ...         &   ...     & ...           & ...       & ...            & ...       & ...           &  ... ,  ... &  ... ,  ...   \\
RXJ1129.2-7546                     & 4730                     &  $-0.141$      &   9.5     & 1.1           &  10       & 1.2            & 6.0       &  1.0          & -0.5 ,  3.5 & -0.1 ,  0.1   \\
$\eta$ Cham  (\mytriangle{yellow}) &                          &                &           &               &           &                &           &               &             &               \\
 \cmidrule(lr){1-1}
RXJ0850.1-7554                     & 5700                     &  $-0.242$      &  $>30$    & 1.0           & $>30$     & 1.0            & $>30$     &  1.0          &  ... ,  ... &  0.0 ,  0.0   \\ 
RXJ0951.9-7901                     & 5630                     &  $-0.005$      &   26      & 1.0           & 28        & 1.0            & 23        &  1.0          & -2.0 ,  3.0 &  0.0 ,  0.0   \\
Tuc Hor  ({\Montriangle{red}})     &                          &                &           &               &           &                &           &               &             &               \\
 \cmidrule(lr){1-1} 
RXJ1140.3-8321$^\dagger$           & 4730                     &  $-0.926$      &   $>30$   & ...           & $>30$     & ...            & $>30$     & ...           &  ... ,  ... &  ... ,  ...   \\
RXJ1140.3-8321$^\ddagger$          & 4590                     &  $-1.030$      &   $>30$   & ...           & $>30$     & ...            & $>30$     & ...           &  ... ,  ... &  ... ,  ...   \\
$\epsilon$ Cham (\ssquare{green})  &                          &                &           &               &           &                &           &               &             &               \\
 \cmidrule(lr){1-1}
RXJ1158.5-7754a                    & 4730                     & ~~0.260        &  2.0      & 1.3           & 2.0       & 1.3            & 1.0       & 1.0           &  0.0 ,  1.0 &  0.0 ,  0.3   \\
RXJ1159.7-7601                     & 4730                     &  $-0.041$      &  6.5      & 1.4           & 7.5       & 1.2            & 4.0       & 1.0           & -1.0 ,  2.5 &  0.2 ,  0.4   \\
RXJ1201.7-7859                     & 5770                     & ~~0.681        &  6.5      & 1.7           & 6.5       & 1.8            & 7.5       & 1.7           &  0.0 , -1.0 & -0.1 ,  0.0   \\
RXJ1239.4-7502$^\dagger$           & 4900                     & ~~0.172        &  4.5      & 1.4           & 5.0       & 1.4            & 3.5       & 1.3           & -0.5 ,  1.0 &  0.0 ,  0.1   \\
RXJ1239.4-7502$^\ddagger$          & 4730                     & ~~0.104        &  4.0      & 1.4           & 4.5       & 1.3            & 2.0       & 1.0           & -0.5 ,  2.0 &  0.1 ,  0.4   \\
Rho Ophiuchus ({\textcolor{red}{$\triangle$}}) &              &                &           &               &           &                &           &               &             &               \\
 \cmidrule(lr){1-1}
RXJ1620.7-2348                     & 4590                     &  $-0.349$      &   15      & 1.0           & 18        & 1.0            & 9.5       & 0.9           & -3.0 ,  5.5 &  0.0 ,  0.1   \\ 
RXJ1621.0-2352                     & 5080                     &  ~~0.081       &  9.5      & 1.3           & 10        & 1.3            & 7.0       & 1.3           & -0.5 ,  2.5 &  0.0 ,  0.0   \\ 
RXJ1621.2-2347                     & 4730                     &  $-0.458$      &   29      & 0.9           & $>30$     & 0.9            & 20        & 0.9           &  ... ,  9.0 &  0.0 ,  0.0   \\ 
RXJ1623.1-2300                     & 4730                     &  $-0.021$      &  6.5      & 1.2           & 7.0       & 1.2            & 3.5       & 1.0           & -0.5 ,  3.0 &  0.0 ,  0.2   \\ 
RXJ1623.4-2425                     & 5860                     &  $-0.250$      &  $>30$    & ...           & $>30$     & 1.0            & $>30$     & 1.1           &  ... ,  ... &  ... ,  ...   \\ 
RXJ1623.5-2523                     & 5250                     &  ~~0.218       &  8.5      & 1.4           & 8.5       & 1.4            & 7.0       & 1.4           &  0.0 ,  1.5 &  0.0 ,  0.0   \\
RXJ1624.0-2456                     & 5250                     &  $-0.072$      &   19      & 1.0           & 20        & 1.1            & 16        & 1.1           & -1.0 ,  3.0 & -0.1 , -0.1   \\
RXJ1624.8-2239                     & 5080                     &  ~~0.511       &  2.5      & 1.7           & 2.5       & 1.8            & 2.0       & 1.7           &  0.0 ,  0.5 & -0.1 ,  0.0   \\ 
RXJ1624.8-2359                     & 4730                     &  $-0.284$      &  16       & 1.0           & 18        & 1.0            & 9.5       & 1.0           & -2.0 ,  6.5 &  0.0 ,  0.0   \\ 
RXJ1625.0-2508                     & 5945                     &  ~~0.205       &  23       & 1.1           & 23        & 1.2            & 21        & 1.2           &  0.0 ,  2.0 & -0.1 , -0.1   \\
RXJ1625.4-2346                     & 5080                     &  $-0.045$      &  15       & 1.3           & 15        & 1.1            & 9.5       & 1.2           &  0.0 ,  5.5 &  0.2 ,  0.1   \\
RXJ1625.6-2613                     & 4060                     &  ...           &  ...      & ...           & ...       & ...            & ...       & ...           &  ... ,  ... &  ... ,  ...   \\
ROXR1 13                           & 5250                     &  ~~0.386       &  5.0      & 1.6           & 5.0       & 1.6            & 4.5       & 1.6           &  0.0 ,  0.5 &  0.0 ,  0.0   \\
RXJ1627.1-2419                     & 5945                     &  $-0.028$      &  $>30$    & 1.1           & $>30$     & 1.1            &  $>30$    & 1.1           &  ... ,  ... &  0.0 ,  0.0   \\
\hline
\end{tabular}\label{HRDs-compareTABLE}
\begin{flushleft}
%Notes: \\
$a -$ Chamaeleon, Tuc-Hor and Rho Ophiuchus candidates triaged into disparate young {\sc sfr} regions, with symbols 
\mycircle{blue}, \mytriangle{yellow}, {\Montriangle{red}}, \ssquare{green} \& {\textcolor{red}{$\triangle$}}, 
matching those data presented in Figures~\ref{HRDs-compareFIG}. \\
$b -$ \lbol calculated using photometric data, \teff, $A_{v}$ extinction values derived from $B-V$ colours and distances 
detailed in Tables~\ref{SFRcandidates-OPTIRphot}$-$\ref{AGEtable}.\\
$c -$ Theoretical ages and masses based on the {\sc parsec} (\citealt{PARSEC}) models.\\
$d -$ Theoretical ages and masses based on the Dartmouth group (\citealt{D08}) models.\\
$e -$ Theoretical ages and masses based on the DAM97 (\citealt{DAM97}).\\
$f -$ Differential values are {\sc parsec} - Dartmouth and {\sc parsec} - DAM97, respectively. \\
\end{flushleft}
\end{table*}

\section{Presence of disc emission: Spectral Energy Distributions}\label{S_seds}

In order to confirm the presence of circumstellar disc material inferred from 
stellar positions in colour-colour space (see \S~\ref{JHK-excess}), we construct 
SEDs and compare them to theoretical radiative transfer models. We combine optical 
photometry derived in this work (\S~\ref{S_photometry}) with additional near- 
and mid-infrared photometry gathered from the literature; these data are presented 
in Tables \ref{SFRcandidates-OPTIRphot}, \ref{T_phot1} and \ref{T_phot2}. Following 
the procedure described in \citet{Aarnio2010}, these photometric data were fit 
using the computed SED grid of \citet{Robitaille:2007}. These $\sim$200,000 
radiative transfer models include a central star surrounded by a flaring disc 
and envelope; the properties of these model components vary as described by 
a set of 14 parameters. Additional parameters used to constrain the fitter 
include $A_{v}$ and distance: the allowed range of $A_{v}$ was set to 0-10mag for 
both SFRs, while the distance range for the different Chamaeleon regions 
was confined to 100-175 pc; Rho Ophiuchus member distances were constrained 
to be within 110-150 pc.

In those cases where substantial infrared excess was seen, we filtered the 
resulting best fit models based on known stellar properties. Spectral types 
reported in Table~\ref{SFRcandidates-UVES} were converted to effective temperatures 
via the KH95 relationships, and only models whose central star has T$_{eff}$ 
within 500K of the estimated effective temperature were included. One should 
retain some level of caution with this approach, as young, solar-type stars 
can show photometric and spectroscopic temporal variations due some combination 
of on-going accretion, rapid rotation and extreme (and stochastic) chromospheric 
and coronal magnetic activity. The sample of best-fit models was further 
narrowed by accepting only cases in which the $\chi^2$ fit value was within 
a 99.73 per cent confidence level ($\Delta \chi^2 =$11.8) of the (post-T$_{eff}$ 
rejection) best fit. Finally, models with disc masses $<$10$^{-10}$ 
M$_{\odot}$ were rejected.

\begin{figure*}
\centering
\epsfig{figure=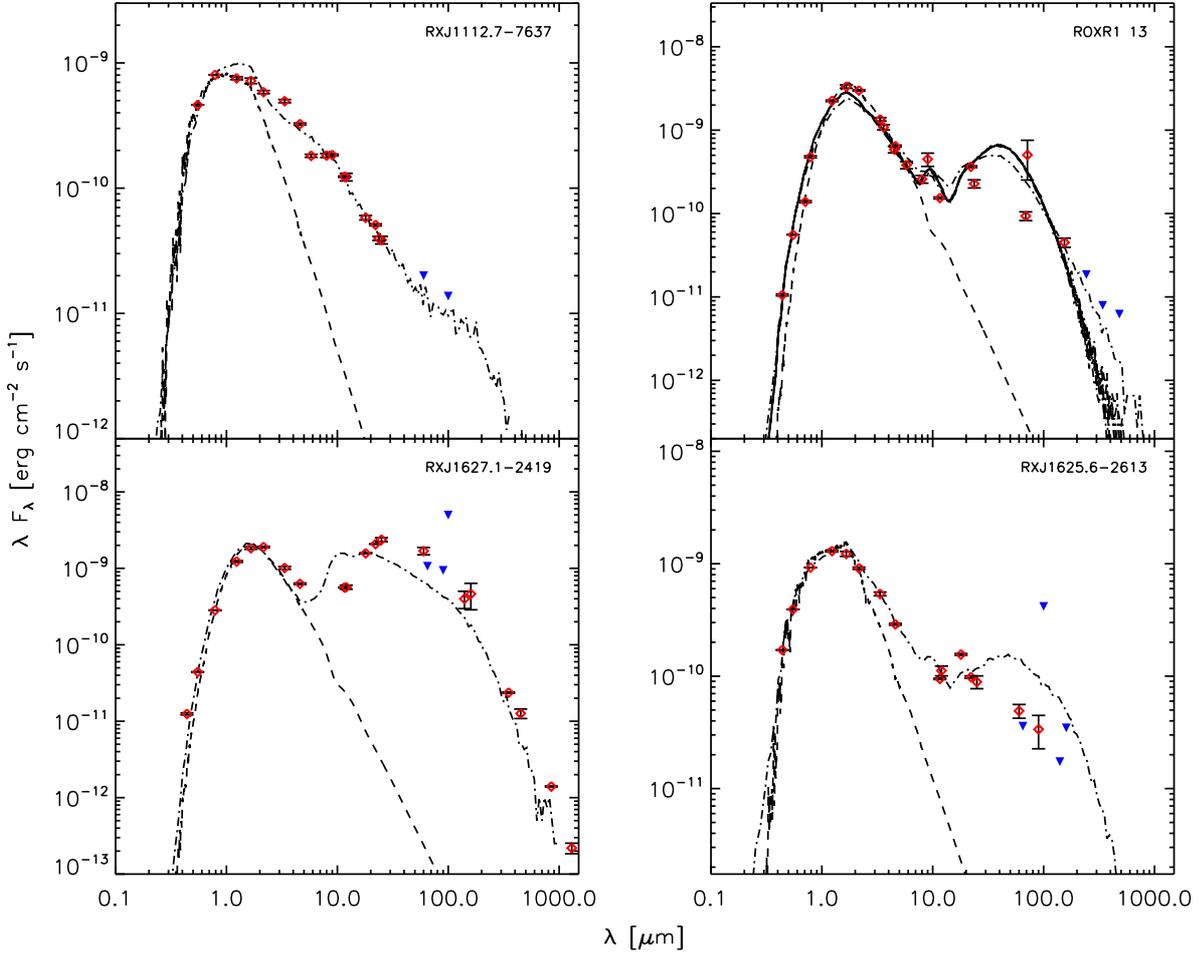,width=145mm,angle=90}
\vspace*{4mm}
\caption{SEDs of Rho Ophiuchus and Chamaeleon candidates show, outwith 
the error budget, that significant departures from a bare photosphere 
are clear, {\em i.e.,} substantial infrared excess is evident. Red diamonds 
and blue triangles denote photometry from  Tables~\ref{SFRcandidates-OPTIRphot}, 
\ref{T_phot1} \& \ref{T_phot2} - where blue triangles are 
upper limits. Dot-dashed lines are the best fit synthetic SEDs as 
determined via criteria defined in \S~\ref{S_seds}, with dashed lines 
representing a stellar photospheric contribution matching spectral-type 
derived effective temperature.}
\label{F_juicyseds}
\end{figure*}

Our SED analysis shows substantial excess for three Rho Ophiuchus objects 
and one Chamaeleon star (see Figure~\ref{F_juicyseds}). In three additional 
Rho Ophiuchus stars, we see low to moderate mid-infrared excess (see 
Figure~\ref{F_moderateseds}); the SEDs for the remainder of the sample 
are consistent with bare photospheres (see 
Figures~\ref{F_bareSEDsCham}~\&~\ref{F_bareSEDsRhoOph}). We compare 
our SED fit Av values to those calculated via extinction laws in 
\S~\ref{Red-Av} and Table~\ref{Reddening-Extinction-Table}, finding that for 
stars with evidence of circumstellar discs, extinction derived from optical 
photometry best matches the SED-fit derived $A_{v}$ values (especially for V-I), 
presumably due to the complications of active accretion or disc radiation in 
the infra-red. For stars without disc signatures, we find the near infrared 
reddening vector E(H-Ks) produces $A_{v}$ values in better agreement with the 
SED results; to our mind, this is attributable to the variable and 
difficult-to-quantify effect (starspots, flares, {\em etc.}) of stellar magnetic 
activity on optical photometry.

In the four star$+$disc systems with the most near- to mid-infrared excess 
(see Figure~\ref{F_juicyseds}), we find that the best-fit model parameters 
have the greatest disc masses, highest accretion rates, and all show some 
sign of emission (or blueshifted absorption) in their H $\alpha$ profiles, 
which are tell-tale indicators of ongoing activity, accretion, or outflow. In 
fact, all four stars exhibit narrow-lined neutral Oxygen [O I] 5577.3 \r{A} 
emission in their {\sc uves} spectra, oftentimes shown to originate in a 
circumstellar disc or envelope. For RXJ1627.1-2419, its O I profile is actually 
complex in nature, and looks double-peaked. Moreover, both RXJ1112.7-7637 and 
RXJ1627.1-2419 also show narrow O I emission lines at 6300.2 \& 6363.9 \r{A}. 
ROXR1 13 only produces a narrow 6363.9 \r{A} emission feature as opposed to 
RXJ1625.6-2613 which only shows 6300.2 \r{A} in emission.

Interestingly, the {\sc uves} spectra of ROXR1 13 and RXJ1627.1-2419 both show 
P Cygni like profiles in the neutral Sodium doublet at $\simeq$ 5900 \r{A} (Fraunhofer 
$D$ lines), indicative of a dense magneto-hydrodynamical driven stellar wind and 
self-absorption through a circumstellar accretion disc ({\em e.g.,} \citealt{pcygni99}). 
For completeness, we note that RXJ1112.7-7637 also exhibits small narrow-lined emission 
features at 6410.1 \r{A}, 6505.4 \r{A}, 6523.5 \r{A} and 6529.0 \r{A}, while RXJ1627.1-2419 
produces an emission feature at 5615.3 \r{A} (all observing-frame wavelengths). 
Finally, RXJ1625.6-2613 has small emission features at 5875.1 \r{A} and 6678.2 \r{A}, 
which may be He I and/or O II species).

In three of four cases, the models indicate disc truncation at the 
dust destruction radius; in the sole case with a best-fit disc truncated 
beyond R$_{\rm trunc}$, RXJ1627.1-2419, there is existing evidence that 
this object is characterized by a transition disc \citep{McClure:2010}. We 
present best-fit model parameters for all seven (7) objects with moderate 
to substantial disc excess in Table~\ref{T_bfmparms}, and we provide 
additional discussion of each object, including a review of extant 
observations garnered from the literature (see \S~\ref{S_sedlitnotes}).

Except in the strongest two cases, we find no correlation between the residual H $\alpha$ 
equivalent width and disc excess indicators; this is perhaps unsurprising, given the youth of 
the sample stars and the wide range of H $\alpha$ fluxes observed in stars $<$100 Myr of 
age \citep[cf.][]{Herbig:1985,Herbig:1988,kenyon1995}.

\section{Conclusions}\label{conclusions}

We have obtained optical $BVIc$ photometry and high-resolution spectroscopy for a modest 
sample of X-ray selected stars in the Chamaeleon and Rho Ophiuchus SFRs. We exploit our 
observational data in order to pursue five principle avenues of investigation with the 
goal of establishing fundamental properties of our target stars; ({\bf I:}) RVs and 
Lithium detections are employed in assigning 1-d kinematic membership of the sample's 
parent associations, as well as in the assessment of stellar youth; ({\bf II:}) in combination 
with 2MASS near-IR photometry and optical spectral types, we use our optical $BVIc$ data 
in order to calculate colour-dependent reddening vectors and $A_{v}$ values; ({\bf III:}) in 
an $J-H$/$H-K$ diagram, we search for evidence of IR-excess in our targets, which is the smoking 
gun evidence normally associated with viscous circumstellar discs; ({\bf IV:}) transforming 
observational plane photometry and spectral types onto the theoretical plane of \lbol and T$_{eff}$, 
we construct HRDs based on four different colour-dependent Av values, thereby allowing us 
to establish model-dependent stellar age and mass; ({\bf V:}) we compare our optical 
photometry, as well as extant near- and mid-IR photometry in the literature, to radiative 
transfer models in order to construct SEDs for all of our targets, identifying and (re-)confirming 
systems consistent with circumstellar accretion discs.

For our Chamaeleon candidates, we {\em ab initio} took advantage of 2-d kinematics in order to 
triage stars into three distinct sub-groups in the region. Two stars, RXJ1112.7-7637 and 
RXJ1129.2-7546, have proper motions consistent with membership of the Cham {\footnotesize I} 
group. Three stars, RXJ0850.1-7554, RXJ0951.9-7901 and RXJ1140.3-8321, have proper motions 
in broad agreement with membership of the $\eta$ Cham population, well-aligned with the 
L\'{o}pez Mart\'{\i} et al. findings. In the more recent Elliott et al. study however, they 
advocate allocating a 30 Myr Carina association membership for RXJ0850.1-7554 and a 45 Myr 
Tuc-Hor membership for both RXJ0951.9-7901 and RXJ1140.3-8321. Moving forward into the 
analysis of the spectroscopic data and theoretical HRDs, we retained $\eta$ Cham tentative 
membership for these three objects, revisiting their membership status as each additional 
layer of analysis was added. RXJ1233.5-7523 is an obvious proper motion field star not 
associated with any of the Chamaeleon groups. The four remaining Chamaeleon stars 
(RXJ1158.5-7754a, RXJ1159.7-7601, RXJ1201.7-7859 and RXJ1239.4-7502) are kinematic members 
of the Eps Cham group.

With one exception, RXJ1303.5-7701, all of our Chamaeleon candidates yield {\sc uves} RVs 
consistent with membership of large samples of low-mass, young stars in the Chamaeleon region. 
Except for RXJ1620.1-2348 and RXJ1624.0-2456, the majority of our Rho Ophiuchus sample 
also reflect this behaviour in presenting 1-d kinematics consistent with membership of 
the parent Rho Ophiuchus cloud. Our {\sc uves} data show that three Chamaeleon group stars 
are RV variable, with confirmation that RXJ1201.7-7859 and RXJ1303.5-7701 are SB1 systems. 
Our spectra do not confirm the J06 finding that RXJ0951.9-7901 is an SB2, and we suggest 
that further RV observations of this object are required to firmly establish its multiplicity 
status.

With four exceptions, our UVES spectra show that our sample stars are young by virtue of 
being Lithium rich, having Lithium {\footnotesize I} 6708\r{A} equivalent widths at least 
as large as their similar T$_{eff}$ counterparts in the 125-Myr Pleiades cluster, oftentimes 
far more. Three of the four Li-poorer stars are either early-type (RXJ1303.5-7701), or are proper 
motion non-members (RXJ1233.5-7523 \& RXJ1620.1-2348) of their parent associations. The low-Lithium 
status of the fourth object, RXJ1140.3-8321, suggests that the star is actually a post T-Tauri 
star considerably older than the $\simeq$ 5 Myr Eta Cha group. This supposition supports the 
Elliott et al. claim that it is a member of the Tuc-Hor group, which going forward we endorse. 
With the exception of the rapidly rotating possible SB2 system RXJ0951.9-7901, there is little 
evidence to suggest that our objects have variable Lithium abundance over a two-year period.

One Chamaeleon star (RXJ1112.7-7637) and four Rho Ophiuchus stars (ROXR1 13, 
RXJ1621.2-2347, RXJ1624.8-2359 \& RXJ1627.1-2419) show evidence of substantial, 
$A_{v} >5$ infrared excess, as judged by their displaced positions compared to 
un-reddened dwarf and giant loci in the $J-H$/$H-K$ colour-colour diagram. 
Nevertheless, for the vast majority of stars in our sample, the Balmer series 
residual H $\alpha$ emission is $<5$\r{A}, except for RXJ1625.6-2613, which has 
an equivalent width $>6$\r{A}. The majority of our sample are therefore consistent, 
in terms of Balmer series H $\alpha$ emission at least, with being weak-lined T-Tauri 
stars.

Employing main sequence models to derive bolometric corrections as well as calculate 
spectral type based photometric colours and effective temperatures, we derive 
stellar ages and masses based on theoretical HRDs for our targets using four separate 
extinction values based on B-V, V-Ic, J-H and H-K colours. By comparison to the 
{\sc parsec} theoretical models, we show that our Cham I and $\epsilon$ Cham stars are 
consistent with membership of young SFRs, having ages \lsi 5 Myr across all four 
colour-based extinction vectors $-$ agreeing with several other studies across the 
literature. For both $\eta$ Cham {\bf candidates}, RXJ0850.1-7554 and RXJ0951.9-7901, 
post T-Tauri like ages of at least 25-30 Myr are calculated, and so in agreement 
with \cite{elliott2014}, we henceforth advocate their membership of the 30 Myr Carina 
and 45 Myr Tuc-Hor associations, respectively. Several stars exhibit a spread 
in ages, based on different colour extinction relations in the \lbol calculation, 
of an order-of-magnitude or more, always decreasing in age as the extinction 
relationships derived go from blue to red reddening vectors ({\em i.e.,} $E(B-V)$ 
$\rightarrow$ $E(H-K)$). With the exception of three stars, RXJ1112.7-7637, 
ROXR1 13 and RXJ1627.1-2419, HRDs stellar masses are consistent across blue 
to red reddening vectors, and reveal that our sample is mostly composed 
of objects $1-2$ times the mass of the Sun. Three objects with determinable 
masses are not well constrained because of strong ($A_{v}$$>$5) and variable 
extinction across the four photometric colours we consider.

We use our modest sample of X-ray selected stars in the Chamaeleon, Carina, 
Tuc-Hor and Rho Ophiuchus regions to benchmark the {\sc parsec} stellar models 
against those of the DAM97 and Dartmouth groups. For extinction based on B-V 
colours, we find that there is good agreement between the {\sc parsec} and Dartmouth 
models at the slightly-older 0.68 ($\pm$ 0.21) Myr level, whereas the DAM97 models 
are younger by 2.86 ($\pm$0.58) Myr $-$ which is $\simeq 30-50$ per cent of the 
group's age for Cham I, $\epsilon$ Cha and Rho Ophiuchus. For extinction based on 
other colours (V-Ic, J-H, H-Ks), we find that Dartmouth models remain closely matched 
to their {\sc parsec} counterparts being only $0.68\pm1.06$ Myr younger, $0.08\pm0.29$ Myr 
older and $0.03\pm0.44$ Myr younger respectively. For the DAM97 model comparisons however, 
the other three colour-based extinctions again yield younger ages albeit at a lower 
level than their B-V colours ($1.32\pm2.06$, $0.68\pm0.98$ and $1.64\pm1.78$ Myr $-$ 
all younger $-$ for V-Ic, J-H and H-Ks colours respectively). In terms of masses, 
{\sc parsec} and Dartmouth agree almost perfectly across all four colour-based 
extinction at the $\simeq$ $2$ per cent level. The DAM97 comparison is also quite good, 
with {\sc parsec} differences of \lsi 0.1 M$_{\odot}$ level across all four extinctions.

Concerned with having employed dwarf-class (main sequence) models to derive effective 
temperatures and bolometric corrections for HRD construction in the analysis of our 
mostly PMS objects, we repeated our analyses using spectral-type, colour and 
bolometric corrections from PMS theoretical models. In calculating differential 
colour-dependent extinction vectors for our sample, the two infrared colours yield 
$A_{v}$ values based on $E(J-H)$ \& $E(H-K)$ $\simeq 3-5$ times their corresponding 
optical colour  $Av$ values based on $E(B-V)$ \& $E(V-I)$ for Chamaeleon stars, 
and $\simeq 4-10$ times in the case of the Rho Ophiuchus ensemble. This implies that 
our stars are considerably more red in the PMS models compared to the MS ones, in 
agreement with the findings of \citet[their figure 4]{PM13}. An comparison of HRD 
ages using the main sequence and PMS model data shows that within the error bars, 
stellar age remains unaffected. In terms of mass however, the PMS input to the HRD 
models consistently yielded $0.1-0.3$ M$_{\odot}$ lower values than for the main 
sequence analysis.

Complementing our $BVIc$ dataset with near- and mid-IR photometry has allowed 
us to construct spectral energy distributions for each star in our sample. 
Comparison of these distributions to radiative transfer models shows that most 
objects are consistent with a bare stellar photosphere. In the event that these 
stars are not too photospherically active ({\em i.e.,} starspots), our bare photosphere 
stars are bright enough, with well-characterized age and mass, that they will make 
prime transiting exoplanet host targets for the upcoming {\em TESS} mission. 

Three stars (RXJ1623.1-2300, RXJ1623.5-2523 \& RXJ1625.0-2508) exhibit marginal 
signs of mid-IR excess emission compared to their bare photosphere models. Of more 
interest however are the one Chamaeleon star (RXJ1112.7-7637) and three Rho Ophiuchus 
stars (ROXR1 13, RXJ1625.6-2613 \& RXJ1627.1-2419) that exhibit strong evidence 
of excess infrared emission; the physical interpretation of this excess emission 
is predicated upon infrared radiation originating in the circumstellar accretion 
disc of a young star. These four stars are southerly enough, bright enough and have 
well-established stellar age and masses to lend themselves favourably to disc-imaging 
surveys using state of the art facilities such as ALMA. High-resolution, spatially 
resolved images of young stars hosting accretion discs are prime candidates for 
the detection and characterization of intra-disc gaps, and sites of active and 
on-going exoplanet formation.

\section*{Acknowledgments}

We are extremely grateful to the anonymous referee whose report was timely, thorough 
and fair. S/he made several important suggestions which have resulted in an improved 
body of work. Her/his neutrality, positive guidance and constructive criticism allowed 
us the academic freedom to fully concentrate on the amelioration of the manuscript, 
for which we are deeply grateful. This manuscript is based on observational data 
acquired using the 1.0m telescope at the Cerro Tololo InterAmerican Observatory 
(Programme ID: VANF-07A-02), operated by the {\sc smarts} consortium, and the UT2 telescope 
operated by the European Southern Observatory at their Paranal site (programme ID: 075.C-0272). 
We are grateful to the staff and scientists at both observatories for their assistance. 
This publication also makes use of data products from the Two Micron All Sky Survey, 
which is a joint project of the University of Massachusetts and the Infrared 
Processing and Analysis Center/California Institute of Technology, funded by the 
National Aeronautics and Space Administration and the National Science Foundation. 
This publication makes use of data products from the Wide-field Infrared Survey Explorer, 
which is a joint project of the University of California, Los Angeles, and the Jet 
Propulsion Laboratory/California Institute of Technology, funded by the National 
Aeronautics and Space Administration. 

Our manuscript also makes use of the the Digitized Sky Surveys, which were produced 
at the Space Telescope Science Institute under U.S. Government grant NAG W-2166. 
The images of these surveys are based on photographic data obtained using the Oschin 
Schmidt Telescope on Palomar Mountain and the UK Schmidt Telescope. The plates were 
processed into the present compressed digital form with the permission of these institutions: 
The National Geographic Society - Palomar Observatory Sky Atlas (POSS-I) was made by the 
California Institute of Technology with grants from the National Geographic Society; The Second 
Palomar Observatory Sky Survey (POSS-II) was made by the California Institute of Technology 
with funds from the National Science Foundation, the National Geographic Society, the Sloan 
Foundation, the Samuel Oschin Foundation, and the Eastman Kodak Corporation. The Oschin 
Schmidt Telescope is operated by the California Institute of Technology and Palomar Observatory. 
The UK Schmidt Telescope was operated by the Royal Observatory Edinburgh, with funding from the 
UK Science and Engineering Research Council (later the UK Particle Physics and Astronomy Research 
Council), until 1988 June, and thereafter by the Anglo-Australian Observatory. The blue plates 
of the southern Sky Atlas and its Equatorial Extension (together known as the SERC-J), as well 
as the Equatorial Red (ER), and the Second Epoch [red] Survey (SES) were all taken with the UK 
Schmidt.

This research has also benefitted from the data, software and/or web tools obtained from the 
High Energy Astrophysics Science Archive Research Center (HEASARC), a service of the Astrophysics 
Science Division at NASA/GSFC and of the Smithsonian Astrophysical Observatory's High Energy 
Astrophysics Division. This research has further made use of the SIMBAD database, operated at 
the CDS, Strasbourg, France. Finally, this research has made use of the VizieR catalogue access 
tool, CDS, Strasbourg, France. The original description of the VizieR service was 
published in \citet{vizier}.

For A.N.A., this work was partially supported by the National Science Foundation grants 
AST-0808072 and AST-1311698, as well as NASA award NNX09AB87G, for which she is grateful. 
For N.C.S., this work was supported by Funda\c{c}\~ao para a Ci\^encia e a Tecnologia (FCT) 
through the research grant UID/FIS/04434/2013. He also acknowledge the support from FCT through 
Investigador FCT contract of reference IF/00169/2012 respectively, and POPH/FSE (EC) by FEDER 
funding through the program {\em Programa Operacional de Factores de Competitividade - COMPETE}, as well as from project reference 
PTDC/FIS-AST/1526/2014. This work results within the collaboration of the COST Action TD 1308. For 
P.A.C., this work was partially supported by NASA grant NNX13AI46G.

\bibliographystyle{mn2e}
\bibliography{bib}

\appendix

\section{Observing Logs for Photometric and Spectroscopic Observing Campaigns}

Logs for the photometric ({\sc y4kcam} - see Table~\ref{YALO-obslog}) and 
spectroscopic ({\sc uves} - see Table~\ref{UVES-obslog}) observing campaigns of 
Chamaeleon and Rho Ophiuchus candidate members are presented, and are referenced 
in \S~\ref{targetselection}. 

\begin{table*}
%\begin{center}
\vspace{3mm} 
\caption{Log of Y4Kcam Chamaeleon and Rho Ophiuchus photometric observations on UT20070630}
\begin{tabular}{lcccccc}
\hline
%                 &          &         &        &        \\
~~~~~~~Object$^{a}$ &  RA(2000)   DEC(2000)    & RA(2000)  DEC(2000)         & Filter  & Exp      & Airmass  & Julian Date \\
                    &  Y4Kcam Field Centre     &       [2MASS]$^{b}$         &         & Time [s] &          &  [days]     \\ \hline
Chamaeleon \\ \hline
 RXJ0850.1$-$7554   & 08 50 14.8   $-$75 56 59 & 08 50 05.41  $-$75 54 38.07 &  B      &   7.0    &  1.73    &  2454282.4678 \\
                    &                          &                             &  V      &   7.0    &  1.73    &  2454282.4697 \\
                    &                          &                             &  I      &   7.0    &  1.74    &  2454282.4707 \\
 RXJ0951.9$-$7901   & 09 51 59.8   $-$79 05 00 & 09 51 50.70  $-$79 01 37.72 &  B      &   7.0    &  1.67    &  2454282.4719 \\
                    &                          &                             &  V      &   7.0    &  1.68    &  2454282.4734 \\
                    &                          &                             &  I      &   7.0    &  1.68    &  2454282.4742 \\
 RXJ1112.7$-$7637   & 11 12 28.9   $-$76 41 00 & 11 12 24.41  $-$76 37 06.41 &  I      &   7.0    &  1.52    &  2454282.4752 \\
                    &                          &                             &  V      &   7.0    &  1.52    &  2454282.4768 \\
                    &                          &                             &  B      &   7.0    &  1.52    &  2454282.4777 \\
 RXJ1129.2$-$7546   & 11 29 16.0   $-$75 50 00 & 11 29 12.62  $-$75 46 26.32 &  B      &   7.0    &  1.49    &  2454282.4791 \\
                    &                          &                             &  V      &   7.0    &  1.49    &  2454282.4805 \\
                    &                          &                             &  I      &   7.0    &  1.49    &  2454282.4814 \\
 RXJ1140.3$-$8321   & 11 40 21.1   $-$83 25 00 & 11 40 16.59  $-$83 21 00.38 &  I      &   7.0    &  1.71    &  2454282.4834 \\
                    &                          &                             &  V      &   7.0    &  1.71    &  2454282.4849 \\
                    &                          &                             &  B      &   7.0    &  1.71    &  2454282.4859 \\
 RXJ1158.5$-$7754a  & 11 58 32.0   $-$77 59 00 & 11 58 28.17  $-$77 54 29.48 &  B      &   7.0    &  1.53    &  2454282.4873 \\
                    &                          &                             &  V      &   7.0    &  1.53    &  2454282.4887 \\
                    &                          &                             &  I      &   7.0    &  1.53    &  2454282.4898 \\
 RXJ1159.7$-$7601   & 11 59 46.0   $-$76 06 00 & 11 59 42.27  $-$76 01 26.08 &  I      &   7.0    &  1.48    &  2454282.4909 \\
                    &                          &                             &  V      &   7.0    &  1.48    &  2454282.4927 \\
                    &                          &                             &  B      &   7.0    &  1.49    &  2454282.4939 \\
 RXJ1233.5$-$7523   & 12 33 34.9   $-$75 26 59 & 12 33 29.81  $-$75 23 11.25 &  B      &   7.0    &  1.45    &  2454282.4956 \\
                    &                          &                             &  V      &   5.0    &  1.45    &  2454282.4999 \\
                    &                          &                             &  I      &   5.0    &  1.45    &  2454282.5008 \\
 RXJ1239.4$-$7502   & 12 39 24.9   $-$75 07 00 & 12 39 21.24  $-$75 02 39.16 &  I      &   5.0    &  1.44    &  2454282.5026 \\
                    &                          &                             &  V      &   7.0    &  1.44    &  2454282.5046 \\
                    &                          &                             &  B      &   7.0    &  1.44    &  2454282.5057 \\ \hline
  Rho Ophiuchus \\ \hline
 RXJ1620.1$-$2348   & 16 20 15.9   $-$23 51 59 & 16 20 10.57  $-$23 48 12.22 &  I      & ~~7.0    &  1.06    &  2454282.6612 \\
                    &                          &                             &  V      & ~~8.0    &  1.07    &  2454282.6642 \\
                    &                          &                             &  B      &  20.0    &  1.07    &  2454282.6651 \\
 RXJ1621.0$-$2352   & 16 21 01.8   $-$23 55 59 & 16 20 57.87  $-$23 52 34.38 &  B      &  20.0    &  1.16    &  2454282.6878 \\
                    &                          &                             &  V      & ~~8.0    &  1.14    &  2454282.6909 \\
                    &                          &                             &  I      & ~~7.0    &  1.14    &  2454282.6918 \\
 RXJ1621.2$-$2347   & 16 21 19.9   $-$23 49 59 & 16 21 16.24  $-$23 47 21.99 &  I      &  10.0    &  1.20    &  2454282.7160 \\
                    &                          &                             &  V      &  25.0    &  1.25    &  2454282.7188 \\
                    &                          &                             &  B      &  40.0    &  1.26    &  2454282.7198 \\
 RXJ1623.1$-$2300   & 16 23 11.9   $-$23 04 59 & 16 23 07.83  $-$23 00 59.67 &  I      &  10.0    &  1.14    &  2454282.6935 \\
                    &                          &                             &  V      &  14.0    &  1.15    &  2454282.6952 \\
                    &                          &                             &  B      &  20.0    &  1.15    &  2454282.6961 \\
 RXJ1623.4$-$2425   & 16 23 25.9   $-$24 28 00 & 16 23 21.81  $-$24 24 57.79 &  B      & ~~7.0    &  1.01    &  2454282.6308 \\
                    &                          &                             &  V      & ~~7.0    &  1.02    &  2454282.6323 \\
                    &                          &                             &  I      & ~~7.0    &  1.02    &  2454282.6340 \\
 RXJ1623.5$-$2523   & 16 23 11.9   $-$23 04 59 & 16 23 32.34  $-$25 23 48.53 &  I      & ~~8.0    &  1.15    &  2454282.6984 \\
                    &                          &                             &  V      &  12.0    &  1.16    &  2454282.7005 \\
                    &                          &                             &  B      &  18.0    &  1.16    &  2454282.7016 \\
 RXJ1624.0$-$2456   & 16 24 11.4   $-$25 00 00 & 16 24 06.32  $-$24 56 46.81 &  B      &  20.0    &  1.02    &  2454282.6355 \\
                    &                          &                             &  V      &  20.0    &  1.02    &  2454282.6371 \\
                    &                          &                             &  I      &  12.0    &  1.02    &  2454282.6385 \\ 
\hline
\end{tabular}
\label{YALO-obslog} 
\end{table*}

\begin{table*}
\vspace{3mm} 
\contcaption{}
\begin{tabular}{lcccccc}
\hline
%                 &          &         &        &        \\
~~~~~~~Object$^{a}$ &  RA(2000)   DEC(2000)    & RA(2000)  DEC(2000)         & Filter  & Exp      & Airmass  & Julian Date \\
                    &  Y4Kcam Field Centre     &       [2MASS]$^{b}$         &         & Time [s] &          &  [days]     \\ \hline
  Rho Ophiuchus \\ \hline
 RXJ1624.8$-$2359   & 16 24 52.4   $-$24 04 00 & 16 24 48.40  $-$23 59 16.02 &  B      &  75.0    &  1.02    &  2454282.6401 \\
                    &                          &                             &  V      &  30.0    &  1.03    &  2454282.6425 \\
                    &                          &                             &  I      &  14.0    &  1.03    &  2454282.6436 \\
 RXJ1624.8$-$2239   & 16 24 55.9   $-$22 42 59 & 16 24 51.36 $-$22 39 32.54  &  B      &  18.0    &  1.18    &  2454282.7035 \\
                    &                          &                             &  V      &  10.0    &  1.19    &  2454282.7051 \\
                    &                          &                             &  I      & ~~7.0    &  1.19    &  2454282.7068 \\
 RXJ1625.0$-$2508   & 16 25 09.9   $-$25 12 59 & 16 25 04.49  $-$25 09 11.49 &  I      & ~~7.0    &  1.03    &  2454282.6452 \\
                    &                          &                             &  V      & ~~7.0    &  1.03    &  2454282.6467 \\
                    &                          &                             &  B      &  25.0    &  1.03    &  2454282.6480 \\
 RXJ1625.4$-$2346   & 16 25 33.0   $-$23 49 59 & 16 25 28.64  $-$23 46 26.55 &  B      &  25.0    &  1.04    &  2454282.6496 \\
                    &                          &                             &  V      &  10.0    &  1.04    &  2454282.6512 \\
                    &                          &                             &  I      & ~~8.0    &  1.04    &  2454282.6527 \\
 RXJ1625.6$-$2613   & 16 25 43.0   $-$26 17 00 & 16 25 38.49  $-$26 13 54.03 &  I      & ~~7.0    &  1.19    &  2454282.7109 \\
                    &                          &                             &  V      &  12.0    &  1.20    &  2454282.7121 \\
                    &                          &                             &  B      &  15.0    &  1.20    &  2454282.7136 \\
 ROXR1 13           & 16 26 09.9   $-$24 27 00 & 16 26 03.02  $-$24 23 36.04 &  B      &  50.0    &  1.01    &  2454282.6206 \\
                    &                          &                             &  I      &  20.0    &  1.01    &  2454282.6255 \\  
                    &                          &                             &  V      &  20.0    &  1.01    &  2454282.6272 \\
 RXJ1627.1$-$2419   & 16 27 14.9   $-$24 22 59 & 16 27 10.28  $-$24 19 12.74 &  I      &  ~~7.0   &  1.04    &  2454282.6542 \\
                    &                          &                             &  V      &  10.0    &  1.04    &  2454282.6556 \\
                    &                          &                             &  B      &  30.0    &  1.04    &  2454282.6573 \\
\hline
\end{tabular}
\begin{flushleft}
%Notes: \\
a $-$ Nomenclature based on Rosat All-Sky Survey detections. 
[http://www.xray.mpe.mpg.de/cgi-bin/rosat/rosat-survey] \\
b $-$  Astrometric data taken from the {\sc 2mass} All-Sky Release Point 
Source catalogue (March 2003). [http://irsa.ipac.caltech.edu/applications/Gator/] \\
\end{flushleft}
\end{table*}

\begin{table*}
\vspace{3mm} 
\caption{Log of {\sc uves} spectroscopic observations of Chamaeleon and Rho Ophiuchus candidate members.}
\begin{tabular}{lccccc}
\hline
~~~~~~~Object       &      UT       &   UT        &   Exp     & Heliocentric Julian & Airmass \\
                    &     Date      &   Time      &  Time [s] &  Date [days]        &         \\ \hline 
Chamaeleon \\ \hline
  RXJ0951.9-7901    &  2005-03-23   &   03:24:07  & ~~282     &  2453452.6444       &  1.729    \\ 
  RXJ1112.7-7637    &  2005-03-24   &   00:56:49  &  1487     &  2453453.5495       &  1.783    \\ 
  RXJ1129.2-7546    &  2005-03-24   &   01:24:59  &  1487     &  2453453.5691       &  1.735    \\ 
  RXJ1140.3-8321    &  2005-03-24   &   04:37:35  & ~~649     &  2453453.6973       &  1.924    \\ 
  RXJ1158.5-7754a   &  2005-03-24   &   04:51:24  & ~~712     &  2453453.7078       &  1.672    \\ 
  RXJ1159.7-7601    &  2005-03-24   &   05:06:38  & ~~540     &  2453453.7176       &  1.607    \\ 
  RXJ1201.7-7859    &  2005-03-24   &   05:31:55  & ~~~~71    &  2453453.7322       &  1.729    \\ 
  RXJ1233.5-7523    &  2005-03-24   &   05:37:23  & ~~163     &  2453453.7368       &  1.584    \\ 
  RXJ1239.4-7502    &  2005-03-24   &   05:44:30  & ~~410     &  2453453.7432       &  1.573    \\ 
  RXJ1303.5-7701    &  2005-03-24   &   05:54:31  & ~~149     &  2453453.7485       &  1.638    \\ \hline
%
%% Rho Oph
  Rho Ophiuchus \\ \hline
  RXJ1620.1-2348    &   2005-03-24  &   09:29:36  &  ~~236    &  2453453.8995       &  1.010   \\
  RXJ1620.7-2348    &   2005-03-27  &   05:12:57  &  2145     &  2453456.7326       &  1.486   \\
  RXJ1621.0-2352    &   2005-03-24  &   09:36:42  &  ~~341    &  2453453.9050       &  1.014   \\
  RXJ1623.1-2300    &   2005-03-27  &   05:52:02  &  1128     &  2453456.7538       &  1.298   \\
  RXJ1623.4-2425    &   2005-03-27  &   06:13:56  &  2585     &  2453456.7774       &  1.213   \\
  RXJ1623.5-2523    &   2005-04-06  &   08:19:39  &  ~~492    &  2453466.8534       &  1.001   \\
  RXJ1624.0-2456    &   2005-04-06  &   08:32:08  &  1961     &  2453466.8706       &  1.005   \\
  RXJ1624.8-2359    &   2005-04-21  &   03:39:35  &  3107     &  2453481.6751       &  1.479   \\
  RXJ1624.8-2239    &   2005-05-04  &   03:00:29  &  ~~781    &  2453494.6352       &  1.416   \\
  RXJ1625.0-2508    &   2005-04-06  &   09:11:05  &  1237     &  2453466.8934       &  1.032   \\
  RXJ1625.4-2346    &   2005-05-04  &   03:23:06  &  ~~125    &  2453494.6471       &  1.303   \\
  RXJ1625.4-2346    &   2005-05-05  &   02:56:40  &  1128     &  2453495.6345       &  1.413   \\
  RXJ1625.6-2613    &   2005-04-16  &   06:31:49  &  ~~781    &  2453476.7809       &  1.027   \\
  ROXR1 13          &   2005-05-15  &   04:35:20  &  2357     &  2453505.7105       &  1.030   \\
  ROXR1 13          &   2005-05-15  &   06:42:51  &  2357     &  2453505.7991       &  1.037   \\
  RXJ1627.1-2419    &   2005-05-15  &   03:46:42  &  2585     &  2453505.6780       &  1.104   \\
\hline
\end{tabular}
\label{UVES-obslog} 
\end{table*}

\section{Reddening Vectors and HRDs using Pre-main Sequence Spectral-type vs Colour Relationships}\label{PMS-types}

Historically, studying properties of star forming region members has 
oftentimes been benchmarked against samples of similar effective 
temperature main sequence stars and mixed suites of theoretical 
pre-main and main-sequence models (e.g. \citealt{neuhauser1995}, 
\citealt{Cov97}, \citealt{M98}, \citealt{Luhman:2004}, 
\citealt{James06}). In our own research programme, we too have adopted 
this philosophy, all the while pondering how appropriate it is to 
employ dwarf-class effective temperatures and bolometric corrections 
for use in HRDs for stars still on the pre-main sequence.

In this section, we therefore provide a reddening/extinction analysis 
for our SFR candidate members based on recent pre-main sequence models 
as opposed to dwarf star properties, and adopt their pre-main sequence 
effective temperatures and bolometric corrections in constructing HRDs. 
We subsequently compare and contrast the derived physical properties 
of our SFR candidate members with those derived from main-sequence 
model parameters. 

In Table~\ref{Reddening-Extinction-PMSTable}, we present the results 
of the reddening/extinction analysis, detailing wavelength-dependent 
vectors derived from the \citet{PM13} pre-main sequence spectral-type 
versus photometric colour values. As in the case for the main sequence 
analysis (see Table~\ref{Reddening-Extinction-Table}), we prefer not to 
derive extinction values for those stars with negative reddening values, 
which occurred five times, and once, for Chamaeleon and Rho Ophiuchus stars, 
respectively, in the main sequence properties section. In this pre-main 
sequence analysis however, such cases jump to thirteen and five 
cases, respectively - representing a marked increase in invalid values. 

While this feature is noteworthy, it is not altogether surprising given 
how photometrically variable these magnetically active stars are (probably 
at the $\simeq$ 10 per cent level). However, when one examines the difference 
in extinction values based on the four photometric colours we employ (see 
Table~\ref{DELTAExtinction-table}), the two infrared colours yield $A_{v}$ 
values based on $E(J-H)$ \& $E(H-K)$ $\simeq 3-5$ times their corresponding 
optical colour  $Av$ values based on $E(B-V)$ \& $E(V-I)$ for Chamaeleon stars, 
and $\simeq 4-10$ times in the case of the Rho Ophiuchus ensemble. This 
suggests that the stars are considerably more red in the PMS models 
compared to the MS ones. We note however that the PM13 PMS models cover 
the $5-30$ Myr age range, which encompasses a broad age range of surface 
stellar conditions during this rapidly variable stage of stellar evolution.

\subsection{HRDs properties using PM13 colours, effective temperatures and bolometric corrections}

We re-construct our series of multi-colour HRDs using the extinction corrections 
(see Table~\ref{Reddening-Extinction-PMSTable}) and bolometric corrections from 
the PM13 empirically-calibrated PMS models, and perform an age-mass analysis 
similar to that presented in \S~\ref{HRDs}. The ages and masses that we derive 
are detailed in Tables~\ref{AGEtablePMS}~\&~\ref{PMSMASStable}. 

A comparison of stellar ages for our sample between the KH95 and PM13 analysis reveals 
the following properties: (a) Chamaeleon I and $\epsilon$ Cha stars show no difference 
in age; (b) For the two former $\eta$ Cha candidates, RXJ0850.1-7554 \& RXJ0951.9-7901, 
we present discussion in \S~\ref{AGES} supporting the supposition that these stars are 
in fact likely members of the older Carina and Tuc-Hor associations, respectively. 
The PM13 HRD analysis, retaining a 97pc distance, confirms this assertion for the 
RXJ0850.1-7554 object returning an age of $>29$ Myr, although the case for RXJ0951.9-7901 
is weaker. Either way, the PM13 mean age of RXJ0951.9-7901 is $16.0 \pm 5.0$ Myr, older 
than the 5-10 Myr age reported in the literature for $\eta$ Cha (\citealt{LawsonEtaCha}; 
\citealt{LuhmanEtaCha}); (c) For the Rho Ophiuchus stars, within the error bars, there 
is no discernible difference between the KH95 and PM13 analysis. Finally, with one or two 
exceptions, Chamaeleon and Rho Ophiuchus stars have stellar masses derived in the PM13 analysis 
consistently $0.1-0.3$ M$_{\odot}$ lower than for the KH95 analysis.

\begin{table*}
\begin{center}
\caption[]{\protect \small Reddening and extinction vectors for Chamaeleon and Rho Ophiuchus targets based 
on empirically calibrated theoretical pre-main sequence photometric colours \citep{PM13}.}
\vspace{3mm} 
\begin{tabular}{lcccccccc}
\hline
~~~~~~~Target         &  $E(B-V)$$^{a}$  & $E(V-Ic)$$^{a}$ & $E(J-H)$$^{a}$ & $E(H-K)$$^{a}$ & $A_{v}$$^{b}$ &  $A_{v}$$^{b}$ & $A_{v}$$^{b}$ & $A_{v}$$^{b}$ \\
                      &                &               &                &                & [$E(B-V)$] & [$E(V-Ic)$] & [$E(J-H)$] & [$E(H-K)$] \\
\hline 
Chamaeleon \\ \hline
RXJ0850.1-7554        &  ~0.007 &  ~0.055 &   ~0.061 &  ~0.024  &   0.022  &    0.136  &   0.540  & ~~0.392  \\
RXJ0951.9-7901        &  ~0.061 &   ...   &   ~0.079 &  -0.022  &   0.189  &    ...    &   0.700  &  ...     \\
RXJ1112.7-7637        &   ...   &  ~0.334 &   ~0.261 &  ~0.385  &   ...    &    0.828  &   2.312  & ~~6.283  \\
RXJ1129.2-7546        &  ~0.387 &  ~0.654 &   ~0.143 &  ~0.086  &   1.200  &    1.622  &   1.267  & ~~1.404  \\
RXJ1140.3-8321$^{c}$  &  ~0.122 &  ~0.169 &   ~0.069 &  -0.086  &   0.378  &    0.419  &   0.611  &  ...     \\
RXJ1140.3-8321$^{c}$  &  ~0.032 &  ~0.019 &   ~0.019 &  -0.096  &   0.099  &    0.047  &   0.168  &  ...     \\
RXJ1158.5-7754a       &  ~0.159 &    ...  &   ~0.113 &  -0.008  &   0.493  &    ...    &   1.001  &  ...     \\
RXJ1159.7-7601        &  ~0.130 &  ~0.187 &   ~0.121 &  ~0.005  &   0.403  &    0.464  &   1.072  &  ~~0.082 \\
RXJ1201.7-7859        &  -0.030 &  -0.010 &   -0.034 &  ~0.009  &     ...  &     ...   &    ...   &  ~~0.147 \\
RXJ1239.4-7502$^{c}$  &  ~0.067 &  ~0.083 &   -0.009 &   ~0.036 &   0.208  &    0.206  &    ...   &  ~~0.588 \\
RXJ1239.4-7502$^{c}$  &  -0.023 &  -0.027 &   -0.069 &   ~0.016 &     ...  &     ...   &    ...   &  ~~0.261 \\ \hline
Rho Ophiuchus \\ \hline
RXJ1620.7-2348        &  ~0.264 &  ~0.375 &   ~0.126 &  ~0.044  &   0.818  &    0.930 &    1.116  & ~~0.718  \\
RXJ1621.0-2352        &  -0.008 &  -0.025 &   -0.085 &  -0.020  &    ...   &     ...  &     ...   &   ...    \\
RXJ1621.2-2347        &  ~1.262 &  ~1.769 &   ~0.680 &  ~0.376  &   3.912  &    4.387 &    6.025  & ~~6.136  \\
RXJ1623.1-2300        &  ~0.293 &  ~0.481 &   ~0.149 &  -0.001  &   0.908  &    1.193 &    1.320  &   ...    \\
RXJ1623.4-2425        &  ~0.533 &  ~0.982 &   ~0.378 &  ~0.198  &   1.652  &    2.435 &    3.349  & ~~3.231  \\
RXJ1623.5-2523        &  ~0.448 &  ~0.704 &   ~0.191 &  ~0.154  &   1.389  &    1.746 &    1.692  & ~~2.513  \\
RXJ1624.0-2456        &  ~0.757 &  ~1.170 &   ~0.418 &  ~0.187  &   2.347  &    2.902 &    3.703  & ~~3.052  \\
RXJ1624.8-2239        &  ~0.114 &  ~0.178 &   ~0.039 &  ~0.056  &   0.353  &    0.441 &    0.346  & ~~0.914  \\
RXJ1624.8-2359        &  ~0.790 &  ~1.469 &   ~0.605 &  ~0.258  &   2.449  &    3.643 &    5.360  & ~~4.211  \\
RXJ1625.0-2508        &  ~0.750 &  ~1.203 &   ~0.418 &  ~0.143  &   2.325  &    2.983 &    3.703  & ~~2.334  \\
RXJ1625.4-2346        &  ~0.378 &  ~0.747 &   ~0.292 &  ~0.119  &   1.172  &    1.853 &    2.587  & ~~1.942  \\
RXJ1625.6-2613        &  ~0.003 &  ~0.014 &   ~0.081 &  ~0.240  &   0.009  &    0.035 &    0.718  & ~~3.917  \\
ROXR1 13              &  ~1.365 &  ~2.166 &   ~0.798 &  ~0.505  &   4.232  &    5.372 &    7.070  & ~~8.242  \\
RXJ1627.1-2419        &  ~1.161 &  ~2.153 &   ~0.988 &  ~0.688  &   3.599  &    5.339 &    8.754  & 11.228   \\  \hline
\end{tabular}\label{Reddening-Extinction-PMSTable} 
\end{center}
\begin{flushleft}
%Notes: \\
a $-$ Reddening vectors are calculated by subtracting appropriate photometric colours for 
each target (see Table~\ref{SFRcandidates-OPTIRphot}) from corresponding theoretical values 
for pre-main sequence stars (from table~6 in \citealt{PM13}) based on spectral types 
(see Table~\ref{SFRcandidates-UVES}). \\
b $-$ Extinction vectors [Av], based on four colour-dependent reddening vectors, are calculated 
to be: For optical data, Av=3.1$\times$$E(B-V)$ and Av=2.48$\times$$E(V-Ic)$ (from \citealt{BB88}); 
For infrared, Av=8.86$\times$$E(J-H)$ and Av=16.32$\times$$E(H-K)$ (from \citealt{RM2005}). 
For negative reddening values, we do not calculate an extinction value.\\
c $-$ Two sets of reddening and extinction values are calculated, one for each of the 
two spectral types. \\
\end{flushleft}
\end{table*}

\begin{table*}
\begin{center}
\caption[]{\protect \small Differential Extinction Vectors for our Chamaeleon and 
Rho Ophiuchus Targets (KH95 main sequence values - PM13 PMS values).}
\vspace{3mm} 
\begin{tabular}{lcccc}
\hline
~~~~~~~Target         & $\Delta$Av$^{a}$ & $\Delta$Av$^{a}$ & $\Delta$Av$^{a}$ & $\Delta$Av$^{a}$ \\
                      & [$E(B-V)$]       & [$E(V-Ic)$]      & [$E(J-H)$]       & [$E(H-K)$] \\ \hline 
Chamaeleon \\ \hline
RXJ0850.1-7554            &   ~~0.186       &  ~~0.050          & ~~0.027         & ~~0.526     \\
RXJ0951.9-7901            &   ~~0.186       &    ...            & ~~0.291         &  ...        \\
RXJ1112.7-7637            &    ...          &  ~~0.000          & ~~0.399         & ~~0.365     \\
RXJ1129.2-7546            &   ~~0.155       &  ~~0.099          & ~~0.583         & ~~0.691     \\
RXJ1140.3-8321$^{b}$      &   ~~0.155       &  ~~0.099          & ~~0.583         &   ...       \\
RXJ1140.3-8321$^{b}$      &   ~~0.124       &  ~~0.298          & ~~0.679         &   ...       \\
RXJ1158.5-7754a           &   ~~0.155       &     ...           & ~~0.583         &   ...       \\
RXJ1159.7-7601            &   ~~0.155       &  ~~0.099          & ~~0.583         & ~~0.691     \\
RXJ1201.7-7859            &    ...          &     ...           &    ...          & ~~0.363     \\
RXJ1239.4-7502$^{b}$      &   ~~0.124       &  ~~0.000          &    ...          & ~~0.364     \\
RXJ1239.4-7502$^{b}$      &    ...          &     ...           &    ...          & ~~0.691     \\ \hline
Mean (all stars)          &   ~~0.155       &  ~~0.092          & ~~0.466         & ~~0.527     \\
($\pm 1\sigma$)           & $\pm 0.023$     & $\pm 0.101$       & $\pm 0.216$     & $\pm 0.164$ \\ \hline
Mean ($A_{v}$$<$5)        &   ~~0.155       &  ~~0.108          & ~~0.476         &  ~~0.554    \\
($\pm 1\sigma$)           & $\pm 0.023$     & $\pm 0.101$       & $\pm 0.232$     & $\pm 0.161$ \\ \hline
Rho Ophiuchus \\ \hline
RXJ1620.7-2348            &   ~~0.124       &  ~~0.298          & ~~0.679         &  ~~0.692    \\
RXJ1621.0-2352            &    ...          &     ...           &    ...          &   ...       \\
RXJ1621.2-2347            &   ~~0.155       &  ~~0.099          & ~~0.583         &  ~~0.692    \\
RXJ1623.1-2300            &   ~~0.155       &  ~~0.099          & ~~0.583         &   ...       \\
RXJ1623.4-2425            &  -0.062         &   -0.049          & ~~0.012         & ~~0.363     \\
RXJ1623.5-2523            &  ~~0.000        &  ~~0.198          & ~~0.302         & ~~0.365     \\
RXJ1624.0-2456            &  ~~0.000        &  ~~0.198          & ~~0.302         & ~~0.364     \\
RXJ1624.8-2239            &  ~~0.031        &  ~~0.075          & ~~0.393         & ~~0.365     \\
RXJ1624.8-2359            &  ~~0.155        &  ~~0.099          & ~~0.583         & ~~0.691     \\
RXJ1625.0-2508            &  -0.031         &   -0.124          &  -0.077         & ~~0.362     \\
RXJ1625.4-2346            &  ~~0.031        &  ~~0.074          & ~~0.394         & ~~0.365     \\
RXJ1625.6-2613            &    ...          &  ~~0.149          & ~~0.515         & ~~0.530     \\
ROXR1 13                  &  ~~0.000        &  ~~0.198          & ~~0.302         & ~~0.364     \\
RXJ1627.1-2419            &  -0.031         &   -0.124          &  -0.078         & ~~0.363     \\ \hline
Mean (all stars)          &   ~~0.044       &  ~~0.092          & ~~0.346         & ~~0.460     \\
($\pm 1\sigma$)           & $\pm 0.081$     & $\pm 0.127$       & $\pm 0.256$     & $\pm 0.148$ \\ \hline
Mean ($A_{v}$$<$5)        &  ~~0.031        &  ~~0.102          & ~~0.345         & ~~0.426     \\
($\pm 1\sigma$)           & $\pm 0.074$     & $\pm 0.130$       & $\pm 0.249$     & $\pm 0.122$ \\ \hline

\end{tabular}\label{DELTAExtinction-table} 
\end{center}
\begin{flushleft}
Notes: \\
a $-$ For each SFR target, differential Av vectors, as a function of photometric colour, are 
calculated by subtracting spectral type-colour dependent Av values derived from main sequence 
(KH95) and pre-main sequence (\citealt{PM13}) theoretical models. \\
b $-$ Two sets of differential extinction values are calculated, one for each of the 
two spectral types. \\
\end{flushleft}
\end{table*}

\begin{table*}
\vspace{3mm}
\caption[]{\protect \small Isochronal ages for Chamaeleon and Rho Ophiuchus candidates stars using four separate 
colour-dependent extinction laws (based on PM13 effective temperatures and bolometric corrections 
for pre-main sequence stars) and {\sc parsec} stellar models.}
\begin{tabular}{lcccccccccc}
\hline
~~~~~~~~Target$^{a}$             & $T_{\mathrm{eff}}$$^{b}$ & [\lbol ]$^{c}$  & Age              & [\lbol]$^{c}$  & Age     & [\lbol]$^{c}$   & Age    & [\lbol]$^{c}$   & Age       & Mean age             \\
                                 &[K]   &           & [Myr]            &          & [Myr]   &           & [Myr]  &           & [Myr]     & [Myr]                \\ \hline 
                                 &      & \multicolumn{2}{c}{$^{b}$$E(B-V)$}  & \multicolumn{2}{c}{$^{b}$$E(V-I)$} & \multicolumn{2}{c}{$^{b}$$E(J-H)$} &\multicolumn{2}{c}{$^{b}$$E(H-K)$}  &  \\
% \cmidrule(lr){1-1}                            \cmidrule(lr){3-4}               \cmidrule(lr){5-6}           \cmidrule(lr){7-8}             \cmidrule(lr){9-10} 
                                               \cmidrule(lr){3-4}               \cmidrule(lr){5-6}           \cmidrule(lr){7-8}             \cmidrule(lr){9-10} 
Chamaeleon I   (\mycircle{blue}) &      &           &                  &          &         &           &        &           &           &                      \\
 \cmidrule(lr){1-1} 

RXJ1112.7-7637                   & 4760 &  ~~9.999  &   ...            & ~~0.177  &  3.5    & ~~0.771   & $<1$   & ~~2.359   &$<1$       &   $<1.8$             \\
RXJ1129.2-7546                   & 4550 & $-0.163$  &   7.0            & ~~0.006  &  4.0    &  $-0.136$ &  6.0   & $-0.081$  & 5.0       &   $5.5 \pm 0.6$      \\  
$\eta$ Cham  (\mytriangle{yellow})  &   &           &                  &          &         &           &        &           &           &                      \\
 \cmidrule(lr){1-1}
  RXJ0850.1-7554                 & 5390 &  $-0.325$ &  $>30$           & $-0.279$ &  $>30$  &  $-0.117$ &   27   & $-0.177$  & $>30$     &   $>29$              \\
  RXJ0951.9-7901                 & 5290 &  $-0.080$ &    21            & ~~9.999  &   ...   & ~~0.125   &   11   & ~~9.999   &  ...      &   $16.0 \pm 5.0$     \\
Tuc Hor  ({\Montriangle{red}})   &      &           &                  &          &         &           &        &           &           &                      \\
 \cmidrule(lr){1-1}
  RXJ1140.3-8321$^\dagger$       & 4550 &  $-0.948$ &   $>30$          & $-0.932$ &  $>30$  &  $-0.855$ & $>30$  & ~~9.999   &  ...      &   $>30$              \\
  RXJ1140.3-8321$^\ddagger$      & 4330 &  $-0.992$ &   $>30$          & $-1.012$ & $>30$   &  $-0.964$ & $>30$  & ~~9.999   &  ...      &   $>30$              \\
$\epsilon$ Cham (\ssquare{green}) &     &           &                  &          &         &           &        &           &           &                      \\
 \cmidrule(lr){1-1}
  RXJ1158.5-7754a                & 4550 & ~~0.238   &  1.5             & ~~9.999  &  ...    & ~~0.441   &   $<1$ & ~~9.999   &  ...      &  $<1.3$              \\
  RXJ1159.7-7601                 & 4550 &  $-0.063$ &   4.5            & $-0.038$ &  4.5    & ~~0.205   &   2.0  &  $-0.191$ &  7.5       &  $4.6 \pm 1.1$       \\
  RXJ1201.7-7859                 & 5500 & ~~9.999   &   ...            & ~~9.999  &  ...    & ~~9.999   &   ...  & ~~0.711   &  3.5      &  $3.5$               \\
  RXJ1239.4-7502$^\dagger$       & 4760 & ~~0.138   &   4.0            & ~~0.137  &  4.0    & ~~9.999   &   ...  & ~~0.290   &  2.0      &  $3.3 \pm 0.7$       \\
  RXJ1239.4-7502$^\dagger$       & 4550 & ~~9.999   &   ...            & ~~9.999  &  ...    & ~~9.999   &   ...  & ~~0.215   &  2.0      &  $2.0$               \\
Rho Ophiuchus ({\textcolor{red}{$\triangle$}}) &  &       &            &          &         &           &        &           &            &                     \\
 \cmidrule(lr){1-1}
RXJ1620.7-2348                   & 4330 &  $-0.310$ &    7.0           & $-0.266$ &  5.0    & $-0.191$  &   4.0  & $-0.350$  &  7.5       &  $5.9 \pm 0.8$      \\
RXJ1621.0-2352                   & 4920 &  ~~9.999  &    ...           & ~~9.999  &  ...    & ~~9.999   &   ...  & ~~9.999   &  ...       &  ...                \\
RXJ1621.2-2347                   & 4550 &  $-0.480$ &     22           & $-0.290$ &   11    & ~~0.366   &  $<1$  & ~~0.410   & $<1$       &  $<8.8$             \\
RXJ1623.1-2300                   & 4550 &  $-0.043$ &    4.5           & ~~0.071  &  3.0    & ~~0.122   &   2.0  & ~~9.999   & ...        &  $3.2 \pm 0.7$      \\
RXJ1623.4-2425                   & 5870 &  $-0.270$ &  $>30$           & ~~0.044  &   30    & ~~0.409   &    14  & ~~0.362   &   16       &  $>23$              \\
RXJ1623.5-2523                   & 5030 &  ~~0.226  &    5.0           & ~~0.369  &  3.5    & ~~0.348   &   4.0  & ~~0.676   &  1.5       &  $3.5 \pm 0.7$      \\
RXJ1624.0-2456                   & 5030 &  $-0.064$ &     14           & ~~0.158  &  6.5    & ~~0.479   &   2.5  & ~~0.218   &  5.0       &  $7.0 \pm 2.5$      \\
RXJ1624.8-2239                   & 4920 &  ~~0.502  &    2.0           & ~~0.538  &  1.5    & ~~0.500   &   2.0  & ~~0.727   & $<1$       &  $<1.6$             \\
RXJ1624.8-2359                   & 4550 &  $-0.306$ &     11           & ~~0.172  &  2.0    & ~~0.859   &  $<1$  & ~~0.399   & $<1$       &  $<3.8$             \\
RXJ1625.0-2508                   & 5970 &  ~~0.169  &     25           & ~~0.432  &   16    & ~~0.720   &   8.5  & ~~0.173   &   25       &  $18.6 \pm 4.0$     \\
RXJ1625.4-2346                   & 4920 &  $-0.053$ &     10           & ~~0.219  &  4.5    & ~~0.513   &   2.0  & ~~0.255   &  4.0       &  $5.1 \pm 1.7$      \\
RXJ1625.6-2613                   & 3970 &  $-0.122$ &    1.5           & $-0.112$ &  1.5    & ~~0.161   &  $<1$  & ~~1.441   & $<1$       &  $<1.3$             \\
ROXR1 13                         & 5030 &  ~~~0.394 &    3.5           & ~~0.850  & $<1$    & ~~1.529   &  $<1$  & ~~1.998   & $<1$       &  $<1.6$             \\
RXJ1627.1-2419                   & 5970 &  $-0.064$ &  $>30$           & ~~0.632  &   10    & ~~1.998   &  $<1$  & ~~2.988   & $<1$       &  $\simeq 10$ (?)    \\
\hline
\end{tabular}\label{AGEtablePMS}
\begin{flushleft}
%Notes: \\
$a -$ Isochronal age determinations are segregated into stars comprising disparate young {\sc sfr} regions, with symbols 
\mycircle{blue}, \mytriangle{yellow}, {\Montriangle{red}}, \ssquare{green} \& {\textcolor{red}{$\triangle$}}, 
matching those data presented in Figures~\ref{PM13pms-HRDAvBV}$-$\ref{PM13pms-HRDAvHK}. \\
$b -$ \lbol data are calculated using \teff and reddening vectors based on PM13 colour-spectral types relationships 
for pre-main sequence stars (see also Tables~\ref{SFRcandidates-UVES}~\&~\ref{Reddening-Extinction-PMSTable}). \\
$c -$ Distances used for \lbol calculations are the same as those used in \S~\ref{HRDs} and Table~\ref{AGEtable}. \\
$\dagger$ $-$ for a given star, calculations are made using earlier spectral type/higher \teff. \\
$\ddagger$ $-$ for a given star, calculations are made using later spectral type/lower \teff. 
\end{flushleft}
\end{table*}

\begin{table*}
\vspace{3mm}
\caption[]{\protect \small Theoretical {\sc parsec} masses for Chamaeleon, Tuc-Hor and Rho Ophiuchus candidates using four 
separate colour-dependent extinction laws (based on PM13 effective temperatures, colours and bolometric corrections 
for PMS stars corresponding to each target's spectral type).}
\begin{tabular}{lcccccc}
\hline
~~~~~~~Target                      & $T_{\mathrm{eff}}$ &   Mass              & Mass                & Mass              & Mass               &  Mean             \\
                                   &  [K]               &  [M$_{\odot}$]      & [M$_{\odot}$]       & [M$_{\odot}$]     & [M$_{\odot}$]      &  Mass [M$_{\odot}$]    \\
                                   &                    &   $E(B-V)$          & $E(V-I)$            & $E(J-H)$          & $E(H-K)$           &                   \\
                                                          \cmidrule(lr){3-3}  \cmidrule(lr){4-4}   \cmidrule(lr){5-5}  \cmidrule(lr){6-6} 
Chamaeleon I   (\mycircle{blue})   &                    &                     &                     &                   &                    &                   \\
 \cmidrule(lr){1-1}   
RXJ1112.7-7637                     & 4760               &    ...              &   1.3             &  1.4               &  $>2.5$             &   $>1.7$          \\
RXJ1129.2-7546                     & 4550               &    1.1              &   1.1             &  1.1               &  1.1                &   1.1 $\pm$ ...    \\
$\eta$ Cham  (\mytriangle{yellow}) &                    &                     &                   &                    &                     &                   \\
 \cmidrule(lr){1-1}
RXJ0850.1-7554                     & 5390               &    0.9              &   0.9             &  1.0               &  0.9                &   0.93 $\pm$ 0.03 \\
RXJ0951.9-7901                     & 5290               &    1.0              &   ...             &  1.3               &  ...                &   1.15 $\pm$ 0.15 \\
Tuc Hor  ({\Montriangle{red}})     &                    &                     &                   &                    &                     &                   \\
 \cmidrule(lr){1-1}
RXJ1140.3-8321$^\dagger$           & 4550               &    ...              &   ...             &  ...               &  ...                &   ...  \\
RXJ1140.3-8321$^\ddagger$          & 4330               &    ...              &   ...             &  ...               &  ...                &   ...  \\
$\epsilon$ Cham (\ssquare{green})  &                    &                     &                   &                    &                     &                   \\
 \cmidrule(lr){1-1}
RXJ1158.5-7754a                    & 4550               &    1.0              &   ...             &  1.0               &  ...                &   1.0  $\pm$ ...  \\
RXJ1159.7-7601                     & 4550               &    1.1              &   1.1             &  1.0               &  1.1                &   1.08 $\pm$ 0.03 \\
RXJ1201.7-7859                     & 5500               &    ...              &   ...             &  ...               &  2.0                &   2.0  $\pm$ ...  \\
RXJ1239.4-7502$^\dagger$           & 4760               &    1.3              &   1.3             &  ...               &  1.3                &   1.3  $\pm$ ...  \\
RXJ1239.4-7502$^\ddagger$          & 4550               &    ...              &   ...             &  ...               &  1.0                &   1.0  $\pm$ ...  \\
Rho Ophiuchus ({\textcolor{red}{$\triangle$}}) &        &                     &                   &                    &                     &                   \\
 \cmidrule(lr){1-1}
RXJ1620.7-2348                     & 4330               &    0.8              &   0.9             &  0.8               &  0.9                &   0.85 $\pm$ 0.03 \\
RXJ1621.0-2352                     & 4920               &    ...              &   ...             &  ...               &  ...                &   ...             \\
RXJ1621.2-2347                     & 4550               &    0.9              &   1.0             &  1.0               &  1.0                &   0.98 $\pm$ 0.03 \\
RXJ1623.1-2300                     & 4550               &    1.1              &   1.1             &  1.0               &  ...                &   1.07 $\pm$ 0.03 \\
RXJ1623.4-2425                     & 5870               &    ...              &   1.1             &  1.3               &  1.3                &   1.23 $\pm$ 0.07 \\
RXJ1623.5-2523                     & 5030               &    1.5              &   1.6             &  1.6               &  2.2                &   1.73 $\pm$ 0.16 \\
RXJ1624.0-2456                     & 5030               &    1.1              &   1.4             &  1.7               &  1.5                &   1.43 $\pm$ 0.13 \\
RXJ1624.8-2239                     & 4920               &    1.6              &   1.6             &  1.6               &  1.7                &   1.63 $\pm$ 0.03 \\
RXJ1624.8-2359                     & 4550               &    1.0              &   1.0             &  1.0               &  1.0                &   1.0 $\pm$ ... \\
RXJ1625.0-2508                     & 5970               &    1.1              &   1.3             &  1.6               &  1.1                &   1.28 $\pm$ 0.12 \\
RXJ1625.4-2346                     & 4920               &    1.2              &   1.4             &  1.6               &  1.5                &   1.43 $\pm$ 0.09 \\
RXJ1625.6-2613                     & 3970               &   $<0.8$            &   $<0.8$          &  $<0.8$            &  $<0.8$             &   $<0.8$          \\
ROXR1 13                           & 5030               &    1.6              &   2.0             &  $>2.5$            &  $>2.5$             &   $>2.2$          \\
RXJ1627.1-2419                     & 5970               &    1.1              &   1.5             &  $>2.5$            &  $>2.5$             &   $>1.9$          \\ \hline
\end{tabular}\label{PMSMASStable}
\begin{flushleft}
Notes: \\
$a -$ {\sc parsec} model mass determinations are segregated into stars comprising disparate young {\sc sfr} regions, 
with symbols \mycircle{blue}, \mytriangle{yellow}, {\Montriangle{red}}, \ssquare{green} \& {\textcolor{red}{$\triangle$}}, matching 
those data presented in Figures~\ref{PM13pms-HRDAvBV}$-$\ref{PM13pms-HRDAvHK}. \\
$b -$ \lbol data are calculated using \teff and reddening vectors based on PM13 colour-spectral types 
relationships for PMS stars (see also Table~\ref{Reddening-Extinction-PMSTable}). \\
$c -$ Distances used for \lbol ~calculations are the same as those presented in the footnotes to 
Table~\ref{AGEtable}. \\
$\dagger$ $-$ for a given star, calculations are made using earlier spectral type/higher \teff. \\
$\ddagger$ $-$ for a given star, calculations are made using later spectral type/lower \teff. 
\end{flushleft}
\end{table*}

\begin{figure}
\epsfig{figure=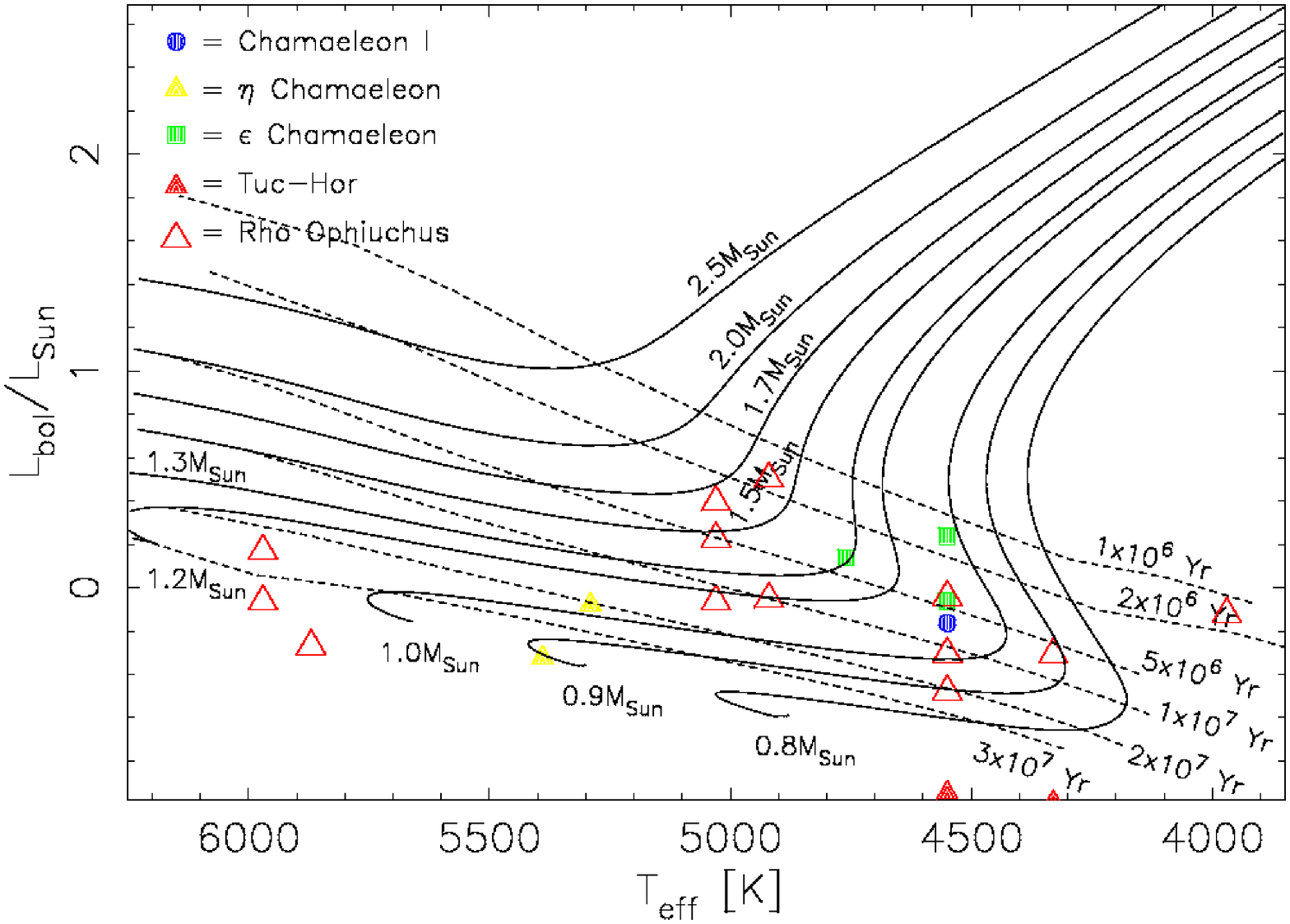,angle=270,width=90mm}
\caption{Similar to Figure~\ref{KH95HRD-AvBV}, we plot an Hertzsprung-Russell diagram, using extinction 
values derived based on $B-V$ colours (see Table~\ref{Reddening-Extinction-PMSTable}), for Cham I 
(\tikzcircle[blue, fill=blue]{2pt}), $\eta$ Cham ({\Montriangle{yellow}}), Tuc-Hor ({\Montriangle{red}}), 
$\epsilon$ Cham ({\Monsquare{green}}) \& Rho Ophiuchus ({\textcolor{red}{$\triangle$}}) objects. 
Effective temperatures, bolometric corrections and reddening vectors are based on the pre-main sequence 
colour-spectral type relationships presented in PM13, with distances to individual stellar groups 
cited in Table~\ref{AGEtable}. Stellar isochrones (dashed tracks) and mass tracks (solid lines) 
are computed from theoretical solar metallicity {\sc parsec} stellar models by \citep{PARSEC}.}
\label{PM13pms-HRDAvBV}
\end{figure}

\begin{figure}
\epsfig{figure=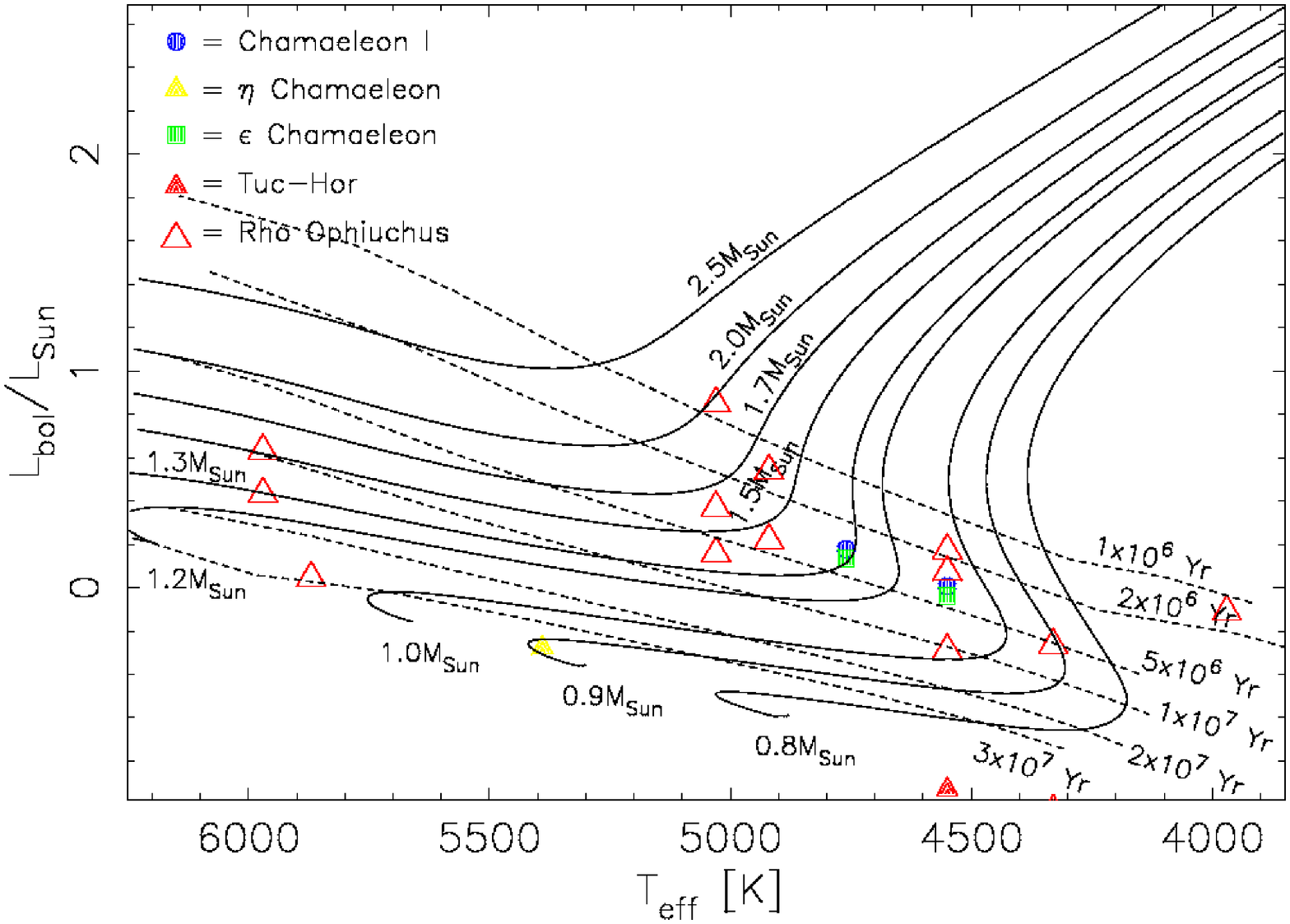,angle=270,width=90mm}
\caption{{\em Idem} to Figure~\ref{PM13pms-HRDAvBV}, except extinction values 
are derived from $V-Ic$ colours.}
\label{PM13pms-HRDAvVI}
\end{figure}

\begin{figure}
\epsfig{figure=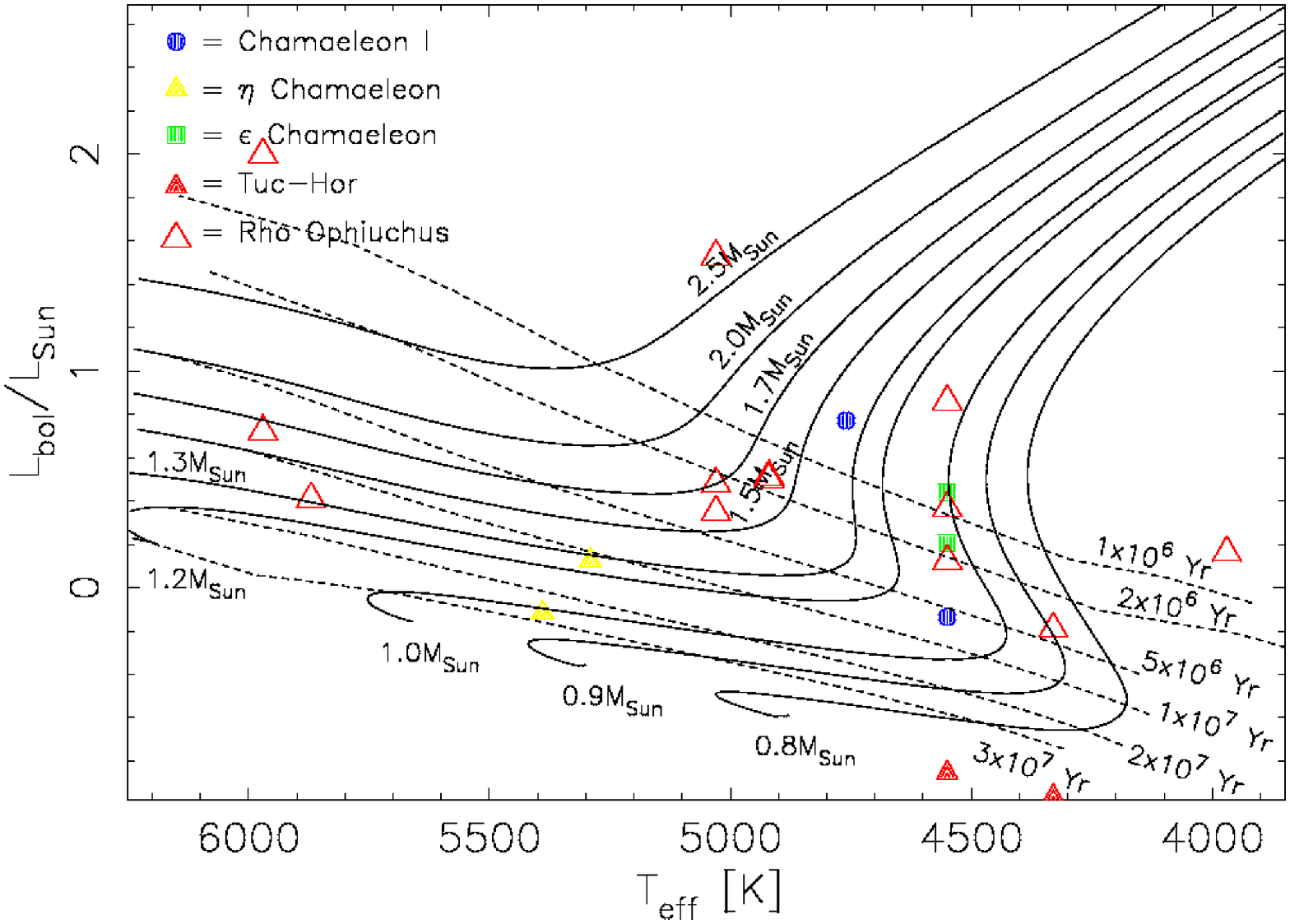,angle=270,width=90mm}
\caption{{\em Idem} to Figure~\ref{PM13pms-HRDAvBV}, except extinction values 
are derived from $J-H$ colours.}
\label{PM13pms-HRDAvJH}
\end{figure}

\begin{figure}
\epsfig{figure=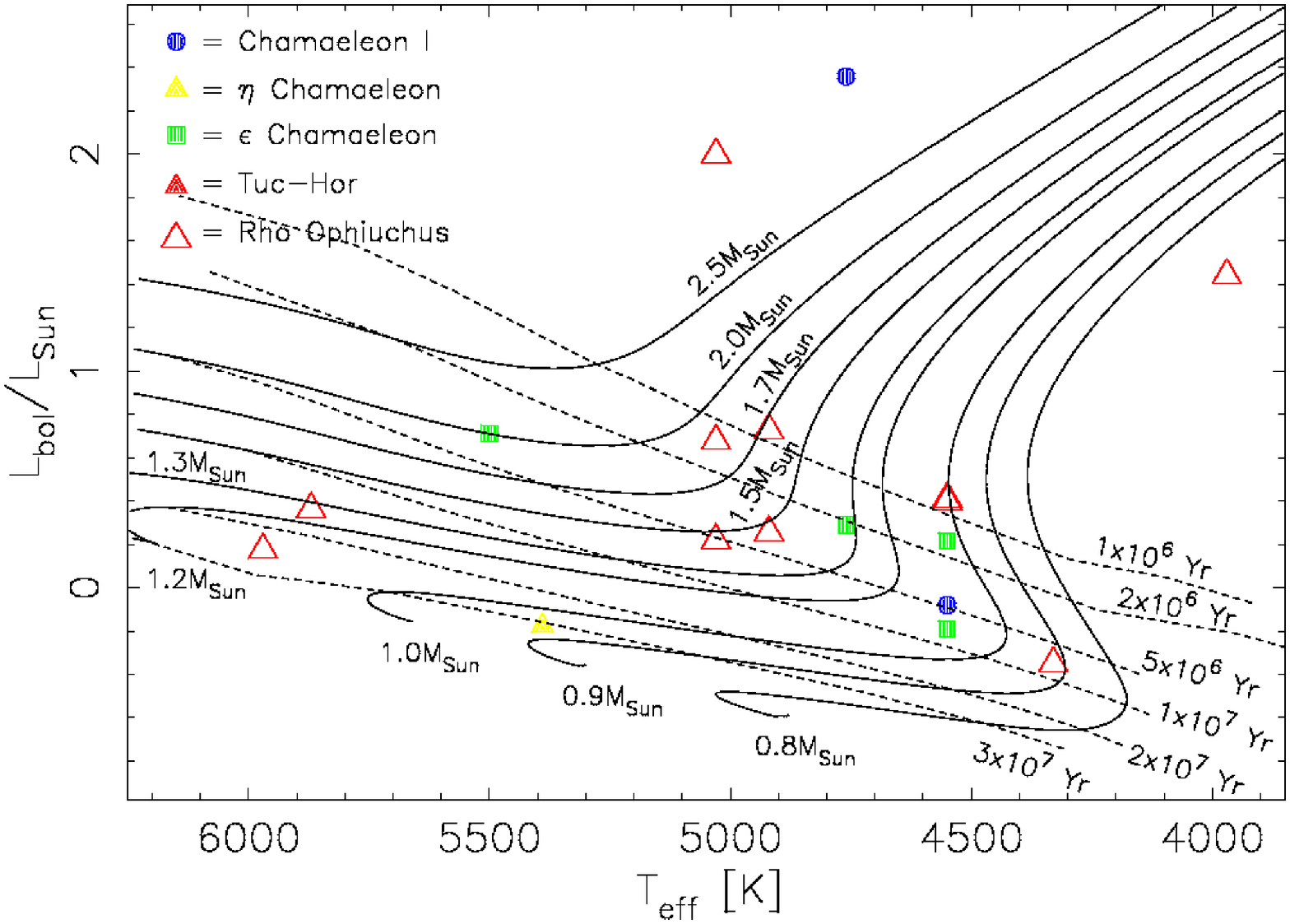,angle=270,width=90mm}
\caption{{\em Idem} to Figure~\ref{PM13pms-HRDAvBV}, except extinction values 
are derived  from $H-K$ colours.}
\label{PM13pms-HRDAvHK}
\end{figure}

\section{Research Notes on Objects with Excess}\label{S_sedlitnotes}

\begin{table*}
\begin{center}
\caption[]{\protect \small Best fitting star$+$disc parameters for our SED models of stars with substantial NIR excess.}
\vspace{3mm} 
\begin{tabular}{llccccc}
\hline
~~~~~~~Target               & Best fit  & T$_{\rm eff}$ & Stellar Mass   & Stellar Radius & Disc Mass & Inclination \\
                            & model ID  & [K]           & [M$_{\odot}$]  & [R$_{\odot}$]  & [M$_{\odot}$]  & [$^{\circ}$] \\ \hline 
RXJ1112.7-7637             & 3018851   & 5090          & 2.52           & 3.60 & $1.16 \times 10^{-2}$ & 87 \\
ROXR1 13                    & 3007238   & 4974          & 3.52           & 6.41 & $4.46 \times 10^{-5}$ & 81 \\ 
RXJ1625.6-2613              & 3015710   & 5683          & 3.02           & 5.53 & $5.16 \times 10^{-3}$ & 87 \\  
RXJ1627.1-2419              & 3011961   & 5865          & 2.46           & 4.18 & $6.25 \times 10^{-2}$ & 32 \\ \hline
RXJ1623.1-2300              & 3001684   & 4274          & 1.01           & 1.79 & $5.62 \times 10^{-8}$ & 87 \\ 
RXJ1623.5-2523              & 3012187   & 5039          & 1.85           & 2.30 & $1.39 \times 10^{-8}$ & 63 \\ 
RXJ1625.0-2508              & 3015211   & 5403          & 1.50           & 1.84 & $5.25 \times 10^{-9}$ & 18 \\ \hline
\end{tabular}\label{T_bfmparms}
\end{center}
\end{table*}

We have gathered below observations and notes from the literature for the seven objects 
which have marginal to high levels of infrared excess emission in their SEDs. Three 
of the four objects with substantial excess have been observed by multiple other 
authors, from sub-mm disc observations/modeling to detailed SED modeling. One object, 
RXJ1625.6-2613 has been relatively unobserved, with only one set of previous 
H $\alpha$/Li {\footnotesize} EW measurements, and its assignation of WTTS status, 
having been reported (\citealt{M98}) is at odds with our analysis 
(see \S~\ref{CTTS-notWTTS}). We have compared our isochrone-derived stellar masses 
with those from our SED models (see Table~\ref{T_bfmparms}), and generally find 
good agreement; the sole exception to such an agreement is RXJ1625.6-2613. In 
this case, our sole model fit which meets the selection criteria (defined in 
\S~\ref{S_seds}) is subject to a degeneracy in stellar luminosity and inclination.

\subsection{ROXR1 13 (a.k.a DoAr 21/RXJ1626.03-2423)}
Being the most extensively, and thoroughly, observed object of our sample abundant 
photometric data are available for ROXR1 13. In our modeling of these data, however, 
we noted inconsistencies likely due to variability; indeed, previous authors have 
noted in the farther wavelengths specifically, there is likely contamination from 
nearby extended sources. ROXR1 13 was observed by \citet{Cieza:2013} with Herschel, 
disc excess seen in both the PACS 70$\mu$m and 160$\mu$m bands. Their H $\alpha$ 
line profile appears to be simply in absorption, but earlier work has shown this 
line to be highly variable \citep{Jensen:2009}. \citet{McClure:2010} obtained 
Spitzer IRS spectra and saw PAH features. The Spitzer c2d survey also observed 
this object, providing IRAC and MIPS photometry \citep{Wahhaj:2010}. Our SED 
for this $A_{v} > 5$ object (see Figure~\ref{F_juicyseds}) clearly shows that 
its photometric data are incompatible with a bare photosphere $>$ 10$\mu$m, 
and it is obviously consistent with a circumstellar disc model (e.g. see 
Table~\ref{T_bfmparms}); such a finding agrees well with its Spitzer c2d 
result \citep{Cieza:2007}, although at odds with the \citet{Valenti2003} 
{\em naked T-Tauri star} (NTTS) status and class III assignment by \citet{Dzib2013}.

\subsection{RXJ1627.1-2419 (a.k.a. SR 21 AB)}
\citet{Furlan:2009} and \citet{McClure:2010} both modeled the SED of this 
object and proposed it is a transitional disc based upon its steep spectral 
slope from 13 to 31 $\mu$m. The spectra of \citet{M98} indicated no ongoing 
accretion, but our spectra show some H $\alpha$ emission; this is unsurprising 
given the high frequency of variability seen in YSO spectral lines generally. 
Submillimeter imaging of \citet{Andrews:2009} resolved a $\sim$37 AU gap 
in the inner disc in the dust continuum, indicating some clearing has taken 
place. We too find that this $A_{v} > 5$ object shows clear evidence of 
infrared excess longward of 5$\mu$m, supporting its class II status reported 
by \citep{Andrews:2007}.

\subsection{RXJ1112.7-7637 (a.k.a. Cha T 2-51/T-51)}
For this $A_{v}$ $> 5$ object, we again find clear evidence in its SED of 
a circumstellar disc, with strong infrared excess longward of a couple of 
microns. This object was also included in the survey of \citet{Furlan:2009}, 
and based on the shallow 13-31$\mu$m spectral slope, it is not likely a 
transition disc object. Both \citet{Furlan:2009} and \citet{Manoj:2011} 
discuss the likeliness, based on the shallow slope in the 13-31$\mu$m 
window, clearing of the outer disc, but RXJ1112.7-7637 lacks a sub-arcsecond 
companion that could do it.

\subsection{RXJ1625.6-2613 (a.k.a. PDS 83, V* V896 Sco, CD-25 1150.4, IRAS 16225-2607)}\label{CTTS-notWTTS}

This target has been previously observed spectroscopically by \citep{M98}, who 
classified as a WTTS, and detected H $\alpha$ emission at the $\simeq$ 4.6\r{A} level as 
well as a strong Li {\footnotesize I} 6708\r{A} absorption line ($\simeq$ 450 m\r{A}). 
Our UVES Li {\footnotesize I} EW measurement agree very well with this earlier datum, 
but we cannot easily, directly at least, compare our H $\alpha$ EWs since we measure the 
residual line profile (observed - template), whereas the study of \citet{M98} solely 
measures the observed profile. We note that in our UVES spectrum however, we observe a 
very strong H $\alpha$ emission feature ($> 6.4$\r{A}), as well as indications of a 
wind in a redshifted self-absorption feature of the H $\alpha$ feature. Similarly, 
\cite{Rojas08} also observe a very strong and complex H $\alpha$ emission feature in 
this object (EW=-12.8\r{A}) as well as compelling evidence for IRAS infra-red emission. 
\cite{Furlan:2009} present a SED of this object which shows clear evidence of mid infra-red 
excess emission and a conspicuous silicon feature at 10$\mu$m, consistent with circumstellar 
material. For this $A_{v} > 4.5$ object, our own SED shows clear evidence of disc-like 
infrared emission; combined with the Rojas et al. and Furlan et al. results, these 
observations are clearly at odds with its Mart\'{\i}n et al. WTTS assignment.

\subsection{Marginal infrared excess targets}

\begin{figure*}
\centering
\epsfig{figure=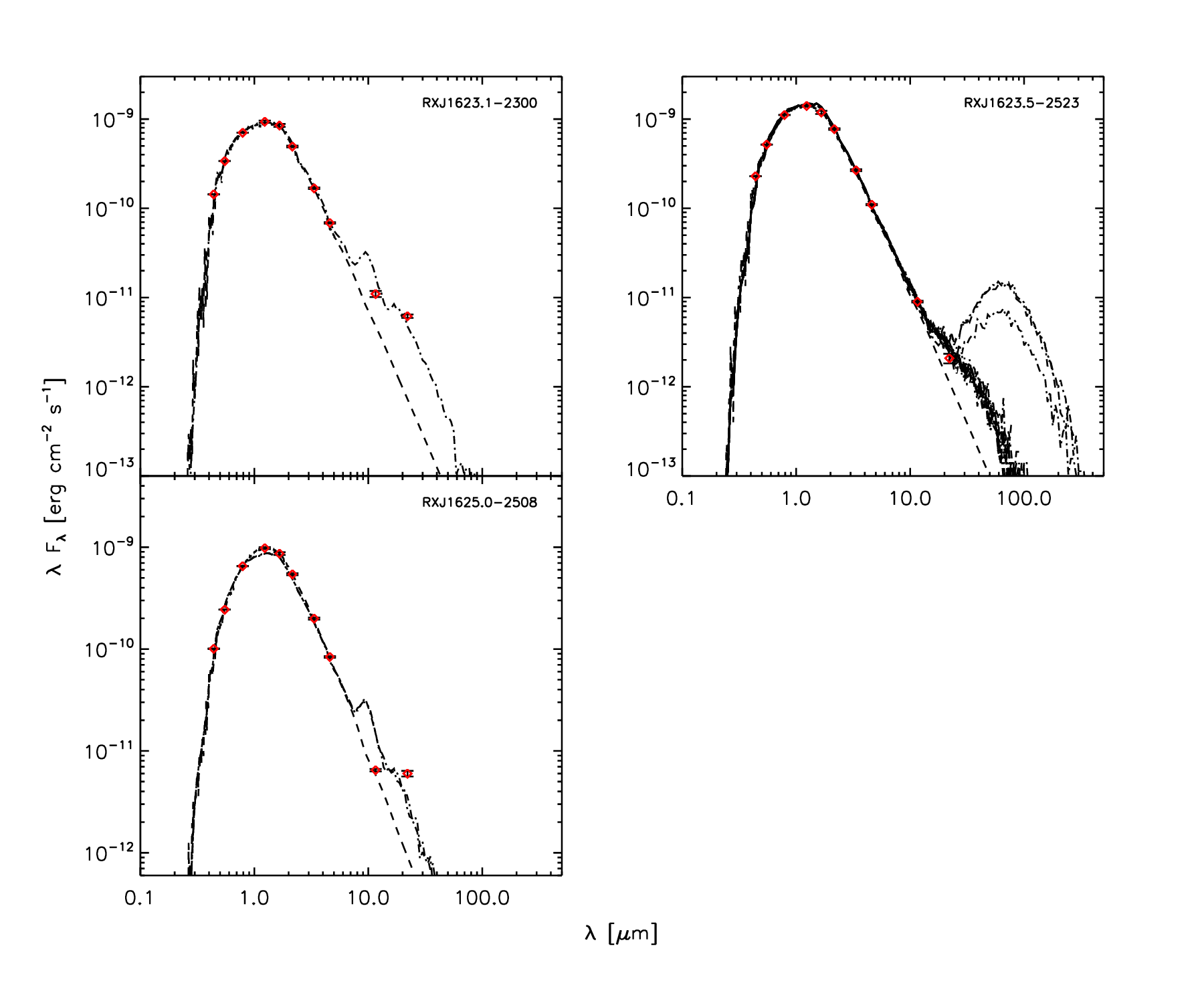,width=125mm,angle=90}
\caption{{\em Idem} to Figure~\ref{F_juicyseds}, however we now plot SEDs 
of Rho Ophiuchus targets with marginal signatures of mid-infrared 
excesses.}
\label{F_moderateseds}
\end{figure*}

RXJ1625.0-2508 and RXJ1623.5-2523, whose SEDs are shown in 
Figure~\ref{F_moderateseds}, were both included in the survey of 
\citet{M98}; our Li {\footnotesize I} EW values are consistent with 
the earlier measurements within error.

\citet{Luhman:2012} suggest RXJ1623.1-2300 is a potential Upper Sco 
member, further noting that excess in the W2 band could mean this 
object has a debris or evolved transitional disc. We see excess in 
W3 and W4, as well a strong, narrow H $\alpha$ emission feature.

\subsection{Remaining $A_{v}>5$ Objects}
SEDs for the two remaining $A_{v}>5$ objects, RXJ1621.2-2347 \& 
RXJ1624.8-2359, are consistent with bare photospheres, and out to 
the WISE W4 band, exhibit essentially no infrared excess emission 
whatsoever. Visual inspection of {\sc dss} optical and {\sc 2mass} 
infrared images show that these two stars lie in heavily-extincted 
regions of the Rho Ophiuchus SFR. We therefore posit that both 
systems are WTTSs experiencing heavy {\em in situ} visual and 
near-infrared extinction.

\section{SEDs for full sample}\label{S_sedappendix}

\begin{figure*}
\centering
\epsfig{figure=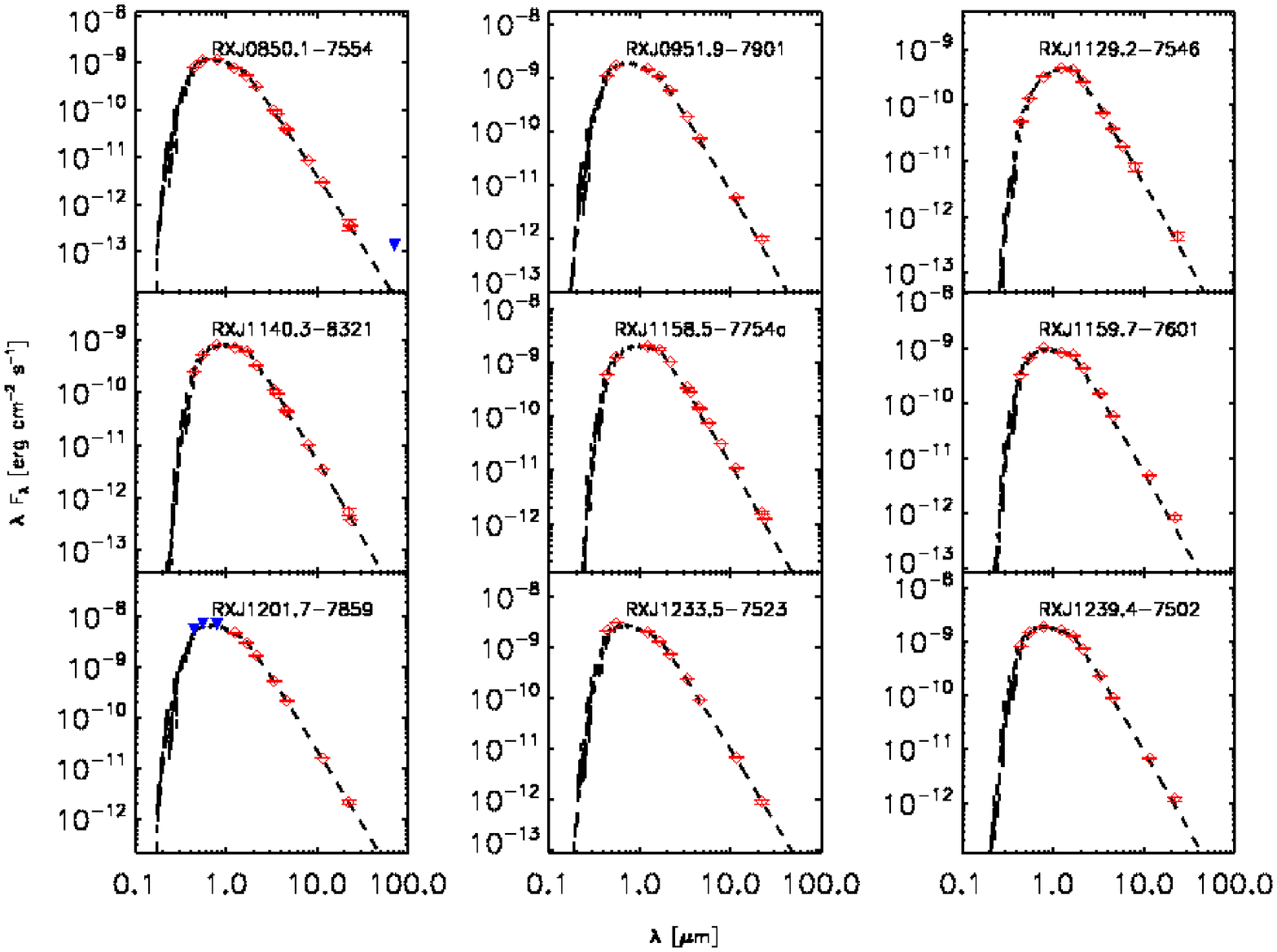,width=150mm,angle=90}
\caption{{\em Idem} to Figure~\ref{F_juicyseds}, we now plot SEDs of 
Chamaeleon candidates that are consistent with bare photospheres.}
\label{F_bareSEDsCham}
\end{figure*}

\begin{figure*}
\centering
\epsfig{figure=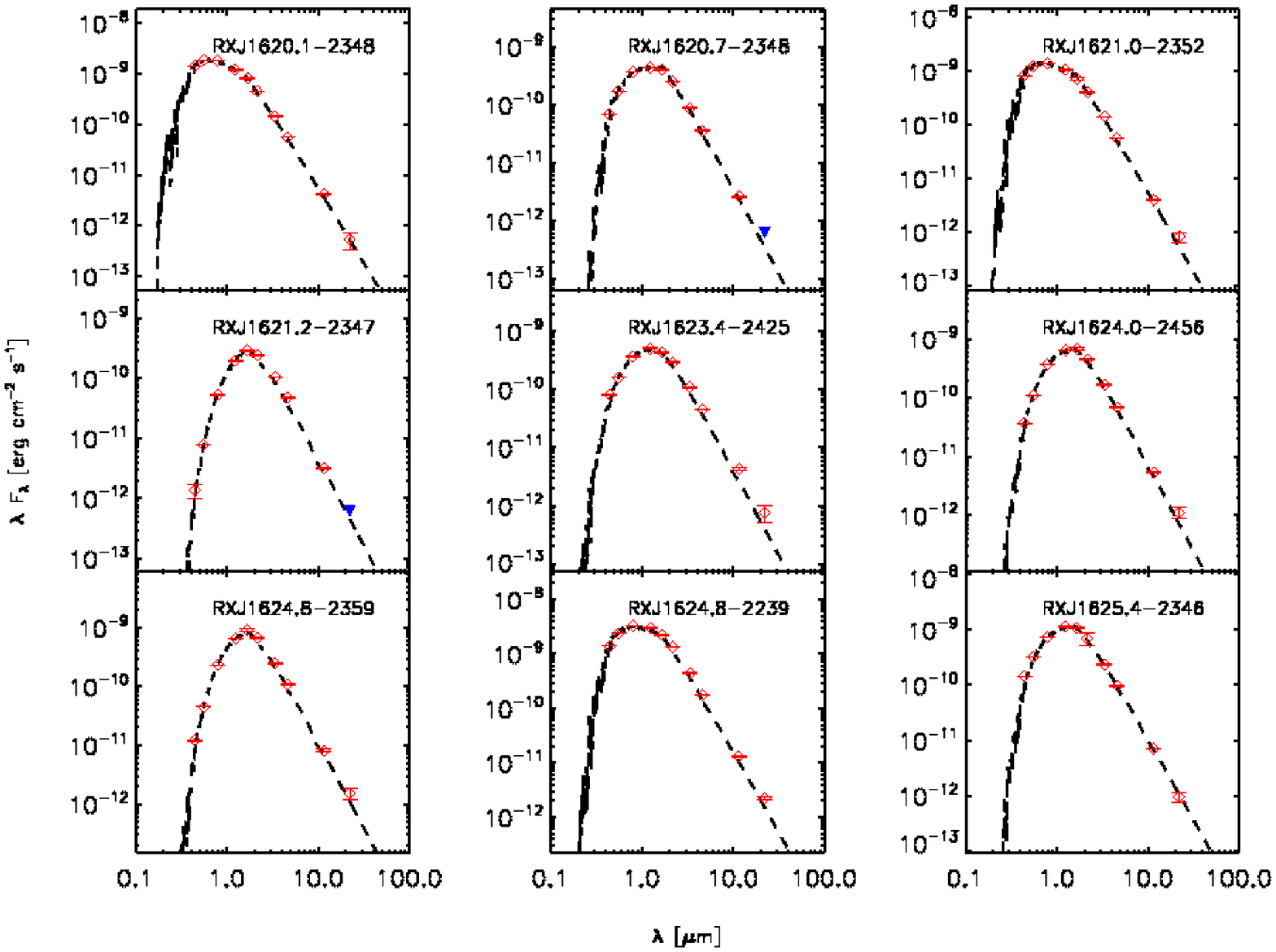,width=150mm, angle=90}
\caption{{\em Idem} to Figure~\ref{F_juicyseds}, however we now plot 
SEDs of Rho Ophiuchus candidates that are consistent with bare 
photospheres.}
\label{F_bareSEDsRhoOph}
\end{figure*}

For completeness, we plot the SEDs of those Chamaeleon and 
Rho Ophiuchus stars that are consistent with WTTSs in 
Figures~\ref{F_bareSEDsCham}~\&~\ref{F_bareSEDsRhoOph}.

\section{Ancillary Infrared Photometric Data}\label{IRdata}

In constructing our SEDs, we exploited mid-infrared and sub-mm photometric 
datasets for the Chamaeleon and Rho Ophiuchus stars obtained using the 
{\sc akari}, Herschel, {\sc iram}, {\sc scuba}, Spitzer, {\sc vla} and {\sc wise} facilities; 
all data are detailed and referenced in Tables~\ref{T_phot1}~\&~\ref{T_phot2}.

\begin{table*}
\vspace{3mm}
\caption[]{\protect \small WISE data for our Chamaeleon and Rho Ophiuchus candidates \citep{Cutri:2012}.}
\begin{tabular}{lccccc}
\hline
~~~~~~~~Target      & WISE 3.4$\mu$m     & WISE 4.6$\mu$m    &  WISE 12$\mu$m    & WISE 22$\mu$m         & Error \\
                    &    [mag]           & [mag]             &     [mag]         & [mag]                 & Code  \\
\hline 
\multicolumn{6}{l}{\textbf{Chamaeleon}} \\ \hline
  RXJ0850.1-7554    &   8.646$\pm$0.022  &  8.684$\pm$0.020  &  8.588$\pm$0.021  &  8.582$\pm$0.227      & AAAB \\
  RXJ0951.9-7901    &   7.995$\pm$0.196  &  7.986$\pm$0.021  &  7.921$\pm$0.019  &  7.833$\pm$0.144      & BAAB \\  
  RXJ1112.7-7637    &   6.611$\pm$0.053  &  6.200$\pm$0.023  &  4.536$\pm$0.014  &  3.365$\pm$0.020      & AAAA \\  
  RXJ1129.2-7546    &   8.815$\pm$0.092  &  8.766$\pm$0.021  &  8.549$\pm$0.022  &  7.476$\pm$0.116      & AAAB \\  
  RXJ1140.3-8321    &   8.509$\pm$0.023  &  8.540$\pm$0.020  &  8.442$\pm$0.022  &  8.371$\pm$0.209      & AAAB \\  
  RXJ1158.5-7754a   &   6.946$\pm$0.024  &  7.166$\pm$0.019  &  7.169$\pm$0.016  &  7.095$\pm$0.092      & AAAA \\  
  RXJ1159.7-7601    &   8.168$\pm$0.023  &  8.198$\pm$0.020  &  8.099$\pm$0.019  &  7.894$\pm$0.155      & AAAB \\  
  RXJ1201.7-7859    &   6.823$\pm$0.060  &  6.792$\pm$0.022  &  6.785$\pm$0.017  &  6.677$\pm$0.065      & AAAA \\  
  RXJ1233.5-7523    &   7.666$\pm$0.027  &  7.738$\pm$0.021  &  7.695$\pm$0.017  &  7.641$\pm$0.143      & AAAB \\  
  RXJ1239.4-7502    &   7.725$\pm$0.026  &  7.753$\pm$0.021  &  7.702$\pm$0.017  &  7.400$\pm$0.092      & AAAA \\ \hline
\multicolumn{6}{l}{\textbf{Rho Ophiuchus}} \\ \hline
  RXJ1620.1-2348    &   8.230$\pm$0.023  &  8.255$\pm$0.021  &  8.241$\pm$0.032  &  7.850$\pm$ ...~~~~~  & AAAU \\
  RXJ1620.7-2348    &   8.786$\pm$0.021  &  8.769$\pm$0.019  &  8.750$\pm$0.042  &  7.953$\pm$ ...~~~~~  & AAAU \\
  RXJ1621.0-2352    &   8.256$\pm$0.024  &  8.263$\pm$0.019  &  8.313$\pm$0.028  &  8.024$\pm$0.279      & AAAB \\
  RXJ1621.2-2347    &   8.555$\pm$0.023  &  8.412$\pm$0.020  &  8.524$\pm$0.033  &  8.506$\pm$ ...~~~~~  & AAAU \\
  RXJ1623.1-2300    &   8.056$\pm$0.024  &  8.036$\pm$0.020  &  7.710$\pm$0.081  &  7.100$\pm$0.120      & AAAB \\
  RXJ1623.4-2425    &   8.528$\pm$0.024  &  8.497$\pm$0.021  &  8.488$\pm$0.042  &  7.878$\pm$ ...~~~~~  & AAAU \\
  RXJ1623.5-2523    &   7.509$\pm$0.030  &  7.509$\pm$0.020  &  7.396$\pm$0.023  &  6.941$\pm$0.161      & AAAB \\
  RXJ1624.0-2456    &   8.073$\pm$0.025  &  8.018$\pm$0.021  &  7.966$\pm$0.021  &  7.354$\pm$0.168      & AAAB \\
  RXJ1624.8-2239    &   6.999$\pm$0.054  &  6.997$\pm$0.020  &  6.983$\pm$0.020  &  6.882$\pm$0.091      & AAAA \\
  RXJ1624.8-2359    &   7.610$\pm$0.030  &  7.533$\pm$0.019  &  7.431$\pm$0.053  &  7.248$\pm$0.156      & AAAB \\
  RXJ1625.0-2508    &   7.846$\pm$0.026  &  7.816$\pm$0.021  &  7.784$\pm$0.030  &  6.708$\pm$0.105      & AAAA \\
  RXJ1625.4-2346    &   7.701$\pm$0.029  &  7.661$\pm$0.020  &  7.690$\pm$0.022  &  8.125$\pm$0.323      & AAAB \\
  RXJ1625.6-2613    &   6.812$\pm$0.072  &  6.493$\pm$0.022  &  4.840$\pm$0.015  &  2.668$\pm$0.024      & AAAA \\
  ROXR1 13          &   5.840$\pm$0.169  &  5.678$\pm$0.070  &  4.431$\pm$0.043  &  1.562$\pm$0.054      & BAAA \\
  RXJ1627.1-2419    &   6.102$\pm$0.083  &  5.499$\pm$0.044  &  2.925$\pm$0.013  & -0.669$\pm$0.010      & AAAA \\
\hline
\end{tabular}
\label{T_phot1}
\end{table*}

{\scriptsize

\begin{landscape}
\begin{table}
\caption[]{\protect \small Supplemental, longer-wavelength photometry for five objects with IR excesses. }
\begin{tabular}{lc|lc|lc|lc|lc}
\hline
Band  & Flux  & Band & Flux  & Band & Flux  & Band & Flux  & Band & Flux  \\
      & [mJy] &      & [mJy] &      & [mJy] &      & [mJy] &      & [mJy] \\  
\hline
\multicolumn{2}{c}{\textbf{ROXR1 13}} &  \multicolumn{2}{c}{\textbf{RXJ1627.1-2419}} & \multicolumn{2}{c}{\textbf{RXJ1625.6-2613}} & \multicolumn{2}{c}{\textbf{RXJ1112.7-7637}} & \multicolumn{2}{c}{\textbf{RXJ1129.2-7546}}      \\
IRAC 3.4$\mu$m	           &   1300$\pm$93     &   IRAS 12$\mu$m 	      &  2270$\pm$136.2  &  IRAS 12$\mu$m 	 &    447$\pm$44.7    &   AKARI 9$\mu$m	         &    552$\pm$8.95   &  IRAC 3.4$\mu$m  &  8.78$\pm$0.02  \\
IRAC 4.5$\mu$m	           &    880$\pm$79     &   IRAS 25$\mu$m 	      & 19800$\pm$1188   &  AKARI 18$\mu$m 	 &  936.1$\pm$18      &   AKARI 18$\mu$m 	 &  348.5$\pm$14.2   &  IRAC 4.5$\mu$m  &  8.76$\pm$0.02  \\
IRAC 5.8$\mu$m	           &    740$\pm$74     &   IRAS 60$\mu$m 	      & 33800$\pm$3718   &  IRAS 25$\mu$m 	 &    741$\pm$96.33   &   IRAS 12$\mu$m 	 &    491$\pm$34.37  &  IRAC 5.8$\mu$m  &  8.79$\pm$0.03  \\
IRAC 8.0$\mu$m	           &    690$\pm$73     &   IRAS 100$\mu$m 	      &   166000 (UL)    &  IRAS 60$\mu$m 	 &    982$\pm$137.48  &   IRAS 25$\mu$m 	 &    321$\pm$22.47  &  IRAC 8.0$\mu$m  &  8.72$\pm$0.04  \\
AKARI 9$\mu$m	           &   1347$\pm$244    &   AKARI 18$\mu$m	      &  9460$\pm$108    &  AKARI 65$\mu$m 	 &     777.4 (UL)     &   IRAS 60$\mu$m 	 &      400 (UL)     &  MIPS 24$\mu$m   &  8.27$\pm$0.2   \\
MIPS 24$\mu$m	           &   1800$\pm$200    &   AKARI 65$\mu$m	      &    23080 (UL)    &  AKARI 90$\mu$m 	 &   1010$\pm$332     &   IRAS 100$\mu$m 	 &      458 (UL)     &  \nophot         &   \nophot       \\
MIPS 70$\mu$m	           &  12000$\pm$6000   &   AKARI 90$\mu$m	      &    28210 (UL)    &  IRAS 100$\mu$m 	 &     13900 (UL)     &   IRAC 5.8$\mu$m 	 &   7.28$\pm$0.02   &  \nophot         &   \nophot       \\
PACS 70$\mu$m              &   2150$\pm$258    &   AKARI 140$\mu$m	      & 18640$\pm$4750   &  AKARI 140$\mu$m      &     809.8 (UL)     &   IRAC 8.0$\mu$m 	 &   6.72$\pm$0.02   &  \nophot         &   \nophot       \\
PACS 160$\mu$m             &   2310$\pm$291    &   AKARI 160$\mu$m	      & 24630$\pm$9330   &  AKARI 160$\mu$m      &      1849 (UL)     &   MIPS 24$\mu$m 	 &   6.29$\pm$0.03   &  \nophot         &   \nophot       \\
SPIRE 250$\mu$m    	   &     1500 (UL)     &   SCUBA 350$\mu$m	      &  2761$\pm$57     &  \nophot              &   \nophot          &   \nophot        	 &   5.30$\pm$0.04   &  \nophot         &   \nophot       \\
SPIRE 350$\mu$m            &      900 (UL)     &   SCUBA 450$\mu$m	      &  1896$\pm$268    &  \nophot              &   \nophot          &   \nophot	         &    3.4$\pm$0.04   &  \nophot         &   \nophot       \\
SPIRE 500$\mu$m            &     1000 (UL)     &   SCUBA 850$\mu$m	      &   397$\pm$6      &  \nophot              &   \nophot          &   \nophot                &   \nophot         &  \nophot         &   \nophot       \\
IRAM 250 GHz (1200$\mu$m)  &     21.0 (UL)     &   IRAM 230 GHz (1.3mm)       &    95$\pm$15     &  \nophot              &   \nophot          &   \nophot                &   \nophot         &  \nophot         &   \nophot       \\
VLA 6cm                    & 12.404$\pm$0.043  &   \nophot                    &    \nophot       &  \nophot              &   \nophot          &   \nophot                &   \nophot         &  \nophot         &   \nophot       \\
VLA 1.4 GHZ (2.1CM)        &      6.1 (UL)     &   \nophot                    &    \nophot       &  \nophot              &   \nophot          &   \nophot                &   \nophot         &  \nophot         &   \nophot       \\
\hline
\end{tabular}\label{T_phot2}
\begin{flushleft}
Notes: \\
AKARI infrared camera all-sky survey 9$\mu$m and 18$\mu$m data from \citet{Ishihara:2010}.\\
90$\mu$m, 140$\mu$m, and 160$\mu$m from \citet{Yamamura:2010}. \\
Spitzer IRAC and MIPS data from \citet{Luhman:2008}. \\
Herschel PACS and SPIRE data from \citet{Cieza:2013}.\\
SCUBA data from \citet{Andrews:2007}. \\
250MHz data from \citet{Altenhoff:1994}, 
1.3mm data from \citet{Motte:1998}.\\
1.4GHz data from \citet{Condon:1998}.\\
6-cm data from \citet{Gagne:2004}.
\end{flushleft}

\end{table}
\end{landscape}
}

\bsp
\label{lastpage}
\end{document}